\documentclass[a4paper,12pt]{article}

\usepackage{verbatim}
\usepackage{amsfonts}
\usepackage{mathrsfs}
\usepackage{amsmath}
\usepackage{amssymb}
\usepackage{framed}
\usepackage[medium]{titlesec}
\usepackage{bm}
\usepackage{cite}
\usepackage[normalem]{ulem}
\usepackage{extarrows}
\usepackage{slashed}
\usepackage{isodateo}
\usepackage{graphicx}
\usepackage{xcolor}
\usepackage[bookmarksnumbered=true,bookmarksopen=true,hyperfootnotes=true]{hyperref}
\hypersetup{colorlinks,%
	linkcolor=[rgb]{0,0.3,0.6}, %
	citecolor=[rgb]{0,0.3,0.6}, %
	urlcolor=[rgb]{0,0.3,0.6}}
\usepackage[hmargin=.7in,vmargin=1.1in]{geometry}
\usepackage{indentfirst}
\usepackage{booktabs}

\usepackage{enumerate}
\usepackage{amsfonts}
\usepackage{mathrsfs}
\usepackage{amsmath}
\usepackage{amssymb}
\usepackage{float}
\usepackage{bm}
\usepackage{slashed}
\usepackage[normalem]{ulem}
\usepackage{extarrows}
\usepackage{slashed}
\usepackage{isodateo}
\usepackage{graphicx}
\usepackage{xcolor}
\usepackage{cancel}

\definecolor{bblue}{rgb}{0,0.2,0.6}

\usepackage{indentfirst}

\graphicspath{{fig/}}

\usepackage{tikz}
\usetikzlibrary{arrows,shapes}
\usetikzlibrary{trees}
\usetikzlibrary{matrix,arrows} 				
\usetikzlibrary{positioning}				
\usetikzlibrary{calc,through}				
\usetikzlibrary{decorations.pathreplacing}  
\usepackage{pgffor}							

\usetikzlibrary{decorations.pathmorphing}	
\usetikzlibrary{decorations.markings}
\tikzset{
	photon/.style={decorate, decoration={snake}, draw=red},
	electron/.style={draw=blue, postaction={decorate},
		decoration={markings,mark=at position .55 with {\arrow[draw=blue]{>}}}},
	gluon/.style={decorate, draw=black,
		decoration={coil,amplitude=4pt, segment length=4pt}} ,
	vector/.style={decorate, decoration={snake}, draw},
	provector/.style={decorate, decoration={snake,amplitude=2.5pt}, draw},
	antivector/.style={decorate, decoration={snake,amplitude=-2.5pt}, draw},
	fermion/.style={draw=black, postaction={decorate},
		decoration={markings,mark=at position .55 with {\arrow[draw=black]{>}}}},
	fermionbar/.style={draw=black, postaction={decorate},
		decoration={markings,mark=at position .55 with {\arrow[draw=black]{<}}}},
	fermionnoarrow/.style={draw=black},
	fermionnoarrowsoft/.style={draw=blue},
	scalar/.style={dashed,draw=black, postaction={decorate},
		decoration={markings,mark=at position .55 with {\arrow[draw=black]{>}}}},
	scalarbar/.style={dashed,draw=black, postaction={decorate},
		decoration={markings,mark=at position .55 with {\arrow[draw=black]{<}}}},
	scalarnoarrow/.style={dashed,draw=black},
	scalarnoarrowsoft/.style={dashed,draw=blue},
	electron/.style={draw=black, postaction={decorate},
		decoration={markings,mark=at position .55 with {\arrow[draw=black]{>}}}},
	bigvector/.style={decorate, decoration={snake,amplitude=4pt}, draw},
}

\tikzstyle{block} = [draw, rectangle, 
minimum height=3em, minimum width=6em]
\usetikzlibrary{arrows,patterns}

\newcommand{\email}[1]{\footnote{Email: \href{mailto:#1}{\nolinkurl{#1}}}}

\def\ma{\mathcal}

\def\be{\begin{equation}}
	\def\ee{\end{equation}}

\def\bar{\overline}

\def\bgm{\begin{matrix}}
	\def\edm{\end{matrix}}

\newcommand{\te}{\text}
\usepackage{mdframed}

\newmdenv[skipabove=0pt,%
skipbelow=5pt,%
leftmargin=0pt,%
rightmargin=0pt,%
innertopmargin=-5pt,%
innerbottommargin=7pt,%
innerleftmargin=2pt,%
innerrightmargin=2pt,%
splittopskip=0pt,%
splitbottomskip=0pt,%
linewidth=0pt,%
nobreak=true]%
{keyeqn2}

\newmdenv[backgroundcolor=gray!15,%
skipabove=0pt,%
skipbelow=5pt,%
leftmargin=0pt,%
rightmargin=0pt,%
innertopmargin=-5pt,%
innerbottommargin=7pt,%
innerleftmargin=2pt,%
innerrightmargin=2pt,%
splittopskip=0pt,%
splitbottomskip=0pt,%
linewidth=0pt,%
nobreak=true]%
{keyeqn}

\usepackage{titlesec}

\begin{document}
	\title{\Large\textbf{
			Hamiltonian formulation of the RS model and field mixing		
			}\\[2mm]}
	
	\author{Qi Chen$^{1,\,}$\email{chenq20@mails.tsinghua.edu.cn}~~~
		Kaixun Tu$^{1,\,}$\email{tkx19@mails.tsinghua.edu.cn}~~~
            Qing Wang$^{1,\,2,\,}$\email{wangq@mail.tsinghua.edu.cn}\\[5mm]
		\normalsize{${}^{1}\,$\emph{Department of Physics, Tsinghua University, Beijing 100084, China}}\\
            \normalsize{${}^{2}\,$\emph{Center for High Energy Physics, Tsinghua University, Beijing 100084, China}}}
	
	\date{}
	\vspace{20mm}
	\maketitle
	\begin{abstract}
		\vspace{10mm}	
While neutrino oscillations have led to attention and research on field mixing arising from quadratic interactions, the field mixing inherent in clothed particles is more fundamental, serving as a significant source of complexity and non-perturbative challenges in quantum field theory.
We present an example of an analytical solution for field mixing involving a three-point interaction between a bosonic field and a fermionic field. 
Specifically, we study the Rothe-Stamatescu (RS) model and utilize lattice regularization to provide a well-defined Hamiltonian which is absent in the original continuous RS model.
Due to the complexity introduced by three-point interactions compared to quadratic interactions, the Fock representation commonly used in discussions of field mixing does not work well; instead, we define a representation based on real space to investigate the physical vacuum and clothed particles.
These eigenstates not only reveal the field mixing between the bosonic and fermionic fields but also allow us to directly observe the spatial entanglement structure.
	\end{abstract}
	\newpage
	\tableofcontents
	
	\newpage

	\section{Introduction}

Traditional perturbative quantum field theory fails in non-perturbative regimes.
To address this issue, lattice quantum field theories have been proposed as a means to regulate quantum field theory before encountering any ill-defined formal calculations. 
These theories have well-defined path integrals or Hamiltonians \cite{Wilson1974,Kogut1975,Susskind1977,Kogut1983,Kogut1979,DeGrand2006,Gattringer2009,Magnifico2021,Davoudi2021,Zohar2022}. While correlation functions in the path integral formulation of a lattice theory can be numerically computed using computers, the calculations are performed in Euclidean spaces after a Wick rotation. This choice limits our ability to directly observe real-time dynamics, such as time-dependent processes like string breaking phenomena, and also introduces challenges associated with the sign problem \cite{Troyer2005,Zohar2022,Fitzpatrick2022}.
The Hamiltonian formulation of a lattice theory, on the other hand, provides a well-defined Hamiltonian and allows for explicit discussions about quantum states, entanglement structure, and time evolution \cite{Davoudi2021,Magnifico2021,Florio2023}. 
In addition, Hamiltonian simulations of relativistic lattice field theories, based on the tool of Hamiltonian lattice field theory, have recently garnered significant attention \cite{Bauer2023}.


When it comes to studying the Hamiltonian formulation of a theory, we immediately encounter two major issues. The first is defining the quantum version of the Hamiltonian, and the second is finding the eigenstates (and eigenvalues) of the Hamiltonian.
Due to the complexity of dealing directly with the full 3+1-dimensional Standard Model, researchers often first consider simpler models in lower dimensions, such as the Schwinger model.    
The Schwinger model which is QED in 1 + 1 dimensions, was initially proposed by Schwinger in 1962\cite{Schwinger1962}. Subsequently, in 1971, Lowenstein and Swieca defined and solved the equations of motion for the Schwinger model\cite{Lowenstein1971}.
%
%
Following this development, the Hamiltonian version of the Schwinger model emerged \cite{Manton1985, Hetrick1988, Link1990, Iso1990}. The energy spectrum and eigenstates of the quantum Hamiltonian were also determined, enabling in-depth exploration of the non-perturbative effects. These studies often utilized techniques such as heat kernel regularization or $\zeta$-function regularization. On the other hand, lattice regularization was employed to construct the Hamiltonian of lattice Schwinger model \cite{Banks1976, Ambjoern1983, Nason1985, Melnikov2000}.
%
%
%
%
 Recently, it became popular to employ lattice regularization based on Kogut-Susskind staggered fermions \cite{Kogut1975,Banks1976,Susskind1977}
 to explore aspects such as the structure of the vacuum state, quark confinement, energy spectra, entanglement structure, gauge symmetries, topology, and real-time dynamics \cite{Byrnes2002,Banuls2013,Hebenstreit2014,Buyens2016,Pichler2016,Buyens2017,Zapp2017,PintoBarros2019,Funcke2020,Butt2020,Rigobello2021,Ikeda2023,Florio2023}.

  However, in the Schwinger model, electromagnetic waves are absent due to Gauss’ law, leaving only a fermion field.
  As for the quadratic interaction between two fields, it has been thoroughly investigated in studies related to neutrino oscillations.
 Considering two different flavor neutrinos, $\nu_e$ and $\nu_\mu$, the Lagrangian for fermion mixing \cite{Blasone1995} is given by    
$$\mathcal{L}_f = \bar\nu_e(i\gamma^\alpha\partial_\alpha-m_e)\nu_e + \bar\nu_\mu(i\gamma^\alpha\partial_\alpha-m_\mu)\nu_\mu - m_{e\mu}(\bar\nu_e\nu_\mu+\bar\nu_\mu\nu_e)\; .$$    
Due to the interaction term $\mathcal{L}_I = -m_{e\mu}(\bar\nu_e\nu_\mu+\bar\nu_\mu\nu_e)$, the creation operator corresponding to an individual flavor field cannot annihilate the vacuum, nor can it generate the eigenstates of the Hamiltonian from the vacuum.
 The creation and annihilation operators for mass eigenstates are a combination of the original flavor creation and annihilation operators.    
In addition to fermion mixing, boson mixing has also been studied\cite{Binger1999}. The Lagrangian for boson mixing is given by    
$$L_b = L_{0,\alpha} + L_{0,\beta} - \lambda(\phi^\dagger_\alpha\phi_\beta+\phi^\dagger_\beta\phi_\alpha)\; ,$$    
where $L_{0,\alpha(\beta)}$ represents the free Lagrangian for the bosonic fields, and the interaction term $L_I = -\lambda(\phi^\dagger_\alpha\phi_\beta+\phi^\dagger_\beta\phi_\alpha)$ leads to mixing of the bosonic fields. 
The non-perturbative vacuum states, non-perturbative effects due to field mixing, and the entanglement resulting from field mixing have all been extensively researched and discussed in the context of both fermion and boson mixing \cite{Blasone1999,Ji2001,Blasone2001,Ji2002,Giunti2005,Blasone2014,Blasone2019,Blasone2021,Smaldone2021}.

   It is worth noting that the previously mentioned $\mathcal{L}_I$ and $L_I$ are simple quadratic interactions involving fields of the same type. However, in general cases, interactions can be of higher order and can involve the coupling between fermionic and bosonic fields, which leads to more complex field mixing.
   Pauli and Fierz introduced a transformation to the fundamental equations of nonrelativistic QED (where the bare electron is described by $\frac{1}{2m}\left(\boldsymbol{p}-\frac{e}{c}\boldsymbol{A}\right)^2$), effectively replacing the electron with its own field plus the electron itself \cite{Pauli1938}. This transformed entity came to be known as the ``dressed electron".
   Ref. \cite{Blum2016} describes how the concept of the dressed electron inspired the birth of renormalization and its significant implications for condensed matter physics.
  With the development of quantum field theory, a concept similar to the dressed electron emerged in relativistic QED known as a ``dressed state", which describes a charged particle ``dressed" with an infrared ``cloud of soft photons" \cite{Chung1965, Kulish1970, Duch2021, Dybalski2017}.  
   In QED, the interaction does not fall off asymptotically, which requires choosing dressed states involving the mixing of bosonic and fermionic fields as in- and out-states for the S-matrices, ensuring the absence of infrared divergences.   
   Additionally, a concept similar to the dressed state is the ``clothed particle".
In quantum field theory, a one-particle state is an eigenstate of the Hamiltonian, and it is referred to as a ``clothed particle" to distinguish it from a bare particle \cite{Greenberg1958,Shebeko2001,Kostylenko2023}. 
The concept of clothed particles demonstrates a more general and fundamental field mixing, which can reflect the process of renormalization non-perturbatively.
Due to the interaction between the fermionic field and bosonic field in the Hamiltonian, the excitation of the one-fermion state involves both the fermionic field and the bosonic field, rather than just the fermionic field alone. The same applies to one-boson state.  
 Furthermore, the physical vacuum encompasses entanglement between various interacting fields. 

For the previously discussed case of quadratic interactions ($\mathcal{L}_I$ and $L_I$), there is a significant difference between the physical vacuum and the bare vacuum, and the unitary inequivalence of the Fock space of base (unmixed) eigenstates and the physical mixed eigenstates has been demonstrated.
The specific structure of the physical vacuum for quadratic interactions has been precisely determined, and the condensate structure of the physical vacuum can lead to non-perturbative effects. For example, approximating the physical vacuum as the bare vacuum results in different neutrino oscillation formulas compared to non-perturbative oscillation formulas.
As for the three-point interactions which are more general interactions, Ref. \cite{Greenberg1958} solves the Hamiltonians of three solvable models to discuss issues related to clothed particles. However, in these models, the free part of the fermionic field is oversimplified to $H_0=m\int \mathrm dp\;b^\dagger_p b_p$ instead of the relativistic form $H_0=\int \mathrm dp\;E_p b^\dagger_p b_p$. These models are non-relativistic, and due to the absence of pair effects, the bare vacuum is equivalent to the physical vacuum, and bare bosons are equivalent to clothed bosons.
Ref. \cite{Dybalski2017} discusses dressed states in a Hamiltonian formulation, where the fermionic field is described by the non-relativistic expression $H_0=\int \mathrm dp\; \frac{p^2}{2m}b^\dagger_p b_p$, and the bare vacuum is equivalent to the physical vacuum.
Therefore, the model we aim to study is the Hamiltonian formulation of a relativistic theory involving a three-point interaction between bosonic and fermionic fields, where the physical vacuum and clothed particles can be solved in a non-perturbative manner to reveal the mixing structure of bosonic and fermionic fields within the eigenstates.

In this paper, we choose the RS model to investigate its eigenstates, explicitly demonstrating the mixing of fermionic and bosonic fields.
The RS model is a solvable (1+1) dimensional model introduced by Rothe and Stamatescu \cite{Rothe1975}. In the original variables of the RS model, there is a bosonic field $\phi_0$ with mass $m_0$ and a massless fermionic field $\Psi_0$ that interact through the term $\Delta\mathcal{L}=-g_0\partial_u\phi_0 \bar\Psi_0 \gamma^5\gamma^\mu\Psi_0$ in the Lagrangian.
 Rothe and Stamatescu regularized the equations of motion for RS model and obtained the correlation functions.       
Later, Ref. \cite{Martinovic2014} propose the possibility of describing the RS model using a Hamiltonian framework. However, it only provides a classical Hamiltonian without the ability to demonstrate the renormalization process and compute energy eigenvalues or eigenstates.
As mentioned earlier, it is currently popular to use lattice regularization with Kogut-Susskind staggered fermions to construct the Hamiltonian formulation of the Schwinger model\cite{Byrnes2002,Banuls2013,Buyens2016,Pichler2016,Buyens2017,Zapp2017,PintoBarros2019,Funcke2020,Butt2020}, which can reveal the real space structure of quantum states and avoid the fermion doubling problem (in 1+1 dimensions). Therefore, we adopt the same regularization method to deal with the RS model.
We provide the Hamiltonian of lattice RS model with staggered fermions, solve the operator equations of motion for the Hamiltonian, and compare them with the original RS model. We also derive the correlation functions and demonstrate that the correlation functions of the lattice RS model can recover those of the original RS model in the continuum limit. This confirms that the lattice RS model presented in this paper is indeed equivalent to the original RS model in the continuum limit.
Since we are dealing with a relativistic theory with a three-point interaction, using the traditional Fock representation would make the representation of quantum states excessively complex, obscuring the structure of the quantum states. Therefore, we introduce a representation based on real space to represent the physical vacuum and clothed particles. This representation not only directly reveals how the bosonic field degrees of freedom mix with the fermionic field degrees of freedom but also illustrates the spatial entanglement structure.

This paper is organized as follows. In Sec. \ref{2}, we provide a brief introduction to the original RS model and some fundamental results related to it using our notation system. 
In Sec. \ref{3}, we present the Hamiltonian of the lattice RS model.
In Sec. \ref{4}, we employ new field variables to decouple the bosonic and fermionic parts of the lattice Hamiltonian. We then derive the eigenstates of the bosonic part in a representation expanded by the eigenstates of new field operator, while the eigenstates of the fermionic part are expressed in a specially defined representation.
In Sec. \ref{5}, we derive the correlation functions for the lattice RS theory and define renormalized fields, masses, and coupling constants. Notably, the field-strength renormalization constant of the fermionic field tends to zero in the continuum limit. We also show that the lattice correlation functions approach those of the original RS model in the continuum limit.
In Sec. \ref{6}, we introduce a representation corresponding to the original field variables and express the physical vacuum and clothed particles in this representation. 
This not only reveals the entanglement between the bosonic and fermionic components of these eigenstates but also allows us to directly observe the spatial entanglement structure of quantum states.
In Appendix \ref{eqomB}, we derive the equations of motion for the bosonic field in the lattice RS model and compare them with those of the original RS model. Similarly, in Appendix \ref{eqomF}, we derive the equations of motion for the fermionic field in the lattice RS model and compare them to the original RS model.
Section \ref{7} concludes the paper with a summary and some discussions, along with an outlook on this work.

\section{A Brief Review of the RS Model}
\label{2}
In this section, we offer a concise overview of the RS model, as proposed in the work by Rothe and Stamatescu\cite{Rothe1975}. 
We will refer to this continuous RS model as the ``original RS model" in the following to distinguish it from the lattice RS model introduced later.
We present the main results using our notation system, and for more detailed derivations and conclusions, please refer to the Ref. \cite{Rothe1975}.


 The Lagrangian of RS model can be written as
\begin{equation}\label{eq:lag}	
	\ma{L}=i\bar\Psi_0 \gamma^\mu\partial_\mu\Psi_0+\frac{1}{2}\partial_\mu\phi_0\partial^\mu\phi_0-\frac{1}{2}m_0^2\phi_0^2-g_0 \bar\Psi_0 \gamma^5\gamma^\mu\Psi_0\partial_{\mu}\phi_0\; .
\end{equation}
In the given expression, $\phi_0$ stands for the bare bosonic field, characterized by its bare mass denoted as $m_0$. The bare fermionic field is identified as $\Psi_0$, while the bare coupling between the bosonic and fermionic fields is represented by $g_0$. Additionally, the $\gamma$ matrices used here are taken as
\begin{equation}\label{gamma0}
		\begin{split}	
			\gamma^0=\gamma_0=\left[
			\begin{array}{ccc}
				0 & 1 \\
				1 & 0 
			\end{array}
			\right]
			\quad,\quad
			\gamma^1=-\gamma_1=\left[
			\begin{array}{ccc}
				0 & -1 \\
				1 & 0 
			\end{array}
			\right]
			\quad,\quad
			\gamma^5=\gamma^0\gamma^1=\left[
			\begin{array}{ccc}
				1 & 0 \\
				0 & -1 
			\end{array}
			\right]
			\; .
		\end{split}
	\end{equation}
It is imperative to emphasize that the Lagrangian \eqref{eq:lag} is only a formal representation, the precise formulation of the theory necessitates a suitable regularization.
Later on, we will perform regularization at the level of Hamiltonian, while the regularization is carried out at the level of equations of motion in the original RS model\cite{Rothe1975}.
The regulated equation of motion governing the bosonic field is expressed as
\begin{equation}\label{eq:phi0}
	(\partial_\mu\partial^\mu+m_0^2)\phi_0(x)=g_0\partial_\mu\lim\limits_{\epsilon\to0}\left\{\bar\Psi_0(x+\epsilon)\gamma^5\gamma^\mu\Psi_0(x)-\frac{g_0}{\pi}\epsilon^{\mu\lambda}\epsilon^{\nu\rho}\frac{\epsilon_\lambda\epsilon_\rho}{\epsilon^2}\partial_\nu\phi_0(x)\right\}\; ,
\end{equation}
where $\epsilon^{\mu}$ is a spacelike vector satisfying $\epsilon^2<0$ and $\epsilon^{\mu\nu}$ is the Levi-Civita tensor in 2d. The regularized fermionic field equation of motion can be written as
\begin{equation}\label{eq:psir}		i\gamma^\mu\partial_\mu\Psi_r(x)=g_r\lim\limits_{\epsilon\to0}\gamma^5\left[\gamma^\mu\Psi_r(x)\partial_\mu\phi_r(x-\epsilon)-i\frac{g_r}{2\pi}\frac{\slashed\epsilon}{\epsilon^2}\Psi_r(x)	\right]\;,
\end{equation}
where we introduce the renormalized fermionic field $\Psi_r$, the renormalized bosonic field $\phi_r$, and the renormalized coupling constant $g_r$, which are connected to the bare parameters as follows:
\begin{eqnarray}
	\phi_r&=& (1-\frac{g_0^2}{\pi})^{\frac{1}{2}}\phi_0\\
	\Psi_r(x)&=& \lim\limits_{\epsilon\to0}\exp\left(\frac{1}{2}g_0^2\langle\Omega|\phi_0(\epsilon)\phi_0(0)|\Omega\rangle\right)\Psi_0(x)\label{czh0}\\
	g_r&=&(1-\frac{g_0^2}{\pi})^{-\frac{1}{2}}g_0
\end{eqnarray}

The regulated equations of motion \eqref{eq:phi0} and \eqref{eq:psir} establish the precise framework for the RS model. Through the utilization of \eqref{eq:phi0} and \eqref{eq:psir}, we gain the ability to compute various observables, including the correlation functions. 
The two-point correlation function for the bosonic field is expressed as
\begin{equation}\label{eq:2ptboson}
	\langle\Omega|\te{T}\{\phi_r(x_1,t_1)\phi_r(x_2,t_2)\}|\Omega\rangle=\int \mathrm dq \mathrm d\omega_q \;\frac{i}{\omega_q^2-q^2-m_r^2+i\epsilon }{\mathrm e}^{-i\omega_q(t_1-t_2) +i(x_1-x_2)q} \; ,
\end{equation}
where we introduce the renormalized mass parameter as $m_r=(1-g_0^2/\pi)^{-1/2}m_0$. The two-point correlation for the fermionic field can be formulated as%
\begin{equation}\label{eq:2ptfermion}
	\langle\Omega|\Psi_r(x)\bar\Psi_r(y)|\Omega\rangle=-\frac{i}{2\pi}\frac{\gamma_\mu (x-y)^\mu}{(x-y)^2}			\exp\left(g_r^2\langle\Omega|\phi_r(x)\phi_r(y)|\Omega\rangle\right)\;.
\end{equation}
It is crucial to highlight that, for the sake of simplicity and without sacrificing generality, we have assumed a particular time order $x^0 > y^0$. Consequently, we have omitted the time order operator $\te{T}$ in this context. Turning to the interaction, the three-point correlation function can be expressed as follows:
%
\begin{equation}\label{eq:3ptcorrelation}
	\langle\Omega|\phi_r(x)\Psi_r(y)\bar\Psi_r(0)|\Omega\rangle=ig_r\gamma^5\langle\Omega|\Psi_r(y)\bar\Psi_r(0)|\Omega\rangle\Big[\langle\Omega|\phi_r(x)\phi_r(y)|\Omega\rangle-\langle\Omega|\phi_r(x)\phi_r(0)|\Omega\rangle\Big]\; .
\end{equation}
Once again, we omit the time-ordering operator,as we are focusing on a specific time-ordering sequence where $x^0 > y^0 > 0$. 
Please note that, for the sake of brevity, we will represent the operator $\hat A$ as $A$. However, if any ambiguity arises, we will include the operator hat symbol for clarity.

Up to this point, we have provided a collection of results for the original RS model.
In the subsequent sections, we adopt lattice regularization to construct the Hamiltonian formulation of the RS model which was originally defined only at the level of equations of motion.
%

\section{The Lattice RS model}
\label{3}
%
%
%
Before delving into the precise formulation of the lattice Hamiltonian, let's first clarify the lattice configuration. In this study, we adopt a real-time lattice approach, which implies that we refrain from employing Wick rotation, instead focusing solely on discretizing the spatial direction. We partition the one-dimensional spatial component into equidistant lattice points with a spacing of $a$, and at each of these sites, we associate a scalar field operator.
For instance, we represent the scalar field on the $n$-th lattice site as $\phi_n$. These discrete field operators, denoted as $\phi_n$, maintain a relationship with the original continuous field $\phi(x)$ as follows:
\begin{equation}\label{eq:phipi}
	\phi_n=\frac{1}{a}\int_{na-\frac{a}{2}}^{na+\frac{a}{2}}\mathrm dx\;\phi(x)\;.
\end{equation}
Similarly, the discrete conjugate momentum density can be seen as $\frac{1}{a}\int_{na-\frac{a}{2}}^{na+\frac{a}{2}}\mathrm dx\;\pi(x)$. However, defining the discrete conjugate momentum density operator $\pi_n$ in this way leads to non-compliance with the canonical commutation relations between the discrete field $\phi_n$ and $\pi_n$. To address this, we define the operator $\pi_n$ as follows:
\begin{equation}
	\pi_n=\int_{na-\frac{a}{2}}^{na+\frac{a}{2}}\mathrm dx\; \pi(x) \; .
\end{equation}
As a result, the continuous commutation relations can be deduced to their discrete counterpart in the following manner:
\begin{equation}\label{eq:discommutationphiphi}
	\left[\phi(x),\pi(y)\right]=i\delta(x-y)  \qquad\Rightarrow\qquad\left[\phi_n,\pi_m\right]=i\delta_{n,m} \; .
\end{equation}

To ensure that the discrete fermionic fields satisfy the canonical commutation relations, we define the discrete fermionic fields as follows:
%
\begin{equation}\label{eq:psi(n)}
	\Psi_\alpha(n)=\frac{1}{\sqrt{a}}\int_{na-\frac{a}{2}}^{na+\frac{a}{2}}\mathrm dx\; \Psi_{c\alpha}(x)
	\;.
\end{equation}
In this context, the subscript $c$ signifies the continuous field. 
So, based on the commutation relations of the continuous fields, we can derive the commutation relations of the discrete fields:
\begin{equation}\label{eq:discommutationpsipsi}
	\left\{\Psi_{c\alpha}(x),\Psi^\dagger_{c\beta}(y)\right\}=\delta_{\alpha,\beta}\delta(x-y)
	\quad\Rightarrow\quad \left\{\Psi_\alpha(n),\Psi^\dagger_\beta(m)\right\}=\delta_{\alpha,\beta}\delta_{n,m}
	\; .
\end{equation}
And we also have the commutation relations between bosonic and fermionic fields:
\begin{equation}\label{eq:psiphi}
	\left[\psi_\alpha(n),\phi_m\right]=0\; ,
	\qquad
	\left[\psi_\alpha(n),\pi_m\right]=0\; . 
\end{equation}

It's worth noting that in (1+1) dimensions, the fermionic field has only two components. We label the components with $\alpha = u, d$, or more explicitly as
\begin{equation}
		\begin{split}			
			\Psi=\left[
			\begin{array}{ccc}
				\Psi_u\\
				\Psi_d
			\end{array}
			\right]
			\; .
		\end{split}
	\end{equation}	

Having completed all the necessary groundwork, we are now ready to present the lattice Hamiltonian for the RS model:
\begin{equation}\label{eq:Hamiltonian}
	\begin{split}			
		H&=\frac{1}{2}\sum\limits_n \Bigg\{ 
		-i\Big[\Psi_u(n)+\Psi_u(n+1)\Big]^\dagger
		\Big[\Psi_u(n+1)-\Psi_u(n) \Big] \frac{1}{a}\\
		&\qquad\qquad\qquad\qquad\qquad
		+i\Big[\Psi_d(n)+\Psi_d(n+1)\Big]^\dagger
		\Big[\Psi_d(n+1)-\Psi_d(n) \Big] \frac{1}{a}\Bigg\}\\
		&+\frac{1}{2}\sum\limits_n \Bigg\{ 
		i\Big[\Psi_d(n+1)-\Psi_d(n)\Big]^\dagger
		\Big[\Psi_u(n+1)-\Psi_u(n) \Big] \frac{1}{a}\\
		&\qquad\qquad\qquad\qquad\qquad
		+i\Big[\Psi_u(n)-\Psi_u(n+1)\Big]^\dagger
		\Big[\Psi_d(n+1)-\Psi_d(n) \Big] \frac{1}{a}\Bigg\}\\
		&
		+\sum\limits_n a \Bigg[\frac{1}{2F}\left(\frac{\pi_n}{a}\right)^2
		+\frac{F}{2}\left(\frac{\phi_{n+1}-\phi_n}{a}\right)^2
		+\frac{1}{2}m^2\phi_n^2\Bigg]\\
		&
		+\sum\limits_n \frac{1}{2aF}
		\Bigg\{-2g
		\Big[\Psi_u^\dagger(n)\Psi_u(n)-\Psi_d^\dagger(n)\Psi_d(n)\Big]
		\pi_n
		+g^2\Big[\Psi_u^\dagger(n)\Psi_u(n)-	\Psi_d^\dagger(n)\Psi_d(n) \Big]^2
		\Bigg\}\\
		&
		+\sum\limits_n \frac{1}{2a}\Bigg[
		i\Big({\mathrm e}^{ig\phi_{n+1}-ig\phi_n}-1\Big)\Psi^\dagger_u(n+1)\Psi_u(n)  
		+i\Big({\mathrm e}^{ig\phi_{n+1}-ig\phi_n}-1\Big)\Psi^\dagger_d(n)\Psi_d(n+1)
		+\text{h.c.} \Bigg]\\
		&
		+\sum\limits_n \frac{1}{2a} \Bigg[2i\Big({\mathrm e}^{-2ig\phi_n}-1\Big)\Psi^\dagger_d(n)\Psi_u(n)+\text{h.c.} \\
		&\qquad  
		+i\Big({\mathrm e}^{ig\phi_{n+1}+ig\phi_n}-1\Big)\Psi^\dagger_u(n+1)\Psi_d(n) +i\Big({\mathrm e}^{ig\phi_n+ig\phi_{n+1}}-1\Big)\Psi^\dagger_u(n)\Psi_d(n+1)
		+\text{h.c.}\Bigg]\\
		&-E_0\; ,
	\end{split}
\end{equation}
where h.c. denotes the Hermitian conjugate of all terms in the same row, and $E_0$ is introduced to ensure that the energy of ground state  is 0. Furthermore, it will be revealed later through the analysis of correlation functions that if we require the continuum limit of the lattice's bare field $\phi_n$ to be the bare field $\phi_0(x)$ of the original RS model, then the coefficient $F$ takes the value of $F=1-\frac{g^2}{\pi}$.

We require that space forms a ring, and let this circle have $N$ lattice points with each lattice point separated by a distance $a$. Thus, the total length of the circle is $L=Na$. Starting from index 0, we number the lattice points consecutively, such that lattice point $n=0$ and lattice point $n=N$ correspond to the same point. This is equivalent to imposing periodic boundary conditions. Therefore, the summation over $n$ is defined to range from $n=0$ to $n=N-1$, denoted as $\sum\limits_n\equiv \sum\limits_{n=0}^{N-1}$.
Keeping the lattice spacing $a$ constant, as the number of lattice points $N$ approaches infinity, the length of the circle becomes infinitely large, and the boundary conditions can be neglected. This returns our model to the lattice field theory on infinite space. Furthermore, if we let the lattice spacing $a$ approach zero, the lattice field theory eventually becomes a continuous field theory. The process of converting a lattice theory with finite volume (length) into a theory of infinite volume is referred to as taking the ``continuum limit", denoted as the limit of $\lim\limits_{a\to0}\lim\limits_{N\to\infty}$. However, for convenience, we sometimes refer to $\lim\limits_{a\to0}$ as the continuum limit in the subsequent text. It's important to note that before taking the limit $\lim\limits_{a\to0}$, we've already taken the limit $\lim\limits_{N\to\infty}$, so referring to both $\lim\limits_{a\to0}$ and $\lim\limits_{a\to0}\lim\limits_{N\to\infty}$ as the ``continuum limit" in the following text won't cause any confusion.

Although the lattice Hamiltonian \eqref{eq:Hamiltonian} may appear complex, it can be shown that, in the continuum limit, the lattice Hamiltonian transforms into the classical Hamiltonian $H_{cl}$ corresponding to the Lagrangian \eqref{eq:lag} when we treat all the field operators of the lattice Hamiltonian as classical fields and set $F=1,m=m_0,g=g_0$.
Specifically, in the continuum limit, the first four lines of \eqref{eq:Hamiltonian} tend towards $H_f = \int\mathrm dx\left(-i\Psi_c^\dagger\gamma^0\gamma^1\partial_1\Psi_c\right)$,
the fifth line tends towards $H_b = \int\mathrm dx\left[\frac{1}{2}\pi^2+\frac{1}{2}(\partial_1\phi)^2+\frac{1}{2}m^2\phi^2\right]$,
the sixth line tends towards $V_1 = \int\mathrm dx\left[gJ_5^0\pi +\frac{1}{2}g^2(J_5^0)^2\right]$,
and the remaining lines tend towards $V_2 = \int\mathrm dx\;g J_5^1\partial_1\phi$,
where $\Psi_c$ is the continuous classical fermionic field and $J_5^\mu\equiv\bar\Psi_c\gamma^5\gamma^\mu\Psi_c$. Formally, the naive continuum limit of the lattice Hamiltonian \eqref{eq:Hamiltonian} retrieves the classical Hamiltonian $H_{cl}\equiv H_f+H_b+V_1+V_2$ of the classical RS model.
However, it is important to note that the fields in the Hamiltonian \eqref{eq:Hamiltonian} are originally operators, and the original RS model is a quantum field theory, not a classical field theory. Therefore, verifying that the continuum limit of the lattice RS model corresponds to the original RS model requires further work. In the following sections, we will demonstrate that the continuum limit of lattice RS model correlation functions matches the correlation functions of the original RS model. Additionally, in the appendixes, we calculate the equations of motion for the lattice RS model and compare them with those of the original RS model.


Starting from the lattice Hamiltonian \eqref{eq:Hamiltonian}, we can derive the equations of motion for the operators $\phi_n$ and $\Psi(n)$ based on the Heisenberg equations. Appendix \ref{eqomB} derives the continuum limit of the equations of motion for the operator $\phi_n$ and compares them to the equations of motion for the original RS model \eqref{eq:phi0}. It is found that they share the same structure with the only difference being the coefficients of one of the terms. In Appendix \ref{eqomF}, the continuum limit of the equations of motion for the operator $\Psi(n)$ is derived and compared to the equations of motion for the original RS model \eqref{eq:psir}. Similarly, they exhibit the same structure with differences in the coefficients of one of the terms between the two theories.
Actually, the continuous theory corresponds to the infrared behavior of the lattice theory. When we take the continuum limit $\lim\limits_{a\to0}\lim\limits_{N\to\infty}$ of the lattice theory, for any non-zero field separation $\epsilon$, the distance between the fields $\phi(x)$ and $\phi(x+\epsilon)$ already encompasses an infinite number of lattice sites. Therefore, the ultraviolet behavior of the lattice theory has been eliminated during this continuum limit process, leaving only the infrared behavior. Even if we let the field separation $\epsilon$ tend to zero later, it will not reveal the ultraviolet behavior of the lattice theory.
However, if we keep the field separation $\epsilon$ at the same order of magnitude as the lattice spacing $a$ (or more directly we set $\epsilon=a$) during the continuum limit $\lim\limits_{a\to0}$, it implies that no matter how small the lattice spacing becomes, the two fields always stay on neighboring lattice sites. Even if we ultimately take the continuum limit, due to the fields $\phi(x)$ and $\phi(x+\epsilon)$ being ``too close", the lattice's ultraviolet behavior continues to play a role. The lattice theory's equations of motion contain ultraviolet behavior that the original RS model's equations of motion do not have, which is why the coefficients of terms representing ultraviolet behavior slightly differ.

However, this does not imply that the continuum limit of our lattice theory is different from the original RS model. A careful analysis reveals that the behavior exhibited by taking the continuum limit and subsequently letting the field separation tend to zero in our lattice theory is consistent with the original RS model (see \eqref{eq:fc2}, \eqref{eq:fc3}, and \eqref{eq:fcuu}). Furthermore, the alternative form of the lattice fermionic field equations of motion \eqref{eq:psi2} does not explicitly manifest the lattice's ultraviolet behavior, and its continuum limit matches the equations of motion for the fermionic field of the original RS model \eqref{eq:opsi1} and \eqref{eq:opsi2} (or see Eq. (3.24) in \cite{Rothe1975}).
For a more detailed derivation and analysis, please refer to Appendix \ref{eqomB} and \ref{eqomF}.



\section{Solving the lattice RS model}
\label{4}
In this section, we focus on resolving the Hamiltonian  \eqref{eq:Hamiltonian}. 
To simplify the Hamiltonian's expression, we introduce a new set of variables:
%
\begin{equation}
	\begin{split}	
		\label{pP}
		\varphi_n&=F^{\frac{1}{2}}\phi_n,~~~~~~ \pi'_n =\left(\frac{1}{F}\right)^{\frac{1}{2}}\left[\pi_n-g\left(\Psi_u^\dagger(n)\Psi_u(n)-\Psi_d^\dagger(n)\Psi_d(n) \right)\right]\; ,\\
		\psi_u(n)&={\mathrm e}^{-ig\phi_n}\Psi_u(n),~~~\psi_d(n)={\mathrm e}^{ig\phi_n}\Psi_d(n)\; .
	\end{split}
\end{equation}
It can be shown that these new variables still satisfy the canonical commutation relations:
\begin{equation}
	\begin{split}\label{eq:commutationnew}
		\Big[\varphi_m,\pi'_n \Big] =i\delta_{nm}~&,~~~
		\Big\{\psi_u(m),\psi_u^\dagger(n)\Big\}=\delta_{nm}
		\; ,\\
		\Big[\psi_u(m),\pi_n'\Big]
		=0~&,~~~
		\Big[\psi_u(m),\varphi_n\Big]=0
		\; ,\\
		\Big\{\psi_d(m),\psi_d^\dagger(n)\Big\}=\delta_{nm}
		\; ~&,~~~
		\Big[\psi_d(m),\pi_n'\Big]=0
		\; ,\\
		\Big[\psi_d(m),\varphi_n\Big]=0
		\; ~&,~~~
		\Big\{\psi_u(m),\psi_d(n)\Big\}=\Big\{{\mathrm e}^{-ig\phi_m}\Psi_u(m),{\mathrm e}^{ig\phi_n}\Psi_d(n)\Big\}=0 \; .
	\end{split}
\end{equation}
With the aid of these recently introduced variables, we can now reformulate the Hamiltonian \eqref{eq:Hamiltonian} into a more concise expression:
\begin{equation}
	\begin{split}\label{eq:Hamiltonian1}	
		H=&\frac{1}{2}\sum\limits_n \Bigg\{ 
		-i\Big[\psi_u(n)+\psi_u(n+1)\Big]^\dagger
		\Big[\psi_u(n+1)-\psi_u(n) \Big] \frac{1}{a}\\
		&\qquad\quad\;\;  +i\Big[\psi_d(n)+\psi_d(n+1)\Big]^\dagger
		\Big[\psi_d(n+1)-\psi_d(n) \Big] \frac{1}{a}\Bigg\}\\
		&+\frac{1}{2}\sum\limits_n \Bigg\{ 
		i\Big[\psi_d(n+1)-\psi_d(n))\Big]^\dagger
		\Big[\psi_u(n+1)-\psi_u(n) \Big] \frac{1}{a}\\
		&\qquad\qquad  +i\Big[\psi_u(n)-\psi_u(n+1)\Big]^\dagger
		\Big[\psi_d(n+1)-\psi_d(n) \Big] \frac{1}{a}\Bigg\}\\
		&
		+\sum\limits_n a\Bigg[\frac{1}{2}\left(\frac{\pi'_n}{a}\right)^2
		+\frac{1}{2}\left(\frac{1}{a}\right)^2(\varphi_{n+1}-\varphi_n)^2
		+\frac{1}{2}\left(F^{-\frac{1}{2}}m\right)^2\varphi_n^2\Bigg]
		-E_0
		\; .	
	\end{split}
\end{equation}
Clearly, the Eq. \eqref{eq:Hamiltonian1} can be decomposed into two separate components: $H = H_B + H_F$. Here, $H_B$ encompasses contributions solely from the bosonic field and its conjugate momentum, while $H_F$ consists of contributions solely from the fermionic fields $\psi_u(n)$ and $\psi_d(n)$. 



\subsection{Solving the Bosonic Part}
\label{subsec:solutionboson}
As previously mentioned, the Hamiltonian has been divided into two distinct components. We will initially focus exclusively on the contributions originating from $H_B$, which can be specifically expressed as
\begin{equation}
	\begin{split}\label{eq:Hboson}	
		H_B=\sum\limits_n a\Bigg[\frac{1}{2}\left(\frac{\pi'_n}{a}\right)^2
		+\frac{1}{2}\left(\frac{1}{a}\right)^2(\varphi_{n+1}-\varphi_n)^2
		+\frac{1}{2}\left(F^{-\frac{1}{2}}m\right)^2\varphi_n^2\Bigg]
		-{E_0}_B
		\;,
	\end{split}
\end{equation}
where ${E_0}_B$ is chosen to ensure the ground state has zero energy.  By employing the Heisenberg equation, we can derive the canonical equations for $\varphi$ and $\pi'$ in the Heisenberg representation:
\begin{equation}\label{eq:1pivarphi}
	\begin{split}	
		\partial_0\varphi_n&=\frac{\pi'_n}{a}  \; ,\\
		\partial_0\pi'_n			
		&=a\left[\frac{1}{a}\left(\frac{\varphi_{n+1}-\varphi_{n}}{a}
		-\frac{\varphi_{n}-\varphi_{n-1}}{a}\right)
		-\left(F^{-\frac{1}{2}}m\right)^2\varphi_n\right]
		\;.
	\end{split}
\end{equation} 
Subsequently, the equation of motion for $\varphi$ can be derived by Eq. \eqref{eq:1pivarphi}:
\begin{equation}\label{eq:eomvarphi}	
	\partial^0\partial_0\varphi_n
	-\frac{1}{a}\left(\frac{\varphi_{n+1}-\varphi_{n}}{a}
	-\frac{\varphi_{n}-\varphi_{n-1}}{a}\right)
	+\left(F^{-\frac{1}{2}}m\right)^2\varphi_n
	=0 \; .
\end{equation}
It is the equation of motion for a free boson with mass $F^{-\frac{1}{2}}m$. 


Again, it is essential to emphasize that the spatial dimension is a circle with circumference $L = Na$. This implies that the initial and final lattice sites are identified to enforce the periodic boundary condition $\varphi_0 = \varphi_N$. As a result, our summation over the variable $n$ spans from $n = 0$ to $n = N-1$. 
For convenience in calculation, we stipulate that $N$ is an odd number. Similar to the solution in continuous field theory, we can express the solution of the equations of motion \eqref{eq:eomvarphi} using creation and annihilation operators that satisfy the commutation relation $[a_q,a^\dagger_l]=\delta_{ql}$ as
\begin{equation}\label{eq:varphi}	
	\varphi_n(t)=\sum_q\sqrt{\frac{1}{2L\omega_q}}
	(a^\dagger_q {\mathrm e}^{i\omega_q t-inaq}+a_q {\mathrm e}^{-i\omega_q t+inaq})  
	\; ,  ~~~\omega_q \equiv\sqrt{\left(\frac{2}{a} \sin \frac{aq}{2}\right)^2+\left(F^{-\frac{1}{2}}m\right)^2 }.		
\end{equation}
Considering the imposed periodic boundary condition, all admissible values for $q$ are determined as follows: $q = \frac{2\pi k}{Na}$, where $k = -\frac{N-1}{2}, -\frac{N-1}{2}+1, \dots, \frac{N-1}{2}-1, \frac{N-1}{2}$. Consequently, the summation over $q$ equivalently translates into a summation over integer $k$, ranging from $k = -\frac{N-1}{2}$ to $k = \frac{N-1}{2}$,  which is expressed as $\sum\limits_q\equiv \sum\limits_{q=-(N-1)\frac{\pi}{Na}}^{(N-1)\frac{\pi}{Na}}=\sum\limits_{k=-\frac{N-1}{2}}^{\frac{N-1}{2}}$. 
During our forthcoming computations, we will frequently make use of a set of handy summation equations, which we outline below:
\begin{equation}\label{eq:eN}
	\sum_n  {\mathrm e}^{inaq}
	=\sum_{n=0}^{N-1}  {\mathrm e}^{inaq}
	=1+{\mathrm e}^{iaq}+\cdots+({\mathrm e}^{iaq})^{N-1}
	=\frac{1-{\mathrm e}^{iaqN}}{1-{\mathrm e}^{iaq}}=\frac{1-{\mathrm e}^{i\frac{2\pi k}{N}N}}{1-{\mathrm e}^{iaq}}=N\delta_{q0}
	\; ,
\end{equation}
\begin{equation}\label{eq:ek}
	\sum_{q}  {\mathrm e}^{inaq}
	=\sum_{k=-\frac{N-1}{2}}^{\frac{N-1}{2}}  {\mathrm e}^{in\frac{2\pi k}{N}}
	={\mathrm e}^{-in\pi}{\mathrm e}^{in\frac{\pi}{N}}
	\frac{1-{\mathrm e}^{in2\pi}}{1-{\mathrm e}^{in\frac{2\pi}{N}}}=N\delta_{n0}
	\; .
\end{equation}

Analogous to our approach with the scalar field $\varphi_n$, we can derive the expression for its conjugate momentum $\pi_n'$ in terms of creation and annihilation operators:
\begin{equation}\label{eq:pivarphi}	
	\pi'_n(t)=ia\sum_q\sqrt{\frac{\omega_q}{2L}}
	(a_q^\dagger {\mathrm e}^{i\omega_q t-inaq}-a_q {\mathrm e}^{-i\omega_q t+inaq})
	\; .
\end{equation}
Substituting \eqref{eq:varphi} and \eqref{eq:pivarphi} into the bosonic Hamiltonian \eqref{eq:Hboson}, we can express the Hamiltonian in terms of creation and annihilation operators as follows:
\begin{equation}\label{eq:Hboson1}
	H_B=\sum_{q}\left(a_q^\dagger a_q+\frac{1}{2}\right)\omega_q-{E_0}_B
	\; .
\end{equation}
Here, we define ${E_0}_B$ as ${E_0}_B\equiv \frac{1}{2}\sum\limits_{q}\omega_q$, such that the ground state energy of $H_B$ is zero.

Having successfully diagonalized $H_B$, we can now proceed to compute the eigenstates associated with this Hamiltonian. We denote the ground state of $H_B$ as $|\Omega\rangle_B$. For any given $q$, the ground state satisfy
\begin{equation}\label{eq:aO}
	a_q|\Omega\rangle_B=0\; .
\end{equation}
And a one-boson state can be represented as
\begin{equation}\label{eq:aOne}
	|q\rangle_B=a^\dagger_q|\Omega\rangle_B\; .
\end{equation}
To illustrate how the bosonic and fermionic fields interact in the upcoming discussion, we need to employ a representation based on real space. For the bosonic part, we can choose the representation expanded by the eigenstates of bosonic field operators.
The eigenstates of the field operator $\hat\varphi_n$ denoted as $|\varphi\rangle_b$ satisfy, for any spatial point $n$,
\begin{equation}
	\hat\varphi_n|\varphi\rangle_b=\varphi_n|\varphi\rangle_b  \; .
\end{equation}
We will subsequently refer to the representation constructed from these field operator eigenstates as $\{|\varphi\rangle_b\}$.
It is worth noting that we intentionally distinguish between the field operators $\hat\varphi_n$ and their corresponding eigenvalues $\varphi_n$, while we will omit the hat symbol for field operators in situations where there is no potential confusion between $\hat\varphi_n$ and $\varphi_n$.


Our next step is to express the ground state $|\Omega\rangle_B$ and the one-particle state $|q\rangle_B$ in terms of the representation $\{|\varphi\rangle_b\}$. According to Eq. \eqref{eq:varphi} and Eq. \eqref{eq:pivarphi}, we can now represent the creation and annihilation operators in terms of the field and its canonical conjugate operators:
\begin{equation}\label{eq:aq}
	\begin{split}
		a_q&=\sqrt{2L\omega_q}\frac{1}{N}\sum\limits_n
		\left(\varphi_n+\frac{i}{a\omega_q}\pi'_n\right){\mathrm e}^{-inaq}
		\; ,\\
		a_q^\dagger&=\sqrt{2L\omega_q}\frac{1}{N}\sum\limits_n
		\left(\varphi_n-\frac{i}{a\omega_q}\pi'_n\right){\mathrm e}^{inaq}
		\; .
	\end{split}
\end{equation}
In the representation $\{|\varphi\rangle_b\}$, any quantum state $|\zeta\rangle_B$ can be expressed in the form of a wave function: $f(\varphi_0, \varphi_1, \ldots, \varphi_N) = \left._b\langle\varphi|\zeta\rangle_B\right.$. Additionally, based on the canonical commutation relations, we can express the field operator $\pi'_n$ as an operator acting on the wave function, specifically, $-i\delta/\delta \varphi_n$. Consequently, relying on Eq. \eqref{eq:aq} and the properties \eqref{eq:aO} of the ground state , the wave function of the ground state, $_b\langle\varphi|\Omega\rangle_B$, satisfies the equation:
%
\begin{equation}\label{eq:eqvacuum}
	\begin{split}
		\sqrt{2L\omega_q}\frac{1}{N}\sum\limits_n
		\left(\varphi_n+\frac{1}{a\omega_q}\frac{\delta}{\delta \varphi_n}\right){\mathrm e}^{-inaq}
		\left._b\langle\varphi|\Omega\rangle_B
		=\left._b\langle\varphi|a_q|\Omega\rangle_B
		=0\right.\right.
		\; .
	\end{split}
\end{equation}
The solution to Eq. \eqref{eq:eqvacuum} can be easily found to be
\begin{equation}\label{eq:vacuum}
	_b\langle\varphi|\Omega\rangle_B
	=\mathcal{N}{\mathrm e}^{-\frac{1}{2}\sum\limits_{n,m}\mathcal{E}_{nm}\varphi_n\varphi_m}
	\; ,~~~\mathcal{E}_{nm}\equiv\frac{1}{N}\sum\limits_q a\omega_q {\mathrm e}^{i(n-m)aq}\; ,
\end{equation}
where $\mathcal{N}$ stands for the normalization coefficient. Utilizing this wavefunction, we can also rephrase the ground state $|\Omega\rangle_B$ in terms of the basis of the representation $\{|\varphi\rangle_b\}$:
\begin{equation}\label{vacuumstate}	
	|\Omega\rangle_B
	=\mathcal{N}\int \mathrm d\varphi\;
	{\mathrm e}^{-\frac{1}{2}\sum\limits_{n,m}\mathcal{E}_{nm}\varphi_n\varphi_m} |\varphi\rangle_b
	\; ,~~~~~~\int \mathrm d\varphi\equiv\int \mathrm d\varphi_1\int \mathrm d\varphi_2 \cdots\int \mathrm d\varphi_N  \; .
\end{equation}

Furthermore, we can simply apply the creation operator $a_q^\dagger$ to the ground state wavefunction in order to construct the corresponding one-particle state wavefunction:
\begin{equation}\label{eq:one}
	\begin{split}
		_b\langle\varphi|q\rangle_B
		&=\left[\sqrt{2L\omega_q}\frac{1}{N}\sum\limits_n
		(\varphi_n-\frac{1}{a\omega_q}\frac{\delta}{\delta \varphi_n}){\mathrm e}^{inaq}\right]
		\mathcal{N}{\mathrm e}^{-\frac{1}{2}\sum\limits_{n,m}\mathcal{E}_{nm}\varphi_n\varphi_m}\\
		&=\mathcal{N}\sqrt{2L\omega_q}\frac{1}{N}\sum\limits_n{\mathrm e}^{inaq}
		\left[\varphi_n
		+\frac{1}{a\omega_q}
		\sum\limits_{m}\mathcal{E}_{nm}\varphi_m
		\right]{\mathrm e}^{-\frac{1}{2}\sum\limits_{n,m}\mathcal{E}_{nm}\varphi_n\varphi_m}
		\; .
	\end{split}
\end{equation}
Subsequently, we can proceed to expand the one-particle state $|q\rangle_B$ within the framework of the representation $\{|\varphi\rangle_b\}$:
\begin{equation}\label{eq:onestate}
	\begin{split}	
		|q\rangle_B
		=\mathcal{N}\sqrt{2L\omega_q}\frac{1}{N}\int \mathrm d\varphi\sum\limits_n{\mathrm e}^{inaq}
		\left[\varphi_n
		+\frac{1}{a\omega_q}
		\sum\limits_{m}\mathcal{E}_{nm}\varphi_m
		\right]{\mathrm e}^{-\frac{1}{2}\sum\limits_{n,m}\mathcal{E}_{nm}\varphi_n\varphi_m}|\varphi\rangle_b
		\; .
	\end{split}
\end{equation}

It's crucial to highlight that our previous discussion exclusively concerns the operator $\hat\varphi$, which is not the scalar field operator featured in the original Hamiltonian \eqref{eq:Hamiltonian}. Instead, $\hat\phi$ corresponds precisely to the field operator present in the original Hamiltonian \eqref{eq:Hamiltonian}, and the relationship between $\hat\varphi$ and $\hat\phi$ is established through the field redefinition outlined in Eq. \eqref{pP}: $\hat\varphi_n=F^{\frac{1}{2}}\hat\phi_n$. 
Let $|\phi\rangle_B$ be an eigenstate of the field operator $\hat\phi$, such that it satisfies $\hat\phi_n|\phi\rangle_B=\phi_n|\phi\rangle_B$, then we have
\begin{equation}\label{Bbzt}
	\hat\varphi_n|\phi\rangle_B
	=F^{\frac{1}{2}}\hat\phi_n|\phi\rangle_B
	=F^{\frac{1}{2}}\phi_n|\phi\rangle_B \; .
\end{equation}
This reveals that $|\phi\rangle_B$ is also an eigenstate of $\hat\varphi$, with the corresponding eigenvalue $F^{\frac{1}{2}}\phi$. Essentially, this implies that $|\phi\rangle_B=|F^{\frac{1}{2}}\phi\rangle_b$. 
We denote the representation expanded by the eigenstates of $\hat\phi$ as $\{|\phi\rangle_B\}$. To determine the representations of the ground state and one-particle states in the representation $\{|\phi\rangle_B\}$, we firstly perform a change of variables on the integration variables $\varphi$ in Eq. \eqref{vacuumstate} and Eq. \eqref{eq:onestate}, substituting $\phi= F^{-\frac{1}{2}}\varphi$. This yields the following expressions:
\begin{equation}\label{eq:vacuumstate2}
	\begin{split}
		|\Omega\rangle_B
		&=\mathcal{N}\int \mathrm d\left(F^{\frac{1}{2}}\phi\right)\;
		{\mathrm e}^{-\frac{1}{2}\sum\limits_{n,m}\mathcal{E}_{nm}F^{\frac{1}{2}}\phi_n F^{\frac{1}{2}}\phi_m} |F^{\frac{1}{2}}\phi\rangle_b\\
		&=\mathcal{N}F^{\frac{1}{2}}\int \mathrm d\phi\;
		{\mathrm e}^{-\frac{F}{2}\sum\limits_{n,m}\mathcal{E}_{nm}\phi_n \phi_m} |F^{\frac{1}{2}}\phi\rangle_b
		\; ,
	\end{split}
\end{equation}	
\begin{equation}\label{eq:onestate2}
	\begin{split}
		|q\rangle_B
		&=\mathcal{N}\sqrt{2L\omega_q}\frac{1}{N}
		\int \mathrm d\left(F^{\frac{1}{2}}\phi\right)
		\sum\limits_n {\mathrm e}^{inaq}
		\left[F^{\frac{1}{2}}\phi_n
		+\frac{1}{a\omega_q}\sum\limits_{m}\mathcal{E}_{nm}F^{\frac{1}{2}}\phi_m
		\right]{\mathrm e}^{-\frac{1}{2}\sum\limits_{n,m}\mathcal{E}_{nm}
			F^{\frac{1}{2}}\phi_n F^{\frac{1}{2}}\phi_m}|F^{\frac{1}{2}}\phi\rangle_b\\
		&=\mathcal{N}F\sqrt{2L\omega_q}\frac{1}{N}
		\int \mathrm d\phi
		\sum\limits_n {\mathrm e}^{inaq}
		\left[\phi_n
		+\frac{1}{a\omega_q}\sum\limits_{m}\mathcal{E}_{nm}\phi_m
		\right]{\mathrm e}^{-\frac{F}{2}\sum\limits_{n,m}\mathcal{E}_{nm}
			\phi_n\phi_m}|F^{\frac{1}{2}}\phi\rangle_b
		\; .
	\end{split}
\end{equation}
Furthermore, due to $|\phi\rangle_B=|F^{\frac{1}{2}}\phi\rangle_b$, Eq. \eqref{eq:vacuumstate2} and Eq. \eqref{eq:onestate2} can be expressed in the following forms:
\begin{equation}\label{eq:vacuumphi}
	|\Omega\rangle_B
	=\mathcal{N}F^{\frac{1}{2}}\int \mathrm d\phi\;
	{\mathrm e}^{-\frac{F}{2}\sum\limits_{n,m}\mathcal{E}_{nm}\phi_n \phi_m} |\phi\rangle_B
	\; ,
\end{equation}
\begin{equation}\label{eq:onephi}
	|q\rangle_B
	=\mathcal{N}F\sqrt{2L\omega_q}\frac{1}{N}
	\int \mathrm d\phi
	\sum\limits_n {\mathrm e}^{inaq}
	\left[\phi_n
	+\frac{1}{a\omega_q}\sum\limits_{m}\mathcal{E}_{nm}\phi_m
	\right]{\mathrm e}^{-\frac{F}{2}\sum\limits_{n,m}\mathcal{E}_{nm}
		\phi_n\phi_m}|\phi\rangle_B
	\; .
\end{equation}
These are the expressions for the ground state and one-boson states in the representation $\{|\phi\rangle_B\}$.

Next, we derive some properties of the field $\phi$. As indicated by \eqref{pP} and \eqref{eq:varphi}, the time evolution of $\phi$ is given by
\begin{equation}\label{eq:phi}	
	\phi_n(t)=F^{-\frac{1}{2}}\varphi_n
	=F^{-\frac{1}{2}}\sum_q\sqrt{\frac{1}{2L\omega_q}}
	(a^\dagger_q {\mathrm e}^{i\omega_q t-inaq}+a_q {\mathrm e}^{-i\omega_q t+inaq}) 
	\; .
\end{equation}
In this context, we can clearly distinguish between the positive and negative frequency components of the scalar field:
\begin{equation}\label{eq:+}
	\begin{split}	
		\phi^+_n(t)
		=F^{-\frac{1}{2}}\sum_q\sqrt{\frac{1}{2L\omega_q}}a_q {\mathrm e}^{-i\omega_q t+inaq}
		\; ,
	\end{split}
\end{equation}
\begin{equation}\label{eq:-}
	\begin{split}	
		\phi^-_n(t)
		=F^{-\frac{1}{2}}\sum_q\sqrt{\frac{1}{2L\omega_q}}a^\dagger_q {\mathrm e}^{i\omega_q t-inaq}
		\; .
	\end{split}
\end{equation}
Employing the commutation relation between creation and annihilation operators, we can compute the commutation relation between $\phi_n^+$ and $\phi_n^-$ at $t=0$:
\begin{equation}\label{[phi]}
	\begin{split}	
		[\phi^+_n,\phi^-_m]
		&=F^{-1}\sum_q  \frac{1}{2L\omega_q} {\mathrm e}^{i(n-m)aq}\\
		&=F^{-1}\sum_{k=-\frac{N-1}{2}}^{\frac{N-1}{2}} \frac{1}{2L\sqrt{\left(\frac{2}{a}\sin\frac{\pi k}{N}\right)^2+\left(F^{-\frac{1}{2}}m\right)^2}} \cos\left[{(n-m)\frac{2\pi k}{N}}\right]\\
		&\equiv f(n-m)
		\; .
	\end{split}
\end{equation}
It is worth noting that $f(n)$ exhibits even symmetry in relation to $n$: $f(-n)=f(n)$. Furthermore, in the continuum limit, both $f(0)$ and $f(1)$ tend towards infinity. However, the disparity between $f(0)$ and $f(1)$ can remain finite:
\begin{equation}\label{eq:f-f}
	\begin{split}		
		\lim_{a\to0}\lim_{N\to\infty}\Big[f(0)-f(1)\Big]&=\lim_{a\to0}\lim_{N\to\infty}\sum_{k=-\frac{N-1}{2}}^{\frac{N-1}{2}} F^{-1}\frac{ 1-\cos\left(\frac{2\pi k}{N}\right)}{2\sqrt{\left(2N\sin\frac{\pi k}{N}\right)^2+N^2a^2\left(F^{-\frac{1}{2}}m\right)^2}}\\
		&=\frac{1}{4\pi F}\int_{-\frac{\pi}{2}}^{\frac{\pi}{2}}  \mathrm d x\; 
		\frac{ 1-\cos\left(2x\right)}{\sqrt{\left(\sin x\right)^2}}\\
		&=\frac{1}{F\pi} 
		\; .
	\end{split}
\end{equation}

\subsection{Solving the Fermionic Part}
\label{subsec:fermion}
In the preceding section, we determined the spectrum and quantum states associated with the bosonic part of the Hamiltonian. In this section, our focus turns to the more complex fermionic component, as fermions on a lattice inherently introduce complications. The Hamiltonian for the fermionic part is
%
\begin{equation}\label{eq:H_0}
	\begin{split}
		H_F=&\frac{1}{2}\sum\limits_n \Bigg\{ 
		-i\Big[\psi_u(n)+\psi_u(n+1)\Big]^\dagger
		\Big[\psi_u(n+1)-\psi_u(n) \Big] \frac{1}{a}\\
		&\qquad\quad\;\;  +i\Big[\psi_d(n)+\psi_d(n+1)\Big]^\dagger
		\Big[\psi_d(n+1)-\psi_d(n) \Big] \frac{1}{a}\Bigg\}\\
		&+\frac{1}{2}\sum\limits_n \Bigg\{ 
		i\Big[\psi_d(n+1)-\psi_d(n))\Big]^\dagger
		\Big[\psi_u(n+1)-\psi_u(n) \Big] \frac{1}{a}\\
		&\qquad\qquad  +i\Big[\psi_u(n)-\psi_u(n+1)\Big]^\dagger
		\Big[\psi_d(n+1)-\psi_d(n) \Big] \frac{1}{a}\Bigg\}\\
		&-{E_0}_F
		\; ,
	\end{split}
\end{equation}
where ${E_0}_F$ is chosen such that the ground state energy is zero.
In fact, $H_F$ is equivalent to the Hamiltonian of free fermionic fields regularized on a lattice using the Kogut-Susskind staggered fermions \cite{Kogut1975,Banks1976,Susskind1977}. We will first transform Eq. \eqref{eq:H_0} into the form of staggered fermions proposed by Kogut and Susskind, and then proceed to solve the corresponding equations of motion to obtain the field operators evolving with time.

The formulation of $H_F$ as presented in Eq. \eqref{eq:H_0} adopts the standard representation for the $\gamma$ matrices given by \eqref{gamma0}.
However, the lattice Hamiltonian with staggered fermions proposed by Kogut and Susskind employs a different representation of the $\gamma$ matrices, denoted as $\gamma'$. These two representations are connected through the transformation:
\begin{equation}
	\gamma'^0=R\gamma^0R^\dagger
	=\left[  \begin{array}{ccc}
		1 & 0 \\
		0 & -1 
	\end{array}   \right]
	\; ,\quad
	\gamma'^1=R\gamma^1R^\dagger
	=\left[  \begin{array}{ccc}
		0 & -1 \\
		1 & 0 
	\end{array}   \right]
	\; ,\quad
	\gamma'^5=\gamma'^0\gamma'^1
	=\left[  \begin{array}{ccc}
		0 & -1 \\
		-1 & 0 
	\end{array}   \right]
\; ,
\end{equation}
where the transformation matrix $R$ is
\begin{equation}
	R=\frac{1}{\sqrt{2}}\left[
	\begin{array}{ccc}
		1 & 1 \\
		-1 & 1 
	\end{array}
	\right]
	\;.
\end{equation}

In the $\gamma'$ representation, we denote the fermionic field operators corresponding to $\psi_u$ and $\psi_d$ as $\psi_u'$ and $\psi_d'$. Their relationship with the original field operators $\psi_u$ and $\psi_d$ can be expressed as follows:
\begin{equation}\label{eq:p'p}
	\begin{split}	
		\psi'_u(n)&=\frac{1}{\sqrt{2}}\Big[\psi_u(n)+\psi_d(n)\Big] \; ,\\
		\psi'_d(n)&=\frac{1}{\sqrt{2}}\Big[-\psi_u(n)+\psi_d(n)\Big]  \; .
	\end{split}
\end{equation}
By employing the $\psi'_u$ and $\psi'_d$ operators, we can offer a more concise representation of the fermionic component of the Hamiltonian:
\begin{equation}\label{eq:H'_0}	
	H_F=\sum\limits_n \left\{ i\psi'^\dagger_u(n) \Big[\psi'_d(n+1)-\psi'_d(n) \Big] \frac{1}{a}
	+i\psi'^\dagger_d(n+1)\Big[\psi'_u(n+1)-\psi'_u(n)\Big]\frac{1}{a}  \right\}
	-{E_0}_F
	\; .
\end{equation}
 This is precisely the lattice Hamiltonian with staggered fermions formulated by Kogut and Susskind. They employed the Jordan-Wigner transformation to rewrite the Hamiltonian \eqref{eq:H'_0} into the Hamiltonian of a one-dimensional quantum antiferromagnetic spin chain. Our approach differs from theirs in that we do not rewrite Hamiltonian \eqref{eq:H'_0} but instead directly solve the equations of motion from \eqref{eq:H'_0}. Utilizing the Heisenberg equations, we can derive the time evolution equation for the fermionic fields as follows:
\begin{equation}\label{eq:u}	
	\partial_0\psi'_u(n,t)=\frac{1}{i}[\psi'_u(n,t),H_0]
	= \left[\psi'_d(n+1,t)-\psi'_d(n,t) \right] \frac{1}{a}
	\; ,
\end{equation}	
\begin{equation}\label{eq:d}	
	\partial_0\psi'_d(n,t)=\frac{1}{i}[\psi'_d(n,t),H_0]
	= \left[\psi'_u(n,t)-\psi'_u(n-1,t) \right] \frac{1}{a} 
	\; .
\end{equation}
By combining the equations of motion \eqref{eq:u} and \eqref{eq:d}, we can express the solutions in terms of creation and annihilation operators that satisfy the anticommutation relations $\{d_q, d^\dagger_l\} = \delta_{ql}$ and $\{b_q, b^\dagger_l\} = \delta_{ql}$:
\begin{equation}\label{eq:psi'u}		
	\psi'_u(n,t)
	=\sum\limits_{q\ne0} \frac{1}{\sqrt{2N}}
	\left[d_q{\mathrm e}^{-i E_q t+i(n+\frac{1}{2})aq}
	+\text{sgn}(q)b_q^\dagger {\mathrm e}^{i E_q t-i(n+\frac{1}{2})aq}\right]
	+\frac{1}{\sqrt{N}}d_0
	\; ,
\end{equation}	
\begin{equation}\label{eq:psi'd}		
	\psi'_d(n,t)
	=\sum\limits_{q\ne0} \frac{1}{\sqrt{2N}}
	\left[-\text{sgn}(q)d_q{\mathrm e}^{-iE_q t+inaq}-b_q^\dagger  {\mathrm e}^{iE_q t-inaq}\right]
	-\frac{1}{\sqrt{N}}b_0^\dagger
	\; ,
\end{equation}
where $E_q$ and $\text{sgn}(q)$ are, respectively,
\begin{equation}	
 	E_q =\left|\frac{2}{a} \sin \frac{aq}{2}\right| \; ,
 \end{equation} 
 \begin{equation}	
 	\text{sgn}(q)=\left\{\begin{array}{ll}		
 		-1 &,\; q<0\\
 		0 &,\; q=0\\
 		1&,\; q>0
 	\end{array}\right.
 \; .
 \end{equation}
The possible values of $q$ are analogous to those of the bosonic field: $q=\frac{2\pi k}{Na} $  , with $k$ ranging from $k=-\frac{N-1}{2},-\frac{N-1}{2}+1,-\frac{N-1}{2}+2,
	\cdots,\frac{N-1}{2}-1,\frac{N-1}{2}$.
By substituting Eq. \eqref{eq:psi'u} and Eq. \eqref{eq:psi'd} into Eq. \eqref{eq:p'p}, we can deduce the expressions for the original fermionic field operators $\psi_u$ and $\psi_d$ in terms of creation and annihilation operators. Notably, these solutions for $\psi_u$ and $\psi_d$ maintain compliance with the canonical commutation relations stated in Eq. \eqref{eq:commutationnew}.

By employing Eq. \eqref{eq:psi'u} and Eq. \eqref{eq:psi'd}, we can reformulate the Hamiltonian \eqref{eq:H_0} using the creation and annihilation operators:
\begin{equation}
	\begin{split}	
		H_F
		=\frac{1}{a}\sum\limits_n \Bigg\{ i&\left[\sum\limits_{q\ne0} \frac{1}{\sqrt{2N}}
		\left[d_q{\mathrm e}^{i(n+\frac{1}{2})aq}
		+\text{sgn}(q)b_q^\dagger {\mathrm e}^{-i(n+\frac{1}{2})aq}\right]
		+\frac{1}{\sqrt{N}}d_0\right]^\dagger\\
		&\quad\Bigg[
		\sum\limits_{q\ne0} \frac{1}{\sqrt{2N}}
		\left[-\text{sgn}(q)d_q{\mathrm e}^{i(n+1)aq}-b_q^\dagger  {\mathrm e}^{-i(n+1)aq}\right]
		-\frac{1}{\sqrt{N}}b_0^\dagger\\
		&\qquad\qquad-
		\sum\limits_{q\ne0} \frac{1}{\sqrt{2N}}
		\left[-\text{sgn}(q)d_q{\mathrm e}^{inaq}-b_q^\dagger  {\mathrm e}^{-inaq}\right]
		+\frac{1}{\sqrt{N}}b_0^\dagger \Bigg] \\
		+
		i&\left[\sum\limits_{q\ne0} \frac{1}{\sqrt{2N}}
		\left[-\text{sgn}(q)d_q{\mathrm e}^{i(n+1)aq}-b_q^\dagger  {\mathrm e}^{-i(n+1)aq}\right]
		-\frac{1}{\sqrt{N}}b_0^\dagger\right]^\dagger\\
		&\quad\Bigg[\sum\limits_{q\ne0} \frac{1}{\sqrt{2N}}
		\left[d_q{\mathrm e}^{i(n+1+\frac{1}{2})aq}
		+\text{sgn}(q)b_q^\dagger {\mathrm e}^{-i(n+1+\frac{1}{2})aq}\right]
		+\frac{1}{\sqrt{N}}d_0\\
		&\qquad\qquad-\sum\limits_{q\ne0} \frac{1}{\sqrt{2N}}
		\left[d_q{\mathrm e}^{i(n+\frac{1}{2})aq}
		+\text{sgn}(q)b_q^\dagger {\mathrm e}^{-i(n+\frac{1}{2})aq}\right]
		-\frac{1}{\sqrt{N}}d_0\Bigg]\Bigg\}
		-{E_0}_F\\ 
		=
		\sum\limits_{q\ne0}
		\Bigg\{ \frac{2}{a}&\left|\sin\left(\frac{a}{2}q\right) \right|d^\dagger_q d_q 
		+\frac{2}{a}\left|\sin\left(\frac{a}{2}q\right)\right| b_q^\dagger b_q \Bigg\}
		-\sum\limits_{q\ne0} \frac{2}{a}\left|\sin\left(\frac{a}{2}q\right)\right|
		-{E_0}_F
		\;.
	\end{split}
\end{equation}
Let ${E_0}_F\equiv-\sum\limits_{q\ne0} \frac{2}{a}\left|\sin\left(\frac{a}{2}q\right)\right|$, then this renders the energy of the ground state zero, and the Hamiltonian $H_F$ can be expressed in a conventional form:
\begin{equation}\label{eq:Hbbdd}
	H_F
	=\sum\limits_{q\ne0}
	E_q \left(d^\dagger_q d_q + b_q^\dagger b_q \right)
	\; .
\end{equation}

In a manner akin to our treatment of the bosonic field, we can employ the diagonalized Hamiltonian \eqref{eq:Hbbdd} to determine the ground state $|\Omega\rangle_F$ that is annihilated by $b_q$ and $d_q$ for all $q$:
\begin{equation}\label{eq:vacuumfermion}			
	b_q|\Omega\rangle_F=d_q|\Omega\rangle_F=0\; .
\end{equation}
In the previous discussion on the quantum state of the bosonic field, we introduced a representation $\{|\phi\rangle_b\}$ and employed this framework to describe quantum states. Now, we will take similar steps for the fermionic field and introduce a new representation. Firstly, define a quantum state $|0\rangle_F$ as follows:
%
\begin{equation}	
	\label{|0>}
	|0\rangle_F\equiv\prod\limits_{q} b_{-q}^\dagger|\Omega\rangle_F
	\; .
\end{equation}
By utilizing the commutation relations between creation and annihilation operators, it is straightforward to demonstrate the normalization of $|0\rangle_F$:
\begin{equation}
	_F\langle 0|0\rangle_F
	= \left._F\langle\Omega|\prod\limits_{q'} b_{q'}
	\prod\limits_{q} b_{-q}^\dagger|\Omega\rangle_F=1\right..
\end{equation}
Based on the properties $b_q^\dagger b_q^\dagger=0$ and $d_q|\Omega\rangle_F=0$, we can infer that $|0\rangle_F$ possesses the following crucial attributes:
\begin{equation}
		\begin{split}	
			\label{eq:u|0>=0}
			\psi'_u(n,t)|0\rangle_F       			
			&=\left\{\sum\limits_{q\ne0} \frac{1}{\sqrt{2N}}
			\left[d_q{\mathrm e}^{-iE_q t+i(n+\frac{1}{2})aq}
			+\text{sgn}(q)b_q^\dagger {\mathrm e}^{iE_q t-i(n+\frac{1}{2})aq}\right]
			+\frac{1}{\sqrt{N}}d_0\right\}
			\prod\limits_{q} b_q^\dagger|\Omega\rangle_F\\
			&=0\; ,
		\end{split}
	\end{equation}	
\begin{equation}\label{eq:d|0>=0}
	\psi'_d(n,t)|0\rangle_F=\left\{\sum\limits_{q\ne0} \frac{1}{\sqrt{2N}}
	\left[-\text{sgn}(q)d_q{\mathrm e}^{-iE_q t+inaq}-b_q^\dagger  {\mathrm e}^{iE_q t-inaq}\right]
	-\frac{1}{\sqrt{N}}b_0^\dagger\right\}
	\prod\limits_{q} b_q^\dagger|\Omega\rangle_F=0\;.
\end{equation}
This implies that $|0\rangle_F$ is the eigenstate of the field operators $\psi'_u$ and $\psi'_d$ with eigenvalue zero. 
Similar to the construction of the traditional Fock representation, here we treat $\psi'^\dagger_u$ and $\psi'^\dagger_d$ as ladder operators acting on the new ``vacuum" state $|0\rangle_F$, resulting in the basis of the entire representation.
To be more precise, we can apply arbitrary fermionic field operators $\psi'^\dagger(n)$ to $|0\rangle_F$ to generate a sequence of quantum states:
\begin{equation}\label{eq:s}
	\psi'^\dagger_{\alpha_1}(n_1)\psi'^\dagger_{\alpha_2}(n_2)\psi'^\dagger_{\alpha_3}(n_3)\cdots\psi'^\dagger_{\alpha_s}(n_s)|0\rangle_F  \; ,
\end{equation}
where
$$s=1,2,\cdots,N \quad,\qquad \alpha_i=u,d  \quad,\qquad   n_i\ne n_j,\forall \alpha_i=\alpha_j\;\;.$$

According to Eq. \eqref{eq:u|0>=0}, Eq. \eqref{eq:d|0>=0}, and the commutation relations, it is easy to prove that the quantum states \eqref{eq:s} are orthogonal and normalized. For instance:
\begin{equation}
	\begin{split}	
		_F\langle 0|\psi'_u(m)\psi'_d(n)\psi'^\dagger_d(n)\psi'^\dagger_u(m)|0\rangle
		&=\langle 0|\psi'_u(m)
		\left[-\psi'^\dagger_d(n)\psi'_d(n)+1\right]
		\psi'^\dagger_u(m)|0\rangle\\
		&=\langle 0|\psi'_u(m)\psi'^\dagger_u(m)|0\rangle\\
		&=1\;.
	\end{split}
\end{equation}
It is worth noting that the expression $\psi'^\dagger_{\alpha_1}(n_1)\psi'^\dagger_{\alpha_2}(n_2)\psi'^\dagger_{\alpha_3}(n_3)\cdots\psi'^\dagger_{\alpha_s}(n_s)|0\rangle_F$ exhibits similarities with the traditional Fock space basis $B^\dagger_{\alpha_1}(q_1)B^\dagger_{\alpha_2}(q_2)B^\dagger_{\alpha_3}(q_3)\cdots B^\dagger_{\alpha_s}(q_s)|\Omega\rangle_F$, where $B_{1}(q)=b_q\;,\;B_{2}(q)=d_q$. However, while momentum labels the creation operators in the Fock space, our ``creation" operators $\psi'^\dagger_{\alpha}(n)$ are labeled by spatial coordinates.
Here, we define a new representation by utilizing the set of quantum states \eqref{eq:s} as our basis, and denote this representation as $\{\prod\psi'^\dagger|0\rangle_F\}$.
Our subsequent calculations will be built upon this representation.


Next, let's derive the representation of the ground state in the representation $\{\prod\psi'^\dagger|0\rangle_F\}$.
According to the definition of $|0\rangle_F$ in Eq. \eqref{eq:vacuumfermion} and the commutation relation $\{b_q,b^\dagger_l\}=\delta_{ql}$, we can express the ground state as
\begin{equation}\label{eq:vacuum1}
	|\Omega\rangle_F=\prod\limits_{q} b_q|0\rangle_F\; .
\end{equation}
Combining Eq. \eqref{eq:psi'u} and Eq. \eqref{eq:psi'd}, we can express the creation and annihilation operators in terms of the field operators $\psi'_u$ and $\psi_d'$ :
\begin{equation}
	\begin{split}	
		\label{eq:dq}
		d_q
		&={\mathrm e}^{-i\frac{1}{2}aq}\frac{1}{\sqrt{2N}}\sum\limits_n {\mathrm e}^{-ianq}\psi'_u(n)
		-\text{sgn}(q)\frac{1}{\sqrt{2N}}\sum\limits_n {\mathrm e}^{-ianq}\psi'_d(n)
		\; ,\\
		d_0&=\frac{1}{\sqrt{N}}\sum\limits_n\psi'_u(n)
		\; ,
	\end{split}
\end{equation}
\begin{equation}
	\begin{split}	
		\label{eq:bq}
		b_q
		&=\text{sgn}(q){\mathrm e}^{-i\frac{1}{2}aq}\frac{1}{\sqrt{2N}}\sum\limits_n {\mathrm e}^{-ianq}\psi'^\dagger_u(n)
		-\frac{1}{\sqrt{2N}}\sum\limits_n {\mathrm e}^{-ianq}\psi'^\dagger_d(n)
		\; ,\\
		b_0&=-\frac{1}{\sqrt{N}}\sum\limits_n \psi'^\dagger_d(n)
		\; .
	\end{split}
\end{equation}
By substituting \eqref{eq:bq} into \eqref{eq:vacuum1}, we can express $|\Omega\rangle_F$ using the basis \eqref{eq:s}:
\begin{equation}
	\begin{split}\label{eq:vacuum2}	
		|\Omega\rangle_F&=\prod\limits_{q} b_q|0\rangle_F\\
		&=
		\prod\limits_{k=-\frac{N-1}{2}}^{-1}\frac{1}{\sqrt{2N}}
		\left[\text{sgn}(k){\mathrm e}^{-i\frac{1}{2}\frac{2\pi k}{N}}\sum\limits_n {\mathrm e}^{-in\frac{2\pi k}{N}}\psi'^\dagger_u(n)
		-\sum\limits_n {\mathrm e}^{-in\frac{2\pi k}{N}}\psi'^\dagger_d(n)\right]\\
		&\times\left(-\frac{1}{\sqrt{N}}\sum\limits_n \psi'^\dagger_d(n)\right)
		\prod\limits_{k=1}^{\frac{N-1}{2}}\frac{1}{\sqrt{2N}}
		\left[\text{sgn}(k){\mathrm e}^{-i\frac{1}{2}\frac{2\pi k}{N}}\sum\limits_n {\mathrm e}^{-in\frac{2\pi k}{N}}\psi'^\dagger_u(n)
		-\sum\limits_n {\mathrm e}^{-in\frac{2\pi k}{N}}\psi'^\dagger_d(n)\right]
		|0\rangle_F\\
		&=\sqrt{2}
		\prod\limits_{k=-\frac{N-1}{2}}^{\frac{N-1}{2}}\frac{1}{\sqrt{2N}}
		\left[\text{sgn}(k){\mathrm e}^{-i\frac{1}{2}\frac{2\pi k}{N}}\sum\limits_n {\mathrm e}^{-in\frac{2\pi k}{N}}\psi'^\dagger_u(n)
		-\sum\limits_n {\mathrm e}^{-in\frac{2\pi k}{N}}\psi'^\dagger_d(n)\right]
		|0\rangle_F
		\;,
	\end{split}
\end{equation}
where the last line employs $\text{sgn}(0)=0$.
Note that the final line of Eq. \eqref{eq:vacuum2} is written in the form of the field operator $\psi'^\dagger$ acting on $|0\rangle_F$, indicating that the vacuum state $|\Omega\rangle_F$ has been expressed in the representation $\{\prod\psi'^\dagger|0\rangle_F\}$.


Fermionic one-particle states can also be presented in the basis \eqref{eq:s}.
The creation operators for fermions and antifermions are $d_{q}^\dagger$ and $b_{q}^\dagger$, respectively. The quantum state of a single fermion is denoted as $|q;+\rangle_F$, while the quantum state of a single antifermion is denoted as $|q;-\rangle_F$. Utilizing the relationships between the creation and annihilation operators and the field operators in Eq. \eqref{eq:dq} and Eq. \eqref{eq:bq}, we can express $|q;+\rangle_F$ and $|q;-\rangle_F$ (where $q\ne 0$) in the representation $\{\prod\psi'^\dagger|0\rangle_F\}$ as follows:
\begin{equation}
	\begin{split}	
		\label{eq:onepsi+}
		|q;+\rangle_F&=d_{q}^\dagger|\Omega\rangle_F\\
		&=
		\frac{1}{\sqrt{N}}\left[
		{\mathrm e}^{i\frac{1}{2}\frac{2\pi k}{N}}
		\sum\limits_n  {\mathrm e}^{in\frac{2\pi k}{N}}\psi'^\dagger_u(n)
		-\text{sgn}(k)
		\sum\limits_n  {\mathrm e}^{in\frac{2\pi k}{N}}\psi'^\dagger_d(n)
		\right]\\
		&\times\prod\limits_{k'=-\frac{N-1}{2}}^{\frac{N-1}{2}}\frac{1}{\sqrt{2N}}
		\left[\text{sgn}(k'){\mathrm e}^{-i\frac{1}{2}\frac{2\pi k'}{N}}
		\sum\limits_n {\mathrm e}^{-in\frac{2\pi k'}{N}}\psi'^\dagger_u(n)
		-\sum\limits_n {\mathrm e}^{-in\frac{2\pi k'}{N}}\psi'^\dagger_d(n)\right]
		|0\rangle_F
		\; ,
	\end{split}
\end{equation}	
\begin{equation}
	\begin{split}	
		\label{eq:onepsi-}
		|q;-\rangle_F&=b_{q}^\dagger|\Omega\rangle_F\\
		&=b_{q}^\dagger\prod\limits_{q'} b_{q'}|0\rangle_F
		=\prod\limits_{q'}^{q'\ne q} b_{q'}|0\rangle_F\\
		&=\sqrt{2}
		\prod\limits_{k'=-\frac{N-1}{2}}^{\frac{N-1}{2};k'\ne k}\frac{1}{\sqrt{2N}}
		\left[\text{sgn}(k'){\mathrm e}^{-i\frac{1}{2}\frac{2\pi k'}{N}}\sum\limits_n {\mathrm e}^{-in\frac{2\pi k'}{N}}\psi'^\dagger_u(n)
		-\sum\limits_n {\mathrm e}^{-in\frac{2\pi k'}{N}}\psi'^\dagger_d(n)\right]
		|0\rangle_F
		\; ,
	\end{split}
\end{equation}
where $k=\frac{N}{2\pi}q$.
It's worth noting that in Eq. \eqref{eq:onepsi-}, the third equality (second line) omits the potential negative sign, as quantum states that differ by a negative sign still describe the same physical system, and quantum states remain normalized.

At this point, we have accomplished the diagonalization of the Hamiltonian \eqref{eq:Hamiltonian}.


\section{
The Correlation Functions and Renormalization}
\label{5}
Correlation functions hold a significant place in quantum field theory, offering valuable insights into the system's behavior. In the upcoming section, we will undertake nonperturbative calculations of correlation functions.
Furthermore, we will explore the concept of the continuum limit for correlation functions, aligning our findings with correlation functions obtained in the original RS model.


As demonstrated in Eq. \eqref{eq:Hamiltonian1}, the Hamiltonian can be divided into two distinct components: $H=H_B+H_F$. Notably, these components are entirely uncoupled, indicating that the vacuum state for the complete Hamiltonian \eqref{eq:Hamiltonian} can be factored into a direct product of the ground state for the bosonic part and the ground state for the fermionic part: $|\Omega\rangle=|\Omega\rangle_F|\Omega\rangle_B$. Therefore, by utilizing the solution of the complete equation of motion for the bosonic field \eqref{eq:phi}, we are ready to compute the two-point correlation function for the bosonic field:
\begin{equation}\label{eq:cphi}
	\begin{split}
		\langle\Omega|\phi_n(t_1)\phi_m(t_2)|\Omega\rangle
		&=\langle\Omega|
		F^{-\frac{1}{2}}\sum_q\sqrt{\frac{1}{2L\omega_q}}
		(a^\dagger_q {\mathrm e}^{i\omega_q t_1-inaq}+a_q {\mathrm e}^{-i\omega_q t_1+inaq})\\
		&\qquad\qquad F^{-\frac{1}{2}}\sum_l\sqrt{\frac{1}{2L\omega_l}}
		(a^\dagger_l {\mathrm e}^{i\omega_l t_2-imal}+a_q {\mathrm e}^{-i\omega_l t_2+imal})
		|\Omega\rangle\\
		&=F^{-1}\sum_q \sum_l
		\sqrt{\frac{1}{2L\omega_q}}\sqrt{\frac{1}{2L\omega_l}}
		{\mathrm e}^{-i\omega_q t_1+inaq}   {\mathrm e}^{i\omega_l t_2-imal}
		\left._B\langle\Omega| a_q a^\dagger_l |\Omega\rangle_B\right.\\
		&=F^{-1}\sum_q \sum_l
		\sqrt{\frac{1}{2L\omega_q}}\sqrt{\frac{1}{2L\omega_l}}
		{\mathrm e}^{-i\omega_q t_1+inaq}   {\mathrm e}^{i\omega_l t_2-imal}
		\delta_{lq}\\
		&=F^{-1}\sum_q  \frac{1}{2L\omega_q}
		{\mathrm e}^{-i\omega_q(t_1-t_2) +i(n-m)aq} 
		\; ,
	\end{split}
\end{equation} 
where $\omega_q =\sqrt{\left(\frac{2}{a} \sin \frac{aq}{2}\right)^2+\left(F^{-\frac{1}{2}}m\right)^2 }$ .

 In order to investigate the continuum limit of \eqref{eq:cphi}, it is beneficial to make some adjustments to the summation. 
 Since $q=\frac{2\pi k}{Na}$, where $k=-\frac{N-1}{2},-\frac{N-1}{2}+1,-\frac{N-1}{2}+2, \ldots, \frac{N-1}{2}-1, \frac{N-1}{2}$, the difference between neighboring $q$ values is $\Delta q=\frac{2\pi}{Na}=\frac{2\pi}{L}$. Consequently, we can express Eq. \eqref{eq:cphi} in the following form:
\begin{equation}\label{eq:cphi1}
	\langle\Omega|\phi_n(t_1)\phi_m(t_2)|\Omega\rangle
	=F^{-1}\sum_q  \Delta q \frac{1}{4\pi\omega_q}
	{\mathrm e}^{-i\omega_q(t_1-t_2) +i(n-m)aq} 
	\; .
\end{equation}
To explore the continuum limit, our initial step involves keeping the lattice spacing $a$ constant while letting the spatial length $L=Na$ tend to infinity. This effectively corresponds to considering the scenario of a large number of lattice sites, i.e. $\lim\limits_{N\to\infty}$. As a result, $\Delta q$ tends to zero, which implies that $q$ becomes continuous. Consequently, the summation over $q$ in Eq. \eqref{eq:cphi1} can be replaced by an integral with respect to $q$:
\begin{equation}\label{eq:Nphi}
	\lim\limits_{N\to\infty}\langle\Omega|\phi_n(t_1)\phi_m(t_2)|\Omega\rangle
	=F^{-1}\int_{-\frac{\pi}{a}}^{\frac{\pi}{a}} \mathrm d q\; \frac{1}{4\pi\omega_q}
	{\mathrm e}^{-i\omega_q(t_1-t_2) +i(n-m)aq} 
	\; .
\end{equation}
According to Eq. \eqref{eq:phipi}, in the continuum limit, the relationship between the continuous bosonic field operator $\phi(x, t)$ and the discrete field operator $\phi_n$ is given by $\phi_n(t) = \phi(x, t)\big|_{x = na}$. Considering two spatial coordinates $x_1 = na$ and $x_2 = ma$, with $x_1$ and $x_2$ fixed, the expression for the continuous two-point correlation function in the limit where the lattice spacing $a$ tends to zero is:
\begin{equation}\label{eq:blim1}
	\begin{split}
		\langle\Omega|\phi(x_1,t_1)\phi(x_2,t_2)|\Omega\rangle
		&=\lim\limits_{a\to 0}\lim\limits_{N\to\infty}
		\langle\Omega|\phi_n(t_1)\phi_m(t_2)|\Omega\rangle\\
		&=\lim\limits_{a\to 0}F^{-1}\int_{-\frac{\pi}{a}}^{\frac{\pi}{a}} \mathrm d q\; \frac{1}{4\pi\omega_q}
		{\mathrm e}^{-i\omega_q(t_1-t_2) +i(x_1-x_2)q}
		\; .
	\end{split}
\end{equation}
In order to arrive at a more explicit form, we can introduce a deformation to the integral. To start, we separate the integral in Eq. \eqref{eq:blim1} into two distinct parts:
\begin{equation}\label{eq:blim2}
	\begin{split}
		&\int_{-\frac{\pi}{a}}^{\frac{\pi}{a}} \mathrm d q\; \frac{1}{\omega_q}
		{\mathrm e}^{-i\omega_q(t_1-t_2) +i(x_1-x_2)q} \\
		&=\int_{0}^{\frac{\pi}{a}} \mathrm d q\; \frac{1}{\omega_q}
		{\mathrm e}^{-i\omega_q(t_1-t_2) +i(x_1-x_2)q}
		+\int_{-\frac{\pi}{a}}^{0} \mathrm d q\; \frac{1}{\omega_q}
		{\mathrm e}^{-i\omega_q(t_1-t_2) +i(x_1-x_2)q}\\
		&=\lim\limits_{\epsilon\to 0}\int_{0}^{\frac{\pi}{a}} \mathrm d q\; \frac{1}{\omega_q}
		{\mathrm e}^{-i\omega_q(t_1-t_2) +i(x_1-x_2)q-\epsilon q}
		+\lim\limits_{\epsilon\to 0}\int_{-\frac{\pi}{a}}^{0} \mathrm d q\; \frac{1}{\omega_q}
		{\mathrm e}^{-i\omega_q(t_1-t_2) +i(x_1-x_2)q+\epsilon q}
		\;.
	\end{split}
\end{equation}
In the last line of Eq. \eqref{eq:blim2}, we have introduced an $\epsilon$ suppression, 
which is a common technique in standard field theory. And we will make a reasonable assumption that, for the complete integral expression, the limits $\lim\limits_{\epsilon\to 0}$ and $\lim\limits_{a\to 0}$ can be interchanged. 
In Eq. \eqref{eq:blim2}, the last two integrals in the final line have identical structures, so we only need to focus on the first integral.
Taking the limit $a \to 0$ of the first integral in the last line of Eq. \eqref{eq:blim2}, we split it into two new integrals:
\begin{equation}\label{eq:blim3}
	\begin{split}
		&\lim\limits_{a\to 0}\int_{0}^{\frac{\pi}{a}} \mathrm d q\; \frac{1}{\omega_q}
		{\mathrm e}^{-i\omega_q(t_1-t_2) +i(x_1-x_2)q-\epsilon q}\\
		&=\lim\limits_{a\to 0}\int_{0}^{\frac{\pi}{\sqrt{a}}} \mathrm d q\; \frac{1}{\omega_q}
		{\mathrm e}^{-i\omega_q(t_1-t_2) +i(x_1-x_2)q-\epsilon q}
		+\lim\limits_{a\to 0}\int_{\frac{\pi}{\sqrt{a}}}^{\frac{\pi}{a}} \mathrm d q\; \frac{1}{\omega_q}
		{\mathrm e}^{-i\omega_q(t_1-t_2) +i(x_1-x_2)q-\epsilon q}
		\; .
	\end{split}
\end{equation}
In Eq. \eqref{eq:blim3}, the second integral in the last line starts from $q = \pi / \sqrt{a}$ rather than zero.
Therefore, based on the expression for $\omega_q$ in Eq. \eqref{eq:varphi}, it is easy to show that for sufficiently small $a$ (implying that $q > \pi / \sqrt{a}$ is large), we have $\omega_q > 1$. Hence, the second integral in the last line of Eq. \eqref{eq:blim3} satisfies the following inequality:
\begin{equation}
	\begin{split}
		\left|\lim\limits_{a\to 0}\int_{\frac{\pi}{\sqrt{a}}}^{\frac{\pi}{a}} \mathrm d q\; \frac{1}{\omega_q}
		{\mathrm e}^{-i\omega_q(t_1-t_2) +i(x_1-x_2)q-\epsilon q}\right|
		&\leqslant\lim\limits_{a\to 0}\int_{\frac{\pi}{\sqrt{a}}}^{\frac{\pi}{a}} \mathrm d q\; \frac{1}{\omega_q}
		\left|{\mathrm e}^{-i\omega_q(t_1-t_2) +i(x_1-x_2)q}\right| {\mathrm e}^{-\epsilon q}\\
		&\leqslant\lim\limits_{a\to 0}\int_{\frac{\pi}{\sqrt{a}}}^{\frac{\pi}{a}} \mathrm d q\;  {\mathrm e}^{-\epsilon q}=\lim\limits_{a\to 0}\frac{{\mathrm e}^{-\epsilon \frac{\pi}{a}}-{\mathrm e}^{-\epsilon \frac{\pi}{\sqrt a}}}{-\epsilon}=0
		\; ,
	\end{split}
\end{equation}
which indicates that the second integral in the last line of Eq. \eqref{eq:blim3} becomes negligible.
Therefore, by substituting the dispersion relation for $\omega_q$ from Eq. \eqref{eq:varphi}, Eq. \eqref{eq:blim3} can be expressed as
\begin{equation}\label{eq:blim5}
	\begin{split}
		&\lim\limits_{a\to 0}\int_{0}^{\frac{\pi}{a}} \mathrm d q\; \frac{1}{\omega_q}
		{\mathrm e}^{-i\omega_q(t_1-t_2) +i(x_1-x_2)q-\epsilon q}\\
		&=\lim\limits_{a\to 0}\int_{0}^{\frac{\pi}{\sqrt{a}}} \mathrm d q\; \frac{1}{\sqrt{\left(\frac{2}{a} \sin \frac{aq}{2}\right)^2+\left(F^{-\frac{1}{2}}m\right)^2 }}
		{\mathrm e}^{-i\sqrt{\left(\frac{2}{a} \sin \frac{aq}{2}\right)^2+\left(F^{-\frac{1}{2}}m\right)^2 }(t_1-t_2) +i(x_1-x_2)q-\epsilon q}\\
		&=\int_{0}^{\infty} \mathrm d q\; \frac{1}{\sqrt{q^2+\left(F^{-\frac{1}{2}}m\right)^2 }}		{\mathrm e}^{-i\sqrt{q^2+\left(F^{-\frac{1}{2}}m\right)^2 }(t_1-t_2) +i(x_1-x_2)q-\epsilon q}
		\; .
	\end{split}
\end{equation}
This is the continuum limit of the first integral in the last line of Eq. \eqref{eq:blim2}. We can apply the same method to analyze the continuum limit of the second integral in the last line of \eqref{eq:blim2}, and the result is very similar to Eq. \eqref{eq:blim5}. Therefore, we obtain the continuum limit of Eq. \eqref{eq:blim2} as
\begin{equation}\label{eq:blim6}
	\begin{split}
		\lim\limits_{a\to 0}\int_{-\frac{\pi}{a}}^{\frac{\pi}{a}} \mathrm d q\; \frac{1}{\omega_q}
		{\mathrm e}^{-i\omega_q(t_1-t_2) +i(x_1-x_2)q} =\int_{-\infty}^{\infty} \mathrm d q\; \frac{1}{\sqrt{q^2+\left(F^{-\frac{1}{2}}m\right)^2 }}		{\mathrm e}^{-i\sqrt{q^2+\left(F^{-\frac{1}{2}}m\right)^2 }(t_1-t_2) +i(x_1-x_2)q}
	\end{split}
\end{equation}
By inserting equation Eq. \eqref{eq:blim6} into Eq. \eqref{eq:blim1}, we can derive the continuum limit of the two-point correlation functions for the bosonic field:
\begin{equation}\label{eq:ccphi}
	\begin{split}		
		\langle\Omega|\phi(x_1,t_1)\phi(x_2,t_2)|\Omega\rangle
		&=\lim\limits_{a\to 0}\lim\limits_{N\to\infty}
		\langle\Omega|\phi_n(t_1)\phi_m(t_2)|\Omega\rangle\\
		&=\lim\limits_{a\to 0}F^{-1}\int_{-\frac{\pi}{a}}^{\frac{\pi}{a}} \mathrm d q\; \frac{1}{4\pi\omega_q}
		{\mathrm e}^{-i\omega_q(t_1-t_2) +i(x_1-x_2)q} \\
		&=F^{-1}\int_{-\infty}^{\infty} \mathrm d q\; 
		\frac{1}{4\pi \sqrt{q^2+\left(F^{-\frac{1}{2}}m\right)^2 }}
		{\mathrm e}^{-i(t_1-t_2)\sqrt{q^2+\left(F^{-\frac{1}{2}}m\right)^2 } +i(x_1-x_2)q}
		\; . 
	\end{split}
\end{equation}
Now, introducing the renormalized bosonic field $\phi_r\equiv F^{\frac{1}{2}}\phi$ and the renormalized mass $m_r\equiv F^{-\frac{1}{2}}m$, the continuum limit of the lattice correlation function \eqref{eq:ccphi} recovers the two-point correlation function of the original RS model \eqref{eq:2ptboson}.W
Please note that we have assumed a specific time ordering $t_1>t_2$, so we have omitted the time ordering operator $T$ in this context.

It is crucial to establish a connection between the bare lattice parameters ($m$, $F$, $g$) and the bare parameters of the original RS model ($m_0$, $g_0$). In the original RS model, the relation between the bare mass and the renormalized mass is given by $m_r=(1-g_0^2/\pi)^{-\frac{1}{2}}m_0$. However, in the context of the lattice RS model, we have $m_r= F^{-\frac{1}{2}}m$. In order to ensure consistency between these results, we must establish the following relationship:
\begin{equation}\label{eq:mm0}
	F^{-\frac{1}{2}}m=\left(1-\frac{g_0^2}{\pi}\right)^{-\frac{1}{2}}m_0
	\; .
\end{equation}
Furthermore, in the original RS model, the relationship between the bare field and the renormalized field is $\phi_r=\left(1-\frac{g_0^2}{\pi}\right)^{-\frac{1}{2}}\phi_0$. Combining this with the previously established relationship in the lattice RS model, $\phi_r= F^{\frac{1}{2}}\phi$, we find that $F^{\frac{1}{2}}\phi=\left(1-\frac{g_0^2}{\pi}\right)^{-\frac{1}{2}}\phi_0$.
If we want to ensure that the continuum limit of the lattice bare field matches the RS model bare field, i.e., $\phi=\phi_0$, we can derive
\begin{equation}\label{eq:FAg}
F=1-\frac{g_0^2}{\pi}\; .
\end{equation}
Combining Eq. \eqref{eq:mm0} and Eq. \eqref{eq:FAg}, we also obtain
\begin{equation}\label{eq:Am}
m=m_0\; .
\end{equation}
To determine the relationship between the lattice bare coupling constant $g$ and the original RS model bare parameters, additional comparisons involving other types of correlation functions are necessary.


Now, we can proceed to calculate the two-point correlation function for the fermionic field $\langle\Omega|\Psi(n,t_1)\bar\Psi(m,t_2)|\Omega\rangle$. Recall that we have defined $\bar\Psi$ as $\bar\Psi=\Psi^\dagger \gamma^0$. This allows us to express the components of $\bar\Psi$ as follows:
\begin{equation}
	\begin{split}
		\left[
		\begin{array}{ccc}
			\bar\Psi_u  &\bar\Psi_d 
		\end{array}
		\right]
		=\left[
		\begin{array}{ccc}
			\Psi_u^\dagger &\Psi_d^\dagger 
		\end{array}
		\right]
		\left[
		\begin{array}{ccc}
			0 & 1 \\
			1 & 0 
		\end{array}
		\right]
		= \left[
		\begin{array}{ccc}
			\Psi_d^\dagger  &\Psi_u^\dagger 
		\end{array}
		\right]
		\; .
	\end{split}
\end{equation}
Considering the relationship between $\Psi$ and $\psi$ as given in Eq. \eqref{pP}, we can express the specific component of the two-point correlation function in terms of $\psi$. 
For example, one component of the two-point correlation function can be written as
\begin{equation}\label{eq:Psiuu}
	\begin{split}
		\langle\Omega|\Psi_u(n,t_1)\bar\Psi_u(m,t_2)|\Omega\rangle
		&=\langle\Omega|\Psi_u(n,t_1)\Psi^\dagger_d(m,t_2)|\Omega\rangle\\
		&=\langle\Omega|
		{\mathrm e}^{ig\phi_n(t_1)}	\psi_u(n,t_1)
		{\mathrm e}^{ig\phi_m(t_2)}    \psi^\dagger_d(m,t_2)
		|\Omega\rangle\\
		&=\left._B\langle\Omega| \left._F\langle\Omega|
		{\mathrm e}^{ig\phi_n(t_1)}	\psi_u(n,t_1)
		{\mathrm e}^{ig\phi_m(t_2)}    \psi^\dagger_d(m,t_2)
		|\Omega\rangle_F|\Omega\rangle_B  \right.\right.\\
		&=
		\left._B\langle\Omega|	{\mathrm e}^{ig\phi_n(t_1)}{\mathrm e}^{ig\phi_m(t_2)}  |\Omega\rangle_B 		
		\left._F\langle\Omega|
		\psi_u(n,t_1) \psi^\dagger_d(m,t_2)
		|\Omega\rangle_F \right.\right.\\
		&= 
		\langle\Omega|	{\mathrm e}^{ig\phi_n(t_1)}{\mathrm e}^{ig\phi_m(t_2)}  |\Omega\rangle	
		\langle\Omega| \psi_u(n,t_1) \psi^\dagger_d(m,t_2) |\Omega\rangle
		\; .
	\end{split}
\end{equation}
We firstly evaluate the correlation function between $\psi_u$ and $\psi_d$ as shown in Eq. \eqref{eq:Psiuu}. 
To do this, we first need to solve for the correlation function of $\psi'$. Using Eq. \eqref{eq:psi'u} and Eq. \eqref{eq:psi'd}, we can express the correlation function of $\psi'$ as
\begin{equation}
	\begin{split}
		\label{eq:psiuu}
		\langle\Omega|\psi'_u(n,t_1)\psi'^\dagger_u(m,t_2)|\Omega\rangle
		&=\frac{1}{2N}\sum\limits_{q\ne0}{\mathrm e}^{-iE_q (t_1-t_2)+i(n-m)aq}
		+\frac{1}{N}
		\; ,
	\end{split}
\end{equation}	
\begin{equation}
	\begin{split}
		\label{eq:psidu}
		\langle\Omega|\psi'_d(n,t_1)\psi'^\dagger_u(m,t_2)|\Omega\rangle
		&=-\frac{1}{2N}\sum\limits_{q\ne0}	
		\text{sgn}(q)  {\mathrm e}^{-iE_q(t_1-t_2)+i(n-m-\frac{1}{2})aq}
		\; ,
	\end{split}
\end{equation}	
\begin{equation}
	\begin{split}
		\label{eq:psiud}
		\langle\Omega|\psi'_u(n,t_1)\psi'^\dagger_d(m,t_2)|\Omega\rangle
		&=-\frac{1}{2N}\sum\limits_{q\ne0}
		\text{sgn}(q) {\mathrm e}^{-iE_q (t_1-t_2)+i(n-m+\frac{1}{2})aq}
		\; ,
	\end{split}
\end{equation}	
\begin{equation}
	\begin{split}
		\label{eq:psidd}
		\langle\Omega|\psi'_d(n,t_1)\psi'^\dagger_d(m,t_2)|\Omega\rangle
		&=\frac{1}{2N}\sum\limits_{q\ne0} {\mathrm e}^{-iE_q(t_1-t_2)+i(n-m)aq}
		\; .
	\end{split}
\end{equation}
According to Eq. \eqref{eq:psiuu} -- Eq. \eqref{eq:psidd} and the field transformation relationship between $\psi$ and $\psi'$ from Eq. \eqref{eq:p'p}, we can calculate the correlation function of $\psi$ as follows:
\begin{equation}
	\begin{split}
		\label{eq:Psiuu.1}
		&\langle\Omega|\psi_u(n,t_1) \psi^\dagger_d(m,t_2)|\Omega\rangle\\		
		&=\frac{1}{2}  \langle\Omega|
		\left[\psi'_u(n,t_1)-\psi'_d(n,t_1)\right]
		\left[\psi'^\dagger_u(m,t_2)+\psi'^\dagger_d(m,t_2)\right]
		|\Omega\rangle\\
		&=\frac{1}{2} \Big[
		\langle\Omega|\psi'_u(n,t_1)\psi'^\dagger_u(m,t_2)|\Omega\rangle
		+
		\langle\Omega|\psi'_u(n,t_1)\psi'^\dagger_d(m,t_2)|\Omega\rangle\\
		&\qquad-
		\langle\Omega|\psi'_d(n,t_1)\psi'^\dagger_u(m,t_2)|\Omega\rangle
		-
		\langle\Omega|\psi'_d(n,t_1)\psi'^\dagger_d(m,t_2)|\Omega\rangle \Big]\\
		&=  \frac{1}{4N}\sum\limits_{q\ne0}{\mathrm e}^{-iE_q (t_1-t_2)+i(n-m)aq}
		+\frac{1}{2N}
		-
		\frac{1}{4N}\sum\limits_{q\ne0}
		\text{sgn}(q) {\mathrm e}^{-iE_q (t_1-t_2)+i(n-m+\frac{1}{2})aq}\\
		&~~~+
		\frac{1}{4N}\sum\limits_{q\ne0}	
		\text{sgn}(q)  {\mathrm e}^{-iE_q(t_1-t_2)+i(n-m-\frac{1}{2})aq}
		-
		\frac{1}{4N}\sum\limits_{q\ne0} {\mathrm e}^{-iE_q(t_1-t_2)+i(n-m)aq}  \\
		&=  -
		\frac{1}{4N}\sum\limits_{q\ne0}
		\text{sgn}(q) {\mathrm e}^{-iE_q (t_1-t_2)+i(n-m+\frac{1}{2})aq}
		+
		\frac{1}{4N}\sum\limits_{q\ne0}	
		\text{sgn}(q)  {\mathrm e}^{-iE_q(t_1-t_2)+i(n-m-\frac{1}{2})aq}+\frac{1}{2N} 
		\; .
	\end{split}
\end{equation}

We can analyze the continuum limit of the correlation function \eqref{eq:Psiuu.1} using a method similar to that used for bosonic fields. While keeping the lattice spacing $a$ constant, we let the length of the spatial direction $L = Na$ approach infinity, which is equivalent to considering a large $N$ limit. Consequently, the sum over $q$ in Eq. \eqref{eq:Psiuu.1} can be replaced by the integral:
\begin{equation}
	\begin{split}
		\label{eq:Psiuu.2}
		&\lim\limits_{N\to\infty}\langle\Omega|\psi_u(n,t_1) \psi^\dagger_d(m,t_2)|\Omega\rangle\\		
		&=\frac{a}{8\pi}\lim\limits_{N\to\infty}\left[
		-
		\sum\limits_{q\ne0}\Delta q\cdot
		\text{sgn}(q) {\mathrm e}^{-iE_q (t_1-t_2)+i(n-m+\frac{1}{2})aq}
		+
		\sum\limits_{q\ne0}	\Delta q \cdot
		\text{sgn}(q)  {\mathrm e}^{-iE_q(t_1-t_2)+i(n-m-\frac{1}{2})aq}\right]\\
		&=\frac{a}{8\pi}\left[
		-
		\int_{-\frac{\pi}{a}}^{\frac{\pi}{a}} \mathrm d q\;
		\text{sgn}(q) {\mathrm e}^{-iE_q (t_1-t_2)+i(n-m+\frac{1}{2})aq}
		+
		\int_{-\frac{\pi}{a}}^{\frac{\pi}{a}} \mathrm d q \;
		\text{sgn}(q)  {\mathrm e}^{-iE_q(t_1-t_2)+i(n-m-\frac{1}{2})aq}\right]
		\; .
	\end{split}
\end{equation}
According to Eq. \eqref{eq:psi(n)}, in the continuum limit, the relationship between the discrete field $\psi(n,t)$ and the continuous fermionic field $\psi_c(x,t)$ is given by $\psi(n,t)=\sqrt a\psi_c(x,t)\Big|_{x=na}$.  Considering $x_1=na$ and $x_2=ma$, we can express the correlation function for the continuous fermionic field as follows:
\begin{equation}
	\begin{split}
		\label{eq:3Psiuu}
		&\langle\Omega|\psi_{cu}(x_1,t_1) \psi^\dagger_{cd}(x_2,t_2)|\Omega\rangle\\	
		&=\lim\limits_{a\to0}\frac{1}{a} \lim\limits_{N\to\infty}
		\langle\Omega|\psi_u(n,t_1) \psi^\dagger_d(m,t_2)|\Omega\rangle\\		
		&=\frac{1}{8\pi}\lim\limits_{a\to0}
		\left[
		-
		\int_{-\frac{\pi}{a}}^{\frac{\pi}{a}} \mathrm d q \;
		\text{sgn}(q) {\mathrm e}^{-iE_q (t_1-t_2)+i(n-m+\frac{1}{2})aq}
		+
		\int_{-\frac{\pi}{a}}^{\frac{\pi}{a}} \mathrm d q \;
		\text{sgn}(q)  {\mathrm e}^{-iE_q(t_1-t_2)+i(n-m-\frac{1}{2})aq}
		\right]\\
		&=\frac{1}{8\pi}\lim\limits_{a\to0}
		\left[
		-
		\int_{-\frac{\pi}{a}}^{\frac{\pi}{a}} \mathrm d q \;
		\text{sgn}(q) {\mathrm e}^{-iE_q (t_1-t_2)+i(x_1-x_2+\frac{1}{2}a)q}
		+
		\int_{-\frac{\pi}{a}}^{\frac{\pi}{a}} \mathrm d q \;
		\text{sgn}(q)  {\mathrm e}^{-iE_q(t_1-t_2)+i(x_1-x_2-\frac{1}{2}a)q}
		\right]
		\; .
	\end{split}
\end{equation}

The two terms in the last line of \eqref{eq:3Psiuu} exhibit similar structures, allowing us to analyze them in unison.
We divide the integral range into two segments:
\begin{equation}\label{eq:4Psiuu}
	\begin{split}
		&\int_{-\frac{\pi}{a}}^{\frac{\pi}{a}} \mathrm d q \;
		\text{sgn}(q)  {\mathrm e}^{-iE_q(t_1-t_2)+i(x_1-x_2\pm\frac{1}{2}a)q} \\
		&=\int_{0}^{\frac{\pi}{a}} \mathrm d q \;
		{\mathrm e}^{-iE_q(t_1-t_2) +i(x_1-x_2\pm\frac{1}{2}a)q}
		-\int_{-\frac{\pi}{a}}^{0} \mathrm d q \;
		{\mathrm e}^{-iE_q(t_1-t_2) +i(x_1-x_2\pm\frac{1}{2}a)q}\\
		&=\lim\limits_{\epsilon\to 0}\int_{0}^{\frac{\pi}{a}} \mathrm d q \;
		{\mathrm e}^{-iE_q(t_1-t_2) +i(x_1-x_2\pm\frac{1}{2}a)q-\epsilon q}
		-\lim\limits_{\epsilon\to 0}\int_{-\frac{\pi}{a}}^{0} \mathrm d q \; 
		{\mathrm e}^{-iE_q(t_1-t_2) +i(x_1-x_2\pm\frac{1}{2}a)q+\epsilon q}
		\; .
	\end{split}
\end{equation}
The subsequent calculation closely resembles the procedure we employed for the bosonic field. Take the limit as the lattice spacing $a$ tends to zero for the first integral in the last line of Eq. \eqref{eq:4Psiuu} and partition the integral into two distinct sections:
\begin{equation}\label{eq:5Psiuu}
	\begin{split}
		&\lim\limits_{a\to 0}\int_{0}^{\frac{\pi}{a}} \mathrm d q \;
		{\mathrm e}^{-iE_q(t_1-t_2) +i(x_1-x_2\pm\frac{1}{2}a)q-\epsilon q}\\
		&=\lim\limits_{a\to 0}\int_{0}^{\frac{\pi}{\sqrt{a}}} \mathrm d q \;
		{\mathrm e}^{-iE_q(t_1-t_2) +i(x_1-x_2\pm\frac{1}{2}a)q-\epsilon q}
		+\lim\limits_{a\to 0}\int_{\frac{\pi}{\sqrt{a}}}^{\frac{\pi}{a}} \mathrm d q \;
		{\mathrm e}^{-iE_q(t_1-t_2) +i(x_1-x_2\pm\frac{1}{2}a)q-\epsilon q}
		\; .
	\end{split}
\end{equation}
To evaluate the second term in Eq. \eqref{eq:5Psiuu}, we can bound it using a series of inequalities:
\begin{equation}
	\begin{split}
		\left|\lim\limits_{a\to 0}\int_{\frac{\pi}{\sqrt{a}}}^{\frac{\pi}{a}} \mathrm d q \;
		{\mathrm e}^{-iE_q(t_1-t_2) +i(x_1-x_2\pm\frac{1}{2}a)q-\epsilon q}\right|&\leqslant\lim\limits_{a\to 0}\int_{\frac{\pi}{\sqrt{a}}}^{\frac{\pi}{a}} \mathrm d q \;
		\left| {\mathrm e}^{-iE_q(t_1-t_2) +i(x_1-x_2\pm\frac{1}{2}a)q} \right|
		{\mathrm e}^{-\epsilon q}\\
		&\leqslant\lim\limits_{a\to 0}\int_{\frac{\pi}{\sqrt{a}}}^{\frac{\pi}{a}} \mathrm d q \;  {\mathrm e}^{-\epsilon q}
		=\lim\limits_{a\to 0}\frac{{\mathrm e}^{-\epsilon \frac{\pi}{a}}-{\mathrm e}^{-\epsilon \frac{\pi}{\sqrt a}}}{-\epsilon}
		=0
		\; .
	\end{split}
\end{equation}
This implies that the second term in Eq. \eqref{eq:5Psiuu} should vanish.  As a result, utilizing the dispersion relation $ E_q =\left|\frac{2}{a} \sin \frac{aq}{2}\right| $, Eq. \eqref{eq:5Psiuu} can be expressed as
\begin{equation}\label{6Psiuu}
	\begin{split}
		\lim\limits_{a\to 0}\int_{0}^{\frac{\pi}{a}} \mathrm d q \;
		{\mathrm e}^{-iE_q(t_1-t_2) +i(x_1-x_2\pm\frac{1}{2}a)q-\epsilon q}
		&=\lim\limits_{a\to 0}\int_{0}^{\frac{\pi}{\sqrt{a}}} \mathrm d q \; 
		{\mathrm e}^{-iE_q(t_1-t_2) +i(x_1-x_2\pm\frac{1}{2}a)q-\epsilon q}\\
		&=\int_{0}^{\infty} \mathrm d q \;
		{\mathrm e}^{-i|q|(t_1-t_2) +i(x_1-x_2)q-\epsilon q}
		\; .
	\end{split}
\end{equation}
Similarly, taking the continuum limit for the the second integral of the last line in Eq. \eqref{eq:4Psiuu} yields a similar result, ultimately leading to the continuum limit of \eqref{eq:4Psiuu} as
\begin{equation}\label{eq:7Psiuu}
	\lim\limits_{a\to 0}\int_{-\frac{\pi}{a}}^{\frac{\pi}{a}} \mathrm d q \;
	\text{sgn}(q)  {\mathrm e}^{-iE_q(t_1-t_2)+i(x_1-x_2\pm\frac{1}{2}a)q} 
	=\int_{-\infty}^{\infty} \mathrm d q \;
	\text{sgn}(q)  {\mathrm e}^{-i|q|(t_1-t_2) +i(x_1-x_2)q}
	\; .
\end{equation} 
Substituting \eqref{eq:7Psiuu} into Eq. \eqref{eq:3Psiuu}, we obtain
\begin{equation}
	\begin{split}
		\label{eq:Psiuu.3}
		&\langle\Omega|\psi_{cu}(x_1,t_1) \psi^\dagger_{cd}(x_2,t_2)|\Omega\rangle\\	
		&=\lim\limits_{a\to0}\frac{1}{a} \lim\limits_{N\to\infty}
		\langle\Omega|\psi_u(n,t_1) \psi^\dagger_d(m,t_2)|\Omega\rangle\\		
		&=\frac{1}{8\pi}
		\left[
		-
		\int_{-\infty}^{\infty} \mathrm d q \;
		\text{sgn}(q) {\mathrm e}^{-i (t_1-t_2)\left|q\right|+i(x_1-x_2)q}
		+
		\int_{-\infty}^{\infty} \mathrm d q \;
		\text{sgn}(q)  {\mathrm e}^{-i(t_1-t_2)\left|q\right|+i(x_1-x_2)q}
		\right]\\
		&=0
		\; .
	\end{split}
\end{equation}
Our discussion has covered the fermionic correlation terms in the last line of Eq. \eqref{eq:Psiuu}. Next, we will explore the bosonic correlation term $\langle\Omega|{\mathrm e}^{ig\phi_n(t_1)}{\mathrm e}^{ig\phi_m(t_2)}|\Omega\rangle$ in the last line of Eq. \eqref{eq:Psiuu}. A detailed calculation of this term yields
\begin{equation}
	\begin{split}
		\label{eq:phiuu}
		\langle\Omega|	{\mathrm e}^{ig\phi_n(t_1)}{\mathrm e}^{ig\phi_m(t_2)}  |\Omega\rangle	
		={\mathrm e}^{-\frac{1}{2}g^2[\phi^+_n,\phi^-_n]}{\mathrm e}^{-\frac{1}{2}g^2[\phi^+_m,\phi^-_m]}
		{\mathrm e}^{-g^2[\phi^+_n(t_1),\phi^-_m(t_2)]}
		\; .
	\end{split}
\end{equation}
The commutation relation for $\phi_n^-$ and $\phi_n^+$ has been computed in the previous section (see \eqref{[phi]}). 
Furthermore, by employing Eq. \eqref{eq:phi}-\eqref{eq:-}, we can express the commutation relation for the bosonic field at distinct spacetime points in terms of the correlation function:
\begin{equation}\label{eq:phi+-}
	\begin{split}		
		\left[\phi^+_n(t_1),\phi^-_m(t_2)\right]
		&=\langle\Omega|\left[\phi^+_n(t_1),\phi^-_m(t_2)\right]|\Omega\rangle=\langle\Omega|\phi^+_n(t_1)\phi^-_m(t_2)|\Omega\rangle\\
		&=\langle\Omega|\left[\phi^-_n(t_1)+\phi^+_n(t_1)\right]\left[\phi^-_m(t_2)+\phi^+_m(t_2)\right]|\Omega\rangle\\
		&=\langle\Omega|\phi_n(t_1)\phi_m(t_2)|\Omega\rangle
		\; .
	\end{split}
\end{equation}
Substituting \eqref{[phi]} and \eqref{eq:phi+-} into Eq. \eqref{eq:phiuu}, we obtain
\begin{equation}
	\label{eq:phiuu1}
	\langle\Omega|	{\mathrm e}^{ig\phi_n(t_1)}{\mathrm e}^{ig\phi_m(t_2)}  |\Omega\rangle	
	={\mathrm e}^{-g^2f(0)}	{\mathrm e}^{-g^2\langle\Omega|\phi_n(t_1)\phi_m(t_2)|\Omega\rangle}
	\; .
\end{equation}
Moving the constant ${\mathrm e}^{-g^2f(0)}$ from the right side of Eq. \eqref{eq:phiuu1} to the left side and using Eq. \eqref{eq:ccphi}, we can take the continuum limit of the equation while keeping $x_1=na$ and $x_2=ma$ unchanged:
\begin{equation}
	\begin{split}
		\label{eq:phiuu2}
		\lim\limits_{a\to0}\lim\limits_{N\to\infty} {\mathrm e}^{g^2f(0)}
		\langle\Omega|	{\mathrm e}^{ig\phi_n(t_1)}{\mathrm e}^{ig\phi_m(t_2)}  |\Omega\rangle	
		=	{\mathrm e}^{-g^2\langle\Omega|\phi(x_1,t_1)\phi(x_2,t_2)|\Omega\rangle}
		\; .
	\end{split}
\end{equation} 
Here we introduce the renormalized fermionic field $\Psi(n,t)_R\equiv{\mathrm e}^{\frac{1}{2}g^2f(0)}\Psi(n,t)$, then according to Eq. \eqref{eq:Psiuu}, \eqref{eq:Psiuu.3}, and \eqref{eq:phiuu2}, the continuum limit of the renormalized correlation function $\langle\Omega|\Psi_u(n,t_1)_R\bar\Psi_u(m,t_2)_R|\Omega\rangle$ is given by
\begin{equation}
	\begin{split}
		\label{eq:Psiuu.4}
		\langle\Omega|\Psi_{ru}(x_1,t_1)\bar\Psi_{ru}(x_2,t_2)|\Omega\rangle
		&=\lim\limits_{a\to0}\frac{1}{a} \lim\limits_{N\to\infty}
		\langle\Omega|\Psi_u(n,t_1)_R\bar\Psi_u(m,t_2)_R|\Omega\rangle\\
		&=\lim\limits_{a\to0}\frac{1}{a} \lim\limits_{N\to\infty} {\mathrm e}^{g^2f(0)}
		\langle\Omega|\Psi_u(n,t_1)\bar\Psi_u(m,t_2)|\Omega\rangle\\
		&=
		{\mathrm e}^{-g^2\langle\Omega|\phi(x_1,t_1)\phi(x_2,t_2)|\Omega\rangle}
		\langle\Omega|\psi_{cu}(x_1,t_1) \psi^\dagger_{cd}(x_2,t_2)|\Omega\rangle\\
		&=0
		\;,
	\end{split}
\end{equation}
where, we utilized the relation between the discrete renormalized fermionic field $\Psi(n,t)_R$ and the continuous renormalized fermionic field $\Psi_r(x,t)$, which is given by $\Psi(n,t)_R=\sqrt a \Psi_r(x,t)\Big|_{x=na}$.

Next, we analyze the other component of $\langle\Omega|\Psi(n,t_1)\bar\Psi(m,t_2)|\Omega\rangle$:
\begin{equation}
	\begin{split}
		\label{eq:Psidu}
		\langle\Omega|\Psi_d(n,t_1)\bar\Psi_u(m,t_2)|\Omega\rangle
		&=\langle\Omega|\Psi_d(n,t_1)\Psi^\dagger_d(m,t_2)|\Omega\rangle\\
		&=\langle\Omega|
		{\mathrm e}^{-ig\phi_n(t_1)}	\psi_d(n,t_1)
		{\mathrm e}^{ig\phi_m(t_2)}    \psi^\dagger_d(m,t_2)
		|\Omega\rangle\\
		&=\left._B\langle\Omega| \left._F\langle\Omega|
		{\mathrm e}^{-ig\phi_n(t_1)}	\psi_d(n,t_1)
		{\mathrm e}^{ig\phi_m(t_2)}    \psi^\dagger_d(m,t_2)
		|\Omega\rangle_F|\Omega\rangle_B  \right.\right.\\
		&=
		\left._B\langle\Omega|	{\mathrm e}^{-ig\phi_n(t_1)}{\mathrm e}^{ig\phi_m(t_2)}  |\Omega\rangle_B 		
		\left._F\langle\Omega|
		\psi_d(n,t_1) \psi^\dagger_d(m,t_2)
		|\Omega\rangle_F \right.\right.\\
		&= 
		\langle\Omega|	{\mathrm e}^{-ig\phi_n(t_1)}{\mathrm e}^{ig\phi_m(t_2)}  |\Omega\rangle	
		\langle\Omega| \psi_d(n,t_1) \psi^\dagger_d(m,t_2) |\Omega\rangle
		\; .
	\end{split}
\end{equation}
Similar to the previous analysis, utilizing the relationship between $\psi$ and $\psi'$ as described in Eq. \eqref{eq:p'p}, and the correlation functions of $\psi'$ \eqref{eq:psiuu}--\eqref{eq:psidd}, we can calculate the two-point correlation function involving $\psi$ in the last line of Eq. \eqref{eq:Psidu} to be
\begin{equation}
	\begin{split}
		\label{eq:Psidu.1}
		&\langle\Omega|\psi_d(n,t_1) \psi^\dagger_d(m,t_2)|\Omega\rangle\\		
		&=\frac{1}{2}  \langle\Omega|
		\left[\psi'_u(n,t_1)+\psi'_d(n,t_1)\right]
		\left[\psi'^\dagger_u(m,t_2)+\psi'^\dagger_d(m,t_2)\right]
		|\Omega\rangle\\
		&=\frac{1}{2} \Bigg[
		\langle\Omega|\psi'_u(n,t_1)\psi'^\dagger_u(m,t_2)|\Omega\rangle
		+
		\langle\Omega|\psi'_u(n,t_1)\psi'^\dagger_d(m,t_2)|\Omega\rangle\\
		&\qquad+
		\langle\Omega|\psi'_d(n,t_1)\psi'^\dagger_u(m,t_2)|\Omega\rangle
		+
		\langle\Omega|\psi'_d(n,t_1)\psi'^\dagger_d(m,t_2)|\Omega\rangle \Bigg]\\
		&=  \frac{1}{4N}\sum\limits_{q\ne0}{\mathrm e}^{-iE_q (t_1-t_2)+i(n-m)aq}
		+\frac{1}{2N}
		-
		\frac{1}{4N}\sum\limits_{q\ne0}
		\text{sgn}(q) {\mathrm e}^{-iE_q (t_1-t_2)+i(n-m+\frac{1}{2})aq}\\
		&~\quad-
		\frac{1}{4N}\sum\limits_{q\ne0}	
		\text{sgn}(q)  {\mathrm e}^{-iE_q (t_1-t_2)+i(n-m-\frac{1}{2})aq}
		+
		\frac{1}{4N}\sum\limits_{q\ne0} {\mathrm e}^{-iE_q(t_1-t_2)+i(n-m)aq}  \\
		&=  
		-
		\frac{1}{4N}\sum\limits_{q\ne0}
		\text{sgn}(q) {\mathrm e}^{-iE_q (t_1-t_2)+i(n-m+\frac{1}{2})aq}
		-
		\frac{1}{4N}\sum\limits_{q\ne0}	
		\text{sgn}(q)  {\mathrm e}^{-iE_q(t_1-t_2)+i(n-m-\frac{1}{2})aq}\\
		&~\quad+
		\frac{1}{2N}\sum\limits_{q\ne0} {\mathrm e}^{-iE_q(t_1-t_2)+i(n-m)aq}	+\frac{1}{2N}
		\; .  
	\end{split}
\end{equation}
By utilizing Eq. \eqref{eq:7Psiuu}, we can calculate the continuum limit of Eq. \eqref{eq:Psidu.1} (keeping $x_1=na$ and $x_2=ma$ invariant):
\begin{equation}
	\begin{split}
		\label{Psidu.3}
		&\langle\Omega|\psi_{cd}(x_1,t_1) \psi^\dagger_{cd}(x_2,t_2)|\Omega\rangle\\	
		&=\lim\limits_{a\to0}\frac{1}{a} \lim\limits_{N\to\infty}
		\langle\Omega|\psi_d(n,t_1) \psi^\dagger_d(m,t_2)|\Omega\rangle\\		
		&=\frac{1}{8\pi}\lim\limits_{a\to0}
		\Bigg[
		-
		\int_{-\frac{\pi}{a}}^{\frac{\pi}{a}} \mathrm d q\;
		\text{sgn}(q) {\mathrm e}^{-i E_q (t_1-t_2)+i(n-m+\frac{1}{2})aq}
		-
		\int_{-\frac{\pi}{a}}^{\frac{\pi}{a}} \mathrm d q\;
		\text{sgn}(q)  {\mathrm e}^{-i E_q(t_1-t_2)+i(n-m-\frac{1}{2})aq}\\
		&\qquad\qquad\quad+
		2\int_{-\frac{\pi}{a}}^{\frac{\pi}{a}} \mathrm d q\; {\mathrm e}^{-i E_q(t_1-t_2)+i(n-m)aq}
		\Bigg]\\
		&=\frac{1}{8\pi}
		\Bigg[
		-
		\int_{-\infty}^{\infty} \mathrm d q\;
		\text{sgn}(q) {\mathrm e}^{-i|q| (t_1-t_2)+i(x_1-x_2)q}
		-
		\int_{-\infty}^{\infty} \mathrm d q\;
		\text{sgn}(q)  {\mathrm e}^{-i|q|(t_1-t_2)+i(x_1-x_2)q}\\
		&\qquad\quad+
		2\int_{-\infty}^{\infty} \mathrm d q\; {\mathrm e}^{-i|q|(t_1-t_2)+i(x_1-x_2)q}
		\Bigg]\\
		&=\frac{1}{4\pi}
		\int_{-\infty}^{\infty} \mathrm d q\;
		\left[1-\text{sgn}(q)\right] {\mathrm e}^{-i (t_1-t_2)\left|q\right|+i(x_1-x_2)q}=\frac{1}{2\pi}
		\int_{-\infty}^{0} \mathrm d q\;  {\mathrm e}^{i (t_1-t_2)q+i(x_1-x_2)q}\\
		&=-\frac{i}{2\pi}\frac{1}{(t_1-t_2)+(x_1-x_2)}
		\;,
	\end{split}
\end{equation}
where the third equality employs Eq. \eqref{eq:7Psiuu}. It's worth mentioning that, for the sake of simplicity and clarity, we have omitted terms related to the small quantity $\epsilon$ in the above equations when they don't lead to confusion.

Next, we investigate the other term in the last line of \eqref{eq:Psidu}, i.e., $\langle\Omega| {\mathrm e}^{-ig\phi_n(t_1)}{\mathrm e}^{ig\phi_m(t_2)} |\Omega\rangle$. With the help of Eq. \eqref{[phi]} and Eq. \eqref{eq:phi+-}, it can be expressed as
\begin{equation}
	\begin{split}
		\label{eq:phidu1}
		\langle\Omega|{\mathrm e}^{-ig\phi_n(t_1)}{\mathrm e}^{ig\phi_m(t_2)}  |\Omega\rangle={\mathrm e}^{-g^2f(0)}{\mathrm e}^{g^2\langle\Omega|\phi_n(t_1)\phi_m(t_2)|\Omega\rangle}
		\; .
	\end{split}
\end{equation}
Utilizing Eq. \eqref{eq:ccphi}, we can derive the continuum limit for \eqref{eq:phidu1}:
\begin{equation}
	\begin{split}
		\label{eq:phidu2}
		\lim\limits_{a\to0}\lim\limits_{N\to\infty} {\mathrm e}^{g^2f(0)}
		\langle\Omega|	{\mathrm e}^{ig\phi_n(t_1)}{\mathrm e}^{ig\phi_m(t_2)}  |\Omega\rangle	
		=	{\mathrm e}^{g^2\langle\Omega|\phi(x_1,t_1)\phi(x_2,t_2)|\Omega\rangle}
		\; .
	\end{split}
\end{equation}
Based on Eq. \eqref{eq:Psidu}, \eqref{Psidu.3}, and \eqref{eq:phidu2}, the continuum limit of the renormalized correlation function $\langle\Omega|\Psi_d(n,t_1)_R\bar\Psi_u(m,t_2)_R|\Omega\rangle$ can be expressed as
\begin{equation}
	\begin{split}
		\label{eq:Psidu.4}
		\langle\Omega|\Psi_{rd}(x_1,t_1)\bar\Psi_{ru}(x_2,t_2)|\Omega\rangle
		&=\lim\limits_{a\to0}\frac{1}{a} \lim\limits_{N\to\infty}
		\langle\Omega|\Psi_d(n,t_1)_R\bar\Psi_u(m,t_2)_R|\Omega\rangle\\
		&=\lim\limits_{a\to0}\frac{1}{a} \lim\limits_{N\to\infty} {\mathrm e}^{g^2f(0)}
		\langle\Omega|\Psi_d(n,t_1)\bar\Psi_u(m,t_2)|\Omega\rangle\\
		&=-\frac{i}{2\pi}\frac{1}{(t_1-t_2)+(x_1-x_2)}
		{\mathrm e}^{g^2\langle\Omega|\phi(x_1,t_1)\phi(x_2,t_2)|\Omega\rangle}
		\; .
	\end{split}
\end{equation}

Using a similar method, we can calculate the other components  $\langle\Omega|\Psi_u(n,t_1)\bar\Psi_d(m,t_2)|\Omega\rangle$ and $\langle\Omega|\Psi_d(n,t_1)\bar\Psi_d(m,t_2)|\Omega\rangle$, and derive the continuum limits of the corresponding renormalized correlation functions:
\begin{equation}\label{eq:Psiud1}		
	\langle\Omega|\Psi_{ru}(x_1,t_1)\bar\Psi_{rd}(x_2,t_2)|\Omega\rangle
	=-\frac{i}{2\pi}\frac{1}{(t_1-t_2)-(x_1-x_2)}
	{\mathrm e}^{g^2\langle\Omega|\phi(x_1,t_1)\phi(x_2,t_2)|\Omega\rangle}
	\; ,
\end{equation}
\begin{equation}\label{eq:Psiud2}
	\langle\Omega|\Psi_{rd}(x_1,t_1)\bar\Psi_{rd}(x_2,t_2)|\Omega\rangle
	=0
	\; .
\end{equation}
Finally, by combining all components of the two-point correlation function \eqref{eq:Psiuu.4}, \eqref{eq:Psidu.4}, \eqref{eq:Psiud1}, and \eqref{eq:Psiud2} (i.e. $\langle\Omega|\Psi_{ru}(x_1,t_1)\bar\Psi_{ru}(x_2,t_2)|\Omega\rangle$, $\langle\Omega|\Psi_{rd}(x_1,t_1)\bar\Psi_{ru}(x_2,t_2)|\Omega\rangle$, $\langle\Omega|\Psi_{ru}(x_1,t_1)\bar\Psi_{rd}(x_2,t_2)|\Omega\rangle$, and $\langle\Omega|\Psi_{rd}(x_1,t_1)\bar\Psi_{rd}(x_2,t_2)|\Omega\rangle$), we can obtain the continuum limit of the two-point correlation function for the fermionic field:
\begin{equation}
	\label{eq:PsiPsi}
	\langle\Omega|\Psi_r(x_1,t_1)\bar\Psi_r(x_2,t_2)|\Omega\rangle
	=-\frac{i}{2\pi}\frac{\gamma_\mu x^\mu}{x^2}
	{\mathrm e}^{g^2\langle\Omega|\phi(x_1,t_1)\phi(x_2,t_2)|\Omega\rangle}
	\;,
\end{equation}
where $x^\mu=(x^0,x^1)=(t_1-t_2,x_1-x_2)$. 
If we further define the renormalized coupling constant as $g_r\equiv F^{-\frac{1}{2}}g$, and combine it with the definition of the renormalized bosonic field, $\phi_r\equiv F^{\frac{1}{2}}\phi$, we obtain
\begin{equation}
	\label{eq:2PsiPsi}
	\langle\Omega|\Psi_r(x_1,t_1)\bar\Psi_r(x_2,t_2)|\Omega\rangle
	=-\frac{i}{2\pi}\frac{\gamma_\mu x^\mu}{x^2}
	{\mathrm e}^{g_r^2\langle\Omega|\phi_r(x_1,t_1)\phi_r(x_2,t_2)|\Omega\rangle}
	\; .
\end{equation}
This is indeed the two-point fermion correlation function \eqref{eq:2ptfermion} in the original RS model. Please note that we assume $t_1>t_2$ and, therefore, omit the time-ordering operator $T$.

It's worth noting that here, we have defined $g_r\equiv F^{-\frac{1}{2}}g$ to match the two-point fermion correlation functions on the lattice with those in the original RS model. However, in practice, the most accurate definition of the renormalized coupling constant typically involves the three-point correlation functions corresponding to interaction vertices. So, it is essential to verify whether $g_r\equiv F^{-\frac{1}{2}}g$ also ensures that the continuum limit of the three-point correlation functions matches those of the original RS model.

Next, we calculate the three-point correlation function $\langle\Omega|\phi_n(t_1)\Psi(m,t_2)\bar\Psi(0,0)|\Omega\rangle$.
Based on the relationship between $\Psi$ and $\psi$ in Eq. \eqref{pP}, one component of the lattice three-point correlation function can be expressed as
\begin{equation}\label{eq:bff1}
	\begin{split}		
		&\langle\Omega|\phi_n(t_1)\Psi_u(m,t_2)\bar\Psi_u(0,0)|\Omega\rangle\\
		&=
		\langle\Omega|\phi_n(t_1)\Psi_u(m,t_2)\Psi^\dagger_d(0,0)|\Omega\rangle\\
         &=
		\langle\Omega|\phi^+_n(t_1)
		{\mathrm e}^{ig\phi_m(t_2)}	\psi_u(m,t_2)
		{\mathrm e}^{ig\phi_0(0)} \psi^\dagger_d(0,0)|\Omega\rangle\\
		&=
		\langle\Omega|\phi_n(t_1)\phi_m(t_2)|\Omega\rangle
		\langle\Omega|ig{\mathrm e}^{ig\phi_m(t_2)}
		\psi_u(m,t_2)
		{\mathrm e}^{ig\phi_0(0)} \psi^\dagger_d(0,0)|\Omega\rangle\\
		&\quad+
		\langle\Omega|\phi_n(t_1)\phi_0(0)|\Omega\rangle 
		\langle\Omega|  {\mathrm e}^{ig\phi_m(t_2)}	\psi_u(m,t_2)ig{\mathrm e}^{ig\phi_0(0)}\psi^\dagger_d(0,0)|\Omega\rangle\\
		&=ig\langle\Omega| \Psi_u(m,t_2)\bar\Psi_u(0,0)|\Omega\rangle
		\Big[\langle\Omega|\phi_n(t_1)\phi_m(t_2)|\Omega\rangle
		+
		\langle\Omega|\phi_n(t_1)\phi_0(0)|\Omega\rangle \Big]
		\;,
	\end{split}
\end{equation}
where the third equality in the above expression is utilized from the fact that
\begin{equation}
	\left[\phi_n^+(t_1),{\mathrm e}^{ig\phi_m(t_2)}\right]=ig{\mathrm e}^{ig\phi_m(t_2)}[\phi^+_n(t_1),\phi^-_m(t_2)]
	=ig{\mathrm e}^{ig\phi_m(t_2)}\langle\Omega|\phi_n(t_1)\phi_m(t_2)|\Omega\rangle
	\; .
\end{equation}
Finding the continuum limit for the renormalized three-point correlation function on the lattice \eqref{eq:bff1} using \eqref{eq:Psiuu.4} is a straightforward process:
\begin{equation}\label{eq:bff2}
	\begin{split}		
		&\langle\Omega|\phi_r(x)\Psi_{ru}(y)\bar\Psi_{ru}(0)|\Omega\rangle\\
		&=ig_r
		\langle\Omega|\Psi_{ru}(y)\bar\Psi_{ru}(0)|\Omega\rangle
		\Big[\langle\Omega|\phi_r(x)\phi_r(y)|\Omega\rangle
		+\langle\Omega|\phi_r(x)\phi_r(0)|\Omega\rangle\Big]\\
		&=0\\
		&=ig_r
		\langle\Omega|\Psi_{ru}(y)\bar\Psi_{ru}(0)|\Omega\rangle
		\Big[\langle\Omega|\phi_r(x)\phi_r(y)|\Omega\rangle
		-\langle\Omega|\phi_r(x)\phi_r(0)|\Omega\rangle\Big]\\
		&=ig_r\sum\limits_{\beta=u,d}(\gamma^5)_{u\beta}
		\langle\Omega|\Psi_{r\beta}(y)\bar\Psi_{ru}(0)|\Omega\rangle
		\Big[\langle\Omega|\phi_r(x)\phi_r(y)|\Omega\rangle
		-\langle\Omega|\phi_r(x)\phi_r(0)|\Omega\rangle\Big]
		\;,
	\end{split}
\end{equation}
where the second equality is derived from the definition of $g_r\equiv F^{-\frac{1}{2}}g$, and the final equality arises from the matrix expression for $\gamma^5$ in \eqref{gamma0}.

In a similar vein, we can compute other components of $\langle\Omega|\phi_n(t_1)\Psi(m,t_2)\bar\Psi(0,0)|\Omega\rangle$ and its continuum limit, for instance,
\begin{equation}\label{eq:bff3}
	\begin{split}		
		&\langle\Omega|\phi_n(t_1)\Psi_d(m,t_2)\bar\Psi_u(0,0)|\Omega\rangle\\
		&=
		\langle\Omega|\phi_n(t_1)\Psi_d(m,t_2)\Psi^\dagger_d(0,0)|\Omega\rangle\\
		&=
		\langle\Omega|\phi^+_n(t_1)
		{\mathrm e}^{-ig\phi_m(t_2)}	\psi_d(m,t_2)
		{\mathrm e}^{ig\phi_0(0)} \psi^\dagger_d(0,0)|\Omega\rangle\\
		&=
		\langle\Omega|\phi_n(t_1)\phi_m(t_2)|\Omega\rangle
		\langle\Omega|-ig{\mathrm e}^{-ig\phi_m(t_2)}
		\psi_d(m,t_2)
		{\mathrm e}^{ig\phi_0(0)} \psi^\dagger_d(0,0)|\Omega\rangle\\
		&\quad ~+
		\langle\Omega|\phi_n(t_1)\phi_0(0)|\Omega\rangle 
		\langle\Omega|  {\mathrm e}^{-ig\phi_m(t_2)}	\psi_d(m,t_2)ig{\mathrm e}^{ig\phi_0(0)}\psi^\dagger_d(0,0)|\Omega\rangle\\
		&=-ig\langle\Omega| \Psi_d(m,t_2)\bar\Psi_u(0,0)|\Omega\rangle
		\Big[\langle\Omega|\phi_n(t_1)\phi_m(t_2)|\Omega\rangle
		-
		\langle\Omega|\phi_n(t_1)\phi_0(0)|\Omega\rangle \Big]
		\;.
	\end{split}
\end{equation}
After renormalization, the corresponding continuum limit of \eqref{eq:bff3} can be expressed as
\begin{equation}\label{eq:bff4}
	\begin{split}		
		&\langle\Omega|\phi_r(x)\Psi_{rd}(y)\bar\Psi_{ru}(0)|\Omega\rangle\\
		&=-ig_r
		\langle\Omega|\Psi_{rd}(y)\bar\Psi_{ru}(0)|\Omega\rangle
		\Big[\langle\Omega|\phi_r(x)\phi_r(y)|\Omega\rangle
		-\langle\Omega|\phi_r(x)\phi_r(0)|\Omega\rangle\Big]\\
		&=ig_r\sum\limits_{\beta=u,d}(\gamma^5)_{d\beta}
		\langle\Omega|\Psi_{r\beta}(y)\bar\Psi_{ru}(0)|\Omega\rangle
		\Big[\langle\Omega|\phi_r(x)\phi_r(y)|\Omega\rangle
		-\langle\Omega|\phi_r(x)\phi_r(0)|\Omega\rangle\Big]
		\; .
	\end{split}
\end{equation}
Similarly, we can calculate the expressions for the other components. Combining the results from all components, we obtain the continuum limit of the lattice three-point correlation function corresponding to the interaction vertex as
\begin{equation}\label{eq:bff5}
	\begin{split}		
		\langle\Omega|\phi_r(x)\Psi_r(y)\bar\Psi_r(0)|\Omega\rangle
		=ig_r\gamma^5
		\langle\Omega|\Psi_r(y)\bar\Psi_r(0)|\Omega\rangle
		\Big[\langle\Omega|\phi_r(x)\phi_r(y)|\Omega\rangle
		-\langle\Omega|\phi_r(x)\phi_r(0)|\Omega\rangle\Big]
		\; .
	\end{split}
\end{equation}
Here, we continue to assume that $x^0>y^0>0$ and omit the time-ordering operators $T$. 
Equation \eqref{eq:bff5} corresponds precisely to the three-point correlation function in the original RS model associated with the interaction vertex, as given in Eq. \eqref{eq:3ptcorrelation}. This also confirms that the renormalized coupling constant can indeed be defined as $g_r\equiv F^{-\frac{1}{2}}g$.


Earlier, we derived some relationships between lattice bare quantities and the corresponding bare quantities in the original RS model, as shown in Eq. \eqref{eq:FAg} and \eqref{eq:Am}. Now, we can further investigate the relationships concerning bare coupling constants. In the original RS model, there is a connection between the renormalized coupling constant and the bare coupling constant: $g_r=(1-g_0^2/\pi)^{-1/2}g_0$. In the lattice RS model, we have a relation between the renormalized coupling constant and the bare coupling constant: $g_r=F^{-1/2}g$. By combining these two relations, we can arrive at
\begin{equation}\label{gg0}
		F^{-\frac{1}{2}}g=(1-\frac{g_0^2}{\pi})^{-\frac{1}{2}}g_0
		\; .
	\end{equation}
Furthermore, by combining Eq. \eqref{eq:FAg} and \eqref{gg0}, we can establish the relationship between the lattice bare coupling constant and the original RS model's bare coupling constant as follows:
\begin{equation}\label{eq:g=g0}
		g=g_0
		\; .
	\end{equation}


 In this section, we have computed the two-point and three-point correlation functions for the RS model on the lattice. We discovered that their continuum limit converges to the correlation functions of the original model after undergoing renormalization. This observation demonstrates that the intricate Hamiltonian we constructed in equation \eqref{eq:Hamiltonian} effectively reproduces the correct behavior of the RS model in the continuum limit. 
It is worth noting that, in both the continuum limit of lattice theory and the original RS model, the bosonic field's correlation functions expressed in terms of the bare field and bare mass do not diverge (This is not the case for fermionic field correlation functions). Therefore, the bare fields $\phi$ and $\phi_0$ are well-defined quantities in both theories. We can directly equate $\phi$ to $\phi_0$, implying that the continuum limit of the lattice theory's bare field is equal to the original RS model's bare field. Consequently, the bare parameters in the lattice theory and the original RS model satisfy Eq. \eqref{eq:FAg}, \eqref{eq:Am}, and \eqref{eq:g=g0}, which can be concisely summarized as
\begin{equation}\label{eq:lcs}
	\begin{split}
		F=1-\frac{g_0^2}{\pi}\quad ,\qquad
		m=m_0\quad ,\qquad
		g=g_0\; .
	\end{split}
\end{equation}
However, for the fermionic field case, the situation becomes more intricate. In the original RS model, the correlation functions of fermionic fields expressed in terms of bare fermionic fields would exhibit divergences, which can also be observed from the computation of lattice correlation functions.
To be more specific, considering the definition of the renormalized lattice fermionic field as $\Psi(n,t)_R={\mathrm e}^{\frac{1}{2}g^2f(0)}\Psi(n,t)$ and the expression for $f(n)$ in equation \eqref{[phi]}, it is apparent that the field-strength renormalization constant $Z={\mathrm e}^{-g^2f(0)}$ for the lattice fermionic field tends to zero in the continuum limit.
To further summarize the relationship between the lattice RS model's renormalization parameters and bare parameters, we have the following:
\begin{equation}\label{rb}
	\begin{split}
	g_r\equiv F^{-\frac{1}{2}}g\quad ,\qquad
	\phi_r\equiv F^{\frac{1}{2}}\phi\quad ,\qquad
	\Psi(n)_R={\mathrm e}^{\frac{1}{2}g^2f(0)}\Psi(n)\; .
\end{split}
\end{equation}

\section{Eigenstates and Field Mixing}
\label{6}
In this section, our attention is drawn to the vacuum state and the clothed particles (one-particle states) of the complete Hamiltonian \eqref{eq:Hamiltonian}. 
Our objective is to understand how the Hamiltonian's original bosonic component and fermionic component combine to give rise to the eigenstates of the full Hamiltonian. 
This mixing of the bosonic field and the fermionic field provides insights into the interactions among the fundamental degrees of freedom, shaping the spectrum of the Hamiltonian and, ultimately, giving rise to the observable degrees of freedom.

The previously defined representations, namely $\{|\phi\rangle_B\}$ and $\{\prod\psi'^\dagger|0\rangle_F\}$, each represent a subspace of the entire system. The complete system encompasses both the fermionic and bosonic fields, and it can be described as a composite system. The basis for this composite system can be formed by taking the direct product of the basis states from the representation $\{\prod\psi'^\dagger|0\rangle_F\}$(as given in equation \eqref{eq:s}) and the basis states from the representation $\{|\phi\rangle_B\}$:
\begin{equation}\label{eq:sphi}
	\begin{split}	
		\psi'^\dagger_{\alpha_1}(n_1)\psi'^\dagger_{\alpha_2}(n_2)\psi'^\dagger_{\alpha_3}(n_3)\cdots\psi'^\dagger_{\alpha_s}(n_s)|0\rangle_F|\phi\rangle_B\; ,
	\end{split}
\end{equation}
where
$$s=1,2,\cdots,N \quad,\qquad \alpha_i=u,d  \quad,\qquad   n_i\ne n_j,\forall \alpha_i=\alpha_j\;\;.$$
We will denote the representation formed by the basis  \eqref{eq:sphi} as $\{\prod\psi'^\dagger|0\rangle_F|\phi\rangle_B\}$.


It is important to highlight that the original Hamiltonian \eqref{eq:Hamiltonian} is formulated in terms of $\Psi$ rather than $\psi'$. As a result, the entire Hilbert space should be a composite of the Hilbert spaces for both $\Psi$ and $\phi$. Consequently, the basis for the complete Hilbert space can also be constructed by taking the direct product of the bases for $\Psi$ and $\phi$, considering the commutation relations \eqref{eq:discommutationphiphi}, \eqref{eq:discommutationpsipsi}, and \eqref{eq:psiphi}.
A representation similar to $\{\prod\psi'^\dagger|0\rangle_F\}$ can also be defined for the fermionic field $\Psi$. We define the quantum state $|0\rangle$ as an eigenstate of the fermionic field $\Psi$ with an eigenvalue of zero:
%
\begin{equation}\label{eq:P0}
	\begin{split}	
		\Psi_u(n)|0\rangle =0\; ,\qquad\quad
		\Psi_d(n)|0\rangle =0\; .
	\end{split}
\end{equation}
Following this, we can apply any number of fermionic field operators $\Psi^\dagger(n)$ to the state $|0\rangle$, yielding a series of quantum states:
\begin{equation}\label{eq:S}
		\begin{split}	
			\Psi^\dagger_{\alpha_1}(n_1)\Psi^\dagger_{\alpha_2}(n_2)\Psi^\dagger_{\alpha_3}(n_3)\cdots\Psi^\dagger_{\alpha_s}(n_s)|0\rangle
			\; ,
		\end{split}
	\end{equation}	
	where $$s=1,2,\cdots,N \quad,\qquad \alpha_i=u,d  \quad,\qquad   n_i\ne n_j,\forall \alpha_i=\alpha_j\;\;.$$
It can be easily verified that the quantum states \eqref{eq:S} are normalized by the commutation relations between $\Psi$ and $\Psi^\dagger$ as well as the properties \eqref{eq:P0}. Using the states \eqref{eq:S} as a basis, we can define a new representation, denoted as $\{\prod\Psi^\dagger|0\rangle\}$.
As for the bosonic field $\phi$ in the original Hamiltonian \eqref{eq:Hamiltonian}, a representation similar to $\{|\phi\rangle_B\}$ can be defined. The eigenstates of $\phi$, denoted as $|\phi\rangle$, satisfy the equation
\begin{equation}\label{bbzt}
\hat\phi |\phi\rangle=\phi|\phi\rangle\; .
\end{equation}
The representation constructed using $|\phi\rangle$ as a basis is denoted as $\{|\phi\rangle\}$.
It's worth noting that even though $|\phi\rangle$ in Eq.\eqref{bbzt} and $|\phi\rangle_B$ in Eq. \eqref{Bbzt} are both eigenstates of the bosonic field $\hat\phi$ with an eigenvalue of $\phi$, they belong to different subsystems, and they are not the same quantum state.
The direct product of the basis from the representation $\{\prod\Psi^\dagger|0\rangle\}$ with the basis from the representation $\{|\phi\rangle\}$ yields a basis for the entire Hilbert space:
\begin{equation}\label{eq:Sphi}
	\begin{split}	
		\Psi^\dagger_{\alpha_1}(n_1)\Psi^\dagger_{\alpha_2}(n_2)\Psi^\dagger_{\alpha_3}(n_3)\cdots\Psi^\dagger_{\alpha_s}(n_s)|0\rangle|\phi\rangle\; ,
	\end{split}
\end{equation}
where $$s=1,2,\cdots,N \quad,\qquad \alpha_i=u,d  \quad,\qquad   n_i\ne n_j,\forall \alpha_i=\alpha_j\;\;.$$
The representation constructed using \eqref{eq:Sphi} as a basis is denoted as $\{\prod\Psi^\dagger|0\rangle|\phi\rangle\}$.



Let's now derive the relationship between the representation $\{\prod\psi'^\dagger|0\rangle_F|\phi\rangle_B\}$ and the representation $\{\prod\Psi^\dagger|0\rangle|\phi\rangle\}$.
In other words, we'll investigate the connection between the quantum states \eqref{eq:sphi} and \eqref{eq:Sphi}.
To begin, let's analyze the most special states in both representations, namely $|0\rangle_F|\phi\rangle_B$ and $|0\rangle|\phi\rangle$. Based on Eq. \eqref{pP} and Eq. \eqref{eq:p'p}, we can derive the relationship between the field operators $\psi'$ and the original field operators $\Psi$:
\begin{equation}
	\begin{split}	
		\label{eq:psiPsi}
		\psi_u'(n)&= \frac{\psi_u(n)+\psi_d(n)}{\sqrt{2}}
		= \frac{1}{\sqrt{2}}
		\left[{\mathrm e}^{-ig\phi_n}\Psi_u(n)+{\mathrm e}^{ig\phi_n}\Psi_d(n)\right]
		\; ,\\
		\psi_d'(n)&=\frac{-\psi_u(n)+\psi_d(n)}{\sqrt{2}}
		=\frac{1}{\sqrt{2}}
		\left[-{\mathrm e}^{-ig\phi_n}\Psi_u(n)+{\mathrm e}^{ig\phi_n}\Psi_d(n)\right]   
		\; ,   
	\end{split}
\end{equation}	
\begin{equation}
	\begin{split}	
		\label{eq:Psipsi}
		\Psi_u(n)&=\frac{1}{\sqrt{2}}{\mathrm e}^{ig\phi_n}\left[\psi_u'(n)-\psi_d'(n)\right]
		\; ,\\
		\Psi_d(n)&=\frac{1}{\sqrt{2}}{\mathrm e}^{-ig\phi_n}\left[\psi_u'(n)+\psi_d'(n)\right]
		\; .
	\end{split}
\end{equation}
Because both $\psi_u'(n)$ and $\psi_d'(n)$ annihilate the state $|0\rangle_F|\phi\rangle_B$,  utilizing the relation given by Eq. \eqref{eq:Psipsi} we have 
\begin{equation}	\Psi_u(n)|0\rangle_F|\phi\rangle_B=\Psi_d(n)|0\rangle_F|\phi\rangle_B=0\; .
\end{equation}
This demonstrates that the quantum state $|0\rangle_F|\phi\rangle_B$ is an eigenstate of the field operator $\Psi$ with an eigenvalue of zero. Additionally, this state is also an eigenstate of the field operator $\hat\phi$ with an eigenvalue of $\phi$. Similarly, the quantum state $|0\rangle|\phi\rangle$ is an eigenstate of field operator $\Psi$ with a zero eigenvalue and is simultaneously an eigenstate of field operator $\hat\phi$ with an eigenvalue of $\phi$. Therefore, we can conclude that 
\begin{equation}\label{eq:00}
	|0\rangle_F|\phi\rangle_B=|0\rangle|\phi\rangle\; .
\end{equation}
It is worth noting the basis of $\{\prod\psi'^\dagger|0\rangle_F|\phi\rangle_B\}$ and $\{\prod\Psi^\dagger|0\rangle|\phi\rangle\}$ are constructed based on $|0\rangle_F|\phi\rangle_B$ and $|0\rangle|\phi\rangle$, respectively. Therefore, we can establish the relationship between these two basis sets using Eq. \eqref{eq:00}. To be precise, by substituting \eqref{eq:psiPsi} and \eqref{eq:00} into \eqref{eq:sphi}, we can derive the explicit expression for the transformation between these two distinct bases:
\begin{equation}\label{eq:sSphi}
	\begin{split}	
		&\psi'^\dagger_{\alpha_1}(n_1)\psi'^\dagger_{\alpha_2}(n_2)\psi'^\dagger_{\alpha_3}(n_3)\cdots\psi'^\dagger_{\alpha_s}(n_s)|0\rangle_F|\phi\rangle_B\\
		&=\frac{1}{\sqrt{2}}
		\left[(-1)^{f(\alpha_1)}{\mathrm e}^{ig\phi_{n_1}}\Psi^\dagger_u(n_1)+{\mathrm e}^{-ig\phi_{n_1}}\Psi^\dagger_d(n_1)\right]\\
		&\quad\;
		\frac{1}{\sqrt{2}}
		\left[(-1)^{f(\alpha_2)}{\mathrm e}^{ig\phi_{n_2}}\Psi^\dagger_u(n_2)+{\mathrm e}^{-ig\phi_{n_2}}\Psi^\dagger_d(n_2)\right]\\
		&\quad\;\;
		\frac{1}{\sqrt{2}}
		\left[(-1)^{f(\alpha_3)}{\mathrm e}^{ig\phi_{n_3}}\Psi^\dagger_u(n_3)+{\mathrm e}^{-ig\phi_{n_3}}\Psi^\dagger_d(n_3)\right]\\
		&\quad\;\; \cdots \\
		&\quad\;\;\;
		\frac{1}{\sqrt{2}}
		\left[(-1)^{f(\alpha_s)}{\mathrm e}^{ig\phi_{n_s}}\Psi^\dagger_u(n_s)+{\mathrm e}^{-ig\phi_{n_s}}\Psi^\dagger_d(n_s)\right]|0\rangle|\phi\rangle
		\;,
	\end{split}
\end{equation}
where
\begin{equation}	
	f(\alpha)=\left\{\begin{array}{ll}		
		0 \; ,&\alpha=u \; ,\\
		1 \; ,& \alpha=d  \; .
	\end{array}\right.
\end{equation}
Upon expanding the product in the Eq. \eqref{eq:sSphi}, we obtain a series of summations over quantum states \eqref{eq:Sphi}. This indicates that we have effectively expressed the basis of $\{\prod\psi'^\dagger|0\rangle_F|\phi\rangle_B\}$ in terms of the basis of $\{\prod\Psi^\dagger|0\rangle|\phi\rangle\}$. With the relation given by Eq. \eqref{eq:sSphi}, we can now express the vacuum state and clothed particles of the Hamiltonian \eqref{eq:Hamiltonian} in the representation $\{\prod\Psi^\dagger|0\rangle|\phi\rangle\}$.

Using Eq. \eqref{eq:vacuumphi} and \eqref{eq:vacuum2}, we can derive the representation of the vacuum state $|\Omega\rangle$ of the total Hamiltonian \eqref{eq:Hamiltonian} in terms of the basis $\{\prod\psi'^\dagger|0\rangle_F|\phi\rangle_B\}$:
\begin{equation}
	\begin{split}
		\label{eq:vacuumpsi}
		|\Omega\rangle&=|\Omega\rangle_F|\Omega\rangle_B\\
		&=\sqrt{2}\mathcal{N}F^{\frac{1}{2}}\int \mathrm d\phi\;
		{\mathrm e}^{-\frac{F}{2}\sum\limits_{n,m}\mathcal{E}_{nm}\phi_n \phi_m} \\
		&\times\prod\limits_{k=-\frac{N-1}{2}}^{\frac{N-1}{2}}\frac{1}{\sqrt{2N}}
		\left[\text{sgn}(k){\mathrm e}^{-i\frac{1}{2}\frac{2\pi k}{N}}\sum\limits_n {\mathrm e}^{-in\frac{2\pi k}{N}}\psi'^\dagger_u(n)
		-\sum\limits_n {\mathrm e}^{-in\frac{2\pi k}{N}}\psi'^\dagger_d(n)\right]
		|0\rangle_F|\phi\rangle_B
		\; .
	\end{split}
\end{equation}
Based on the transformation relations between the bases of the representations $\{\prod\psi'^\dagger|0\rangle_F|\phi\rangle_B\}$ and $\{\prod\Psi^\dagger|0\rangle|\phi\rangle\}$ given by Eq. \eqref{eq:sSphi}, we can express the vacuum state \eqref{eq:vacuumpsi} in the representation $\{\prod\Psi^\dagger|0\rangle|\phi\rangle\}$ as follows:
\begin{equation}
	\begin{split}
		\label{eq:vacuumPsi}
		|\Omega\rangle
		&=\sqrt{2}\mathcal{N}F^{\frac{1}{2}}\int \mathrm d\phi\;
		{\mathrm e}^{-\frac{F}{2}\sum\limits_{n,m}\mathcal{E}_{nm}\phi_n \phi_m} \\
		&\qquad\times\prod\limits_{k=-\frac{N-1}{2}}^{\frac{N-1}{2}}
		\frac{1}{2\sqrt{N}}
		\Bigg\{\text{sgn}(k){\mathrm e}^{-i\frac{1}{2}\frac{2\pi k}{N}}\sum\limits_n {\mathrm e}^{-in\frac{2\pi k}{N}}
		\left[{\mathrm e}^{ig\phi_n}\Psi^\dagger_u(n)+{\mathrm e}^{-ig\phi_n}\Psi^\dagger_d(n)\right]\\
		&\qquad\qquad\qquad\qquad\qquad
		-\sum\limits_n {\mathrm e}^{-in\frac{2\pi k}{N}}
		\left[-{\mathrm e}^{ig\phi_n}\Psi^\dagger_u(n)+{\mathrm e}^{-ig\phi_n}\Psi^\dagger_d(n)\right]
		\Bigg\}
		|0\rangle|\phi\rangle\\
		&=\sqrt{2}\mathcal{N}F^{\frac{1}{2}}\int \mathrm d\phi\;
		{\mathrm e}^{-\frac{F}{2}\sum\limits_{n,m}\mathcal{E}_{nm}\phi_n \phi_m} \\
		&\qquad\times\prod\limits_{k=-\frac{N-1}{2}}^{\frac{N-1}{2}}
		\frac{1}{2\sqrt{N}}
		\Bigg[
		\left(\text{sgn}(k){\mathrm e}^{-i\frac{1}{2}\frac{2\pi k}{N}}+1\right)
		\sum\limits_n {\mathrm e}^{-in\frac{2\pi k}{N}}{\mathrm e}^{ig\phi_n}\Psi^\dagger_u(n)\\
		&\qquad\qquad\qquad\qquad\qquad
		+\left(\text{sgn}(k){\mathrm e}^{-i\frac{1}{2}\frac{2\pi k}{N}}-1\right)
		\sum\limits_n {\mathrm e}^{-in\frac{2\pi k}{N}}{\mathrm e}^{-ig\phi_n}\Psi^\dagger_d(n)
		\Bigg]
		|0\rangle|\phi\rangle
		\; .
	\end{split}
\end{equation}
	This is the physical vacuum state represented using the original field degrees of freedom in the Hamiltonian \eqref{eq:Hamiltonian}.
	Note that each fermionic field operator $\Psi^\dagger(n)$ is first attached by a value ${\mathrm e}^{ig\phi_n}$ related to the bosonic field eigenstate before being summed. This makes the vacuum state an entangled state, where fermionic and bosonic field degrees of freedom are entangled, exhibiting a mixing between the bosonic and fermionic fields, rather than a simple direct product of the bare fermion vacuum and bare boson vacuum.
%

Likewise, we can derive the one-boson state $|q;B\rangle$ in the representation $\{\prod\Psi^\dagger|0\rangle|\phi\rangle\}$ using \eqref{eq:onephi}, \eqref{eq:vacuum2} and \eqref{eq:sSphi}:
\begin{equation}\label{eq:B}
	\begin{split}
		|q;B\rangle
		&=\mathcal{N}F 2\sqrt{L\omega_q}\frac{1}{N}
		\int \mathrm d\phi\; {\mathrm e}^{-\frac{F}{2}\sum\limits_{n,m}\mathcal{E}_{nm}
			\phi_n\phi_m}
		\sum\limits_n {\mathrm e}^{inaq}
		\left[\phi_n
		+\frac{1}{a\omega_q}\sum\limits_{m}\mathcal{E}_{nm}\phi_m
		\right]
		\\
		&\times\prod\limits_{k=-\frac{N-1}{2}}^{\frac{N-1}{2}}
		\frac{1}{2\sqrt{N}}
		\Bigg[
		\left(\text{sgn}(k){\mathrm e}^{-i\frac{1}{2}\frac{2\pi k}{N}}+1\right)
		\sum\limits_n {\mathrm e}^{-in\frac{2\pi k}{N}}{\mathrm e}^{ig\phi_n}\Psi^\dagger_u(n)\\
		&\qquad\qquad\qquad\qquad\qquad
		+\left(\text{sgn}(k){\mathrm e}^{-i\frac{1}{2}\frac{2\pi k}{N}}-1\right)
		\sum\limits_n {\mathrm e}^{-in\frac{2\pi k}{N}}{\mathrm e}^{-ig\phi_n}\Psi^\dagger_d(n)
		\Bigg]
		|0\rangle|\phi\rangle
		\; .
	\end{split}
\end{equation}
This is the one-boson state represented using the original field degrees of freedom in the Hamiltonian \eqref{eq:Hamiltonian}. 
Once again, we can observe that the one-boson state also exhibits entanglement, mixing the fermionic and bosonic degrees of freedom. 
	The only difference between the one-boson state and the vacuum state is the additional term in the one-boson state, which is given by $\sum\limits_n {\mathrm e}^{inaq}
	\left[\phi_n
	+\frac{1}{a\omega_q}\sum\limits_{m}\mathcal{E}_{nm}\phi_m
	\right]$. However, this term does not involve fermionic content. Therefore, it may be said that the entanglement between the fermionic and bosonic fields in the one-boson state arises from the entanglement in the vacuum state.

Similarly, we can apply a nearly identical approach to derive the one-fermion state in the basis of $\{\prod\Psi^\dagger|0\rangle|\phi\rangle\}$. By employing \eqref{eq:vacuumphi}, \eqref{eq:onepsi+}, and \eqref{eq:onepsi-}, we can deduce the one-fermion state $|q;F+\rangle$ as well as the one-antifermion state $|q;F-\rangle$ of the total Hamiltonian \eqref{eq:Hamiltonian}:
\begin{equation}\label{eq:F+}
	\begin{split}
		|q;F+\rangle
			=&\mathcal{N}F^{\frac{1}{2}}\int \mathrm d\phi\;
			{\mathrm e}^{-\frac{F}{2}\sum\limits_{n,m}\mathcal{E}_{nm}\phi_n \phi_m} \\
			\times 		
			\frac{1}{\sqrt{N}}\Bigg[     		
			&\left({\mathrm e}^{i\frac{1}{2}\frac{2\pi k}{N}}+\text{sgn}(k)\right)
			\sum\limits_n  {\mathrm e}^{in\frac{2\pi k}{N}}
			{\mathrm e}^{ig\phi_n}\Psi^\dagger_u(n)
			+
			\left({\mathrm e}^{i\frac{1}{2}\frac{2\pi k}{N}}-\text{sgn}(k)\right)
			\sum\limits_n  {\mathrm e}^{in\frac{2\pi k}{N}}
			{\mathrm e}^{-ig\phi_n}\Psi^\dagger_d(n)  
			\Bigg]
			\\
			&\times \prod\limits_{k'=-\frac{N-1}{2}}^{\frac{N-1}{2}}
			\frac{1}{2\sqrt{N}}
			\Bigg[
			\left(\text{sgn}(k'){\mathrm e}^{-i\frac{1}{2}\frac{2\pi k'}{N}}+1\right)
			\sum\limits_n {\mathrm e}^{-in\frac{2\pi k'}{N}}{\mathrm e}^{ig\phi_n}\Psi^\dagger_u(n)\\
			&\qquad\qquad\qquad\qquad\qquad
			+\left(\text{sgn}(k'){\mathrm e}^{-i\frac{1}{2}\frac{2\pi k'}{N}}-1\right)
			\sum\limits_n {\mathrm e}^{-in\frac{2\pi k'}{N}}{\mathrm e}^{-ig\phi_n}\Psi^\dagger_d(n)
			\Bigg]
			|0\rangle|\phi\rangle
			\; ,
	\end{split}
\end{equation}
\begin{equation}\label{eq:F-}
	\begin{split}
		|q;F-\rangle
			=&\sqrt{2}\mathcal{N}F^{\frac{1}{2}}\int \mathrm d\phi\;
			{\mathrm e}^{-\frac{F}{2}\sum\limits_{n,m}\mathcal{E}_{nm}\phi_n \phi_m} \\
			&\times \prod\limits_{k'=-\frac{N-1}{2}}^{\frac{N-1}{2};k'\ne k}\frac{1}{2\sqrt{N}}
			\Bigg[
			\left(\text{sgn}(k'){\mathrm e}^{-i\frac{1}{2}\frac{2\pi k'}{N}}+1\right)
			\sum\limits_n {\mathrm e}^{-in\frac{2\pi k'}{N}}{\mathrm e}^{ig\phi_n}\Psi^\dagger_u(n)\\
			&\qquad\qquad\qquad\qquad\qquad
			+\left(\text{sgn}(k'){\mathrm e}^{-i\frac{1}{2}\frac{2\pi k'}{N}}-1\right)
			\sum\limits_n {\mathrm e}^{-in\frac{2\pi k'}{N}}{\mathrm e}^{-ig\phi_n}\Psi^\dagger_d(n)
			\Bigg]
			|0\rangle|\phi\rangle
			\; ,
	\end{split}
\end{equation}
where $k\equiv\frac{aN}{2\pi}q$.
This is the one-fermion state represented using the original field degrees of freedom in the Hamiltonian \eqref{eq:Hamiltonian}. It also exhibits entanglement between fermionic and bosonic degrees of freedom. However, the structure of fermionic(antifermionic) one-particle states is vastly different from that of one-boson states.
In addition, equations \eqref{eq:B}, \eqref{eq:F+}, and \eqref{eq:F-}  indicate that there is a significant distinction in the structure between clothed particles (real one-particle states) and bare particles. Due to the interaction between the fermionic field and bosonic field in the Hamiltonian, the excitation of the fermionic clothed particles involves both the fermionic field and the bosonic field (considering Eq. \eqref{eq:dq}, \eqref{eq:bq}, and \eqref{eq:psiPsi} together), rather than just the fermionic field alone.  Although the boson creation operator is composed entirely of the bosonic field (considering Eq. \eqref{pP} and \eqref{eq:aq} together), due to the vacuum entanglement between bosonic and fermionic field degrees of freedom, bosonic clothed particles also exhibit a mixture of fermionic and bosonic content.
%

Furthermore, the representation $\{\prod\Psi^\dagger|0\rangle|\phi\rangle\}$ can also reveal the spatial entanglement structure of quantum states. As seen from \eqref{eq:Sphi}, the basis in the representation $\{\prod\Psi^\dagger|0\rangle|\phi\rangle\}$ is formed by the direct product of the fermionic part $\Psi^\dagger_{\alpha_1}(n_1)\Psi^\dagger_{\alpha_2}(n_2)\Psi^\dagger_{\alpha_3}(n_3)\cdots\Psi^\dagger_{\alpha_s}(n_s)|0\rangle$ and the bosonic part $|\phi\rangle$. The quantum states of the bosonic part are eigenstates of the bosonic field operator $\hat\phi|\phi\rangle=\phi|\phi\rangle$.
	The field operator $\hat\phi$ at different points are independent of each other and satisfy the commutation relations $[\hat\phi_n,\hat\phi_m ]=0$. As a result, the eigenstate of $\hat\phi$ can be written in a direct product formulation as follows:
	$$|\phi\rangle=
	|\phi_1\rangle_1 |\phi_2\rangle_2
	|\phi_3\rangle_3  \cdots
	=\prod_{n}  |\phi_n\rangle_n\; ,$$
	where $|\phi\rangle_n$ is the eigenstate of $\hat\phi_n$ but in a Hilbert space constructed only for the lattice point $n$.
	
	Similarly, due to the anticommutation of fermionic fields at different spatial points, one can define the following four quantum states at a specific spatial point $n$:
	$$|00\rangle_n\; ,\quad 
	|10\rangle_n=\Psi^\dagger_{u}(n)|00\rangle_n\;,\quad 
	|01\rangle_n=\Psi^\dagger_{d}(n)|00\rangle_n \; ,\quad
	|11\rangle_n=\Psi^\dagger_{u}(n)\Psi^\dagger_{d}(n)|00\rangle_n\; .$$
	Then, the quantum state $|0\rangle$ can be expressed as the direct product of quantum states at different spatial points
	\begin{equation}
		\begin{split}			
			|0\rangle
			=|00\rangle_0|00\rangle_1  \cdots |00\rangle_{N-1}
			\; .
		\end{split}
	\end{equation}	
For quantum states of the form like \eqref{eq:S}, they can also be expressed as a direct product of quantum states at different spatial points: 
\begin{equation}
	\begin{split}		&\Psi^\dagger_u(n_1)\Psi^\dagger_u(n_2)\Psi^\dagger_d(n_2)\cdots\Psi^\dagger_d(n_s
		)
		|0\rangle\\
		&=\cdots
		|00\rangle_{n_1-1}|10\rangle_{n_1}|00\rangle_{n_1+1}\cdots
		|00\rangle_{n_2-1}|11\rangle_{n_2}|00\rangle_{n_2+1}\cdots
		|00\rangle_{n_s-1}|01\rangle_{n_s}|00\rangle_{n_s+1}\cdots
	\end{split}
\end{equation}	
This signifies that the basis of the representation $\{\prod\Psi^\dagger|0\rangle|\phi\rangle\}$ can be expressed as a direct product of quantum states at different spatial points. Consequently, employing the representation $\{\prod\Psi^\dagger|0\rangle|\phi\rangle\}$ not only highlights the entanglement between fermionic and bosonic fields, as mentioned earlier, but also reveals the spatial entanglement structure of the vacuum state through \eqref{eq:vacuumPsi}, while \eqref{eq:B}, \eqref{eq:F+}, and \eqref{eq:F-} also illustrate the spatial entanglement structure of the clothed particles.

	\section{Conclusions and Discussions}
	\label{7}
	We have presented the lattice Hamiltonian of the RS model, diagonalized it, and subsequently derived lattice correlation functions, as well as the physical vacuum and clothed particles. The continuum limit of the lattice correlation functions matches the original RS model's correlation functions, affirming that the continuum limit of the lattice theory corresponds to the original RS model.
	In order to gain a more intuitive understanding of the complex Hamiltonian, we have analyzed the equations of motion for the lattice theory in the appendices. In Appendix \ref{eqomB}, we obtained the equations of motion for the bosonic field in the lattice RS model and compared them with those of the original RS model. Similarly, in Appendix \ref{eqomF}, we derived the equations of motion for the fermionic field in the lattice RS model and compared them to the original RS model.	
	It is worth noting that the equations of motion for the lattice RS model share the same structure as those of the original RS model, with only some differences in the coefficients of regularization terms. These differences arise because the original RS model describes the infrared behavior of the lattice RS model. Even the ultraviolet behavior of the original RS model falls under the infrared behavior of the lattice model. Consequently, the coefficients of the regularization terms in the equations of motion for the original RS model differ slightly from those of the lattice model.	
	However, this discrepancy does not imply that the continuum limit of the lattice theory is not the original RS model. As mentioned earlier, the continuum limit of the lattice correlation functions matches the original RS model's correlation functions, and the behavior exhibited by taking the continuum limit first and then letting the field spacing tend to zero aligns with that of the original RS model (see discussions in the appendices regarding \eqref{eq:fc2}, \eqref{eq:fc3}, and \eqref{eq:fcuu}).

	Creation and annihilation operators directly associated with the bare fields are referred to as ``bare operators", denoted as $a_p$. One-particle states that are eigenstates of the Hamiltonian are called ``clothed particles", and the creation and annihilation operators that produce clothed particles from the physical vacuum are called ``clothed operators", denoted as $\alpha_p$.	
	Then, clothed operators $\alpha$ can be expressed in terms of bare operators $a$, with the specific ``clothing transformation" given by $\alpha_p=W^\dagger a_p W$, where the transformation operator $W$ is a function of all bare operators $a$ and satisfies $W^\dagger W=W W^\dagger=1$ \cite{Shebeko2001}.
	It is worth noting that, the transformation operators $W$ induced by interactions such as $\mathcal{L}_I=-m_{e\mu}(\bar\nu_e\nu_\mu+\bar\nu_\mu\nu_e)$ and $L_I=-\lambda(\phi^\dagger_\alpha\phi_\beta+\phi^\dagger_\beta\phi_\alpha)$ have been studied in previous works\cite{Blasone1995, Ji2001}.


	 Although we did not adopt the Fock representation in this paper, we can still convert the results of the paper into the Fock representation and express the clothed operators in terms of bare operators. In fact, the operators $a_p$ in Eq. \eqref{eq:aq}, $d_p$ in Eq. \eqref{eq:dq}, and $b_p$ in Eq. \eqref{eq:bq} are already clothed operators. If we denote the corresponding bare operators as $A_p$, $D_p$, and $B_p$, then by combining Eqs. \eqref{eq:aq}, \eqref{eq:dq}, \eqref{eq:bq}, and \eqref{pP}, we can express the clothed operators ($a$, $d$, $b$) in terms of bare operators ($A$, $D$, $B$), i.e., $x_p = f(A, D, B)$, where $x = a, d, b$.	
	Since we are considering the interaction $\Delta\mathcal{L}=-g\partial_u\phi \bar\Psi \gamma^5\gamma^\mu\Psi$, which is more complex than the quadratic interactions mentioned above, expressing it in the form $x_p=W(A, D, B)^\dagger X_p W(A, D, B)$ (where $X = A, D, B$) would require further derivations and calculations. 
	However, it can be anticipated that $W(A, D, B)$ will be highly complex and challenging to intuitively understand.   
         Therefore, in this paper we did not employ the conventional Fock representation but instead chose the representation defined by the basis \eqref{eq:Sphi}, denoted as $\{\prod\Psi^\dagger|0\rangle|\phi\rangle\}$. In this representation, the specific form of the physical vacuum of the RS model is given by Eq. \eqref{eq:vacuumPsi}, the specific form of the one-boson state is given by Eq. \eqref{eq:B}, and the one-fermion states and one-antifermion states are given by Eq. \eqref{eq:F+} and Eq. \eqref{eq:F-}, respectively. It can be observed that both the physical vacuum and the clothed particles exhibit entanglement between the bosonic and fermionic fields.


In addition to the entanglement between the fermionic and bosonic fields, the basis in the representation $\{\prod\Psi^\dagger|0\rangle|\phi\rangle\}$ can all be expressed as direct products of quantum states at different spatial points, allowing us to directly observe the spatial entanglement structure of the quantum states.
The vacuum in classical theory is local, meaning that if space is divided into many parts, each part is still a vacuum. However, in quantum field theory, due to the non-locality of quantum states and the entanglement between spatial points, strictly speaking, the vacuum state cannot be simply divided into two parts. Nevertheless, since the vacuum state corresponds to the classical vacuum, it should exhibit locality on large scales.
Let's first ignore the entanglement between the bosonic and fermionic fields and focus solely on the wave function of the bosonic part in the vacuum state \eqref{eq:vacuumPsi}, which is given by $\int \mathrm d\phi\; {\mathrm e}^{-\frac{F}{2}\sum\limits_{n,m}\mathcal{E}_{nm}\phi_n \phi_m}$. If $\mathcal{E}_{nm}\propto\delta_{nm}$, then there would be no entanglement between different spatial points in the bosonic vacuum state, and it would exhibit the same locality as the classical vacuum. However, in reality, $\mathcal{E}_{nm}=\frac{1}{N}\sum\limits_q a\omega_q {\mathrm e}^{i(n-m)aq}$ is not proportional to $\delta_{nm}$, indicating entanglement between different spatial points and preventing the vacuum from being arbitrarily divided into two parts. This seems contradictory to the locality of the classical vacuum.
Nevertheless, it can be easily proven that $\mathcal{E}_{nm}\propto e^{-m_r|na-ma|}\to0$ as $|na-ma|\to\infty$. This implies that the entanglement between points with large spatial separations becomes weak. Consequently, from a macroscopic perspective,
the bosonic part of the vacuum state \eqref{eq:vacuumPsi} does indeed exhibit locality similar to the classical vacuum.

          
          However, for the fermionic part of the vacuum state \eqref{eq:vacuumPsi}, it is challenging to intuitively discern locality. Moreover, and the complete quantum state exhibits entanglement between fermions and bosons. Therefore, we need a more quantitative analysis of this entanglement. Specifically, we can choose two regions, denoted as region A and region B, with their union referred to as region A+B.
          Since ${\prod\Psi^\dagger|0\rangle|\phi\rangle}$ is a representation based on real space, the entanglement entropy of regions A, B, and A+B can be computed in this representation.         
          If the sum of the entanglement entropy of region A and the entanglement entropy of region B, minus the entanglement entropy of region A+B, decreases rapidly as the distance between the regions A and B increases, it confirms that the vacuum state exhibit locality similar to the classical vacuum from a macroscopic perspective.         
          
          In the future, in addition to computing the entanglement entropy of quantum states, we can consider introducing external sources to the RS model to make the system non-uniform and study the possible emergence of spatial cloud structures. We can also introduce a fermion mass term to the RS model and develop perturbation theory based on this work to compute quantum states. Essentially, this paper provides a solvable Hamiltonian containing a three-point interaction, from which Feynman rules can be derived in a well-defined manner, demonstrating how the bare parameters of the Hamiltonian evolve into various parameters of the lower-level Feynman diagram.



%


\section{Acknowledgments}
We would like to thank Weijun Kong and Yi-Da Li for helpful discussions.

\appendix

\section{The Equation of Motion For the bosonic field}
\label{eqomB}

For the sake of simplicity, we won't explicitly indicate the time dependence and will focus only on spatial coordinate dependence. In the subsequent analysis, we denote the positive and negative frequency components of the fermionic field $\psi'_u(n)$ as $\psi'^+_u(n)$ and $\psi'^-_u(n)$, respectively. Similarly, the positive and negative frequency parts of $\psi'^-_d(n)$ are denoted as $\psi'^+_d(n)$ and $\psi'^-_d(n)$. By employing the commutation relations for creation and annihilation operators, as well as \eqref{eq:psi'u} and \eqref{eq:psi'd}, we can deduce the following commutation relations:
\begin{equation}\label{psi+-1}
	\begin{split}	
		\left\{ (\psi'^-_u(n))^\dagger,\psi'^-_u(n+m)\right\}
		&=\left\{\sum\limits_{q\ne0} \frac{1}{\sqrt{2N}}\text{sgn}(q)b_q {\mathrm e}^{i(n+\frac{1}{2})aq}
		,\sum\limits_{l\ne0}\frac{1}{\sqrt{2N}}\text{sgn}(l)b_l^\dagger {\mathrm e}^{-i(n+m+\frac{1}{2})al}\right\}\\
		&=\frac{1}{2N}\sum\limits_{q\ne0} {\mathrm e}^{-imaq}\\
		&=\frac{1}{2}\delta_{m,0}-\frac{1}{2N}
		\; ,
	\end{split}
\end{equation}
\begin{equation}
	\begin{split}	
		\left\{ (\psi'^-_d(n))^\dagger,\psi'^-_d(n+m)\right\}
		&=\left\{\sum\limits_{q\ne0} \frac{1}{\sqrt{2N}}b_q {\mathrm e}^{inaq}+\frac{1}{\sqrt{N}}b_0
		,\sum\limits_{l\ne0} \frac{1}{\sqrt{2N}}b_l^\dagger {\mathrm e}^{-i(n+m)al}+\frac{1}{\sqrt{N}}b_0^\dagger\right\}\\
		&=\frac{1}{2N}\sum\limits_{q\ne0} {\mathrm e}^{-imaq}+\frac{1}{N}\\
		&=\frac{1}{2}\delta_{m,0}+\frac{1}{2N}
		\; ,
	\end{split}
\end{equation}
\begin{equation}
	\begin{split}	
		\left\{ (\psi'^-_d(n))^\dagger,\psi'^-_u(n+m)\right\}
		&=\left\{-\sum\limits_{q\ne0} \frac{1}{\sqrt{2N}}b_q {\mathrm e}^{inaq}-\frac{1}{\sqrt{N}}b_0
		,\sum\limits_{l\ne0}\frac{1}{\sqrt{2N}}\text{sgn}(l)b_l^\dagger {\mathrm e}^{-i(n+m+\frac{1}{2})al}\right\}\\
		&=-\frac{1}{2N}\sum\limits_{q\ne0} \text{sgn}(q) {\mathrm e}^{-i(m+\frac{1}{2})aq}\\
		&=i\frac{1}{2N}\cot\left[(m+\frac{1}{2})\frac{\pi}{N}\right]
		\; ,
	\end{split}
\end{equation}
\begin{equation}\label{psi+-4}
	\begin{split}	
		\left\{ (\psi'^-_u(n))^\dagger,\psi'^-_d(n+m)\right\}
		=\left\{ (\psi'^-_d(n+m))^\dagger,\psi'^-_u(n)\right\}^\dagger
		=i\frac{1}{2N}\cot\left[(m-\frac{1}{2})\frac{\pi}{N}\right]
		\; .
	\end{split}
\end{equation}

In a similar manner, we represent the positive and negative frequency components of $\psi_u(n)$ as $\psi^+_u(n)$ and $\psi^-_u(n)$, respectively, and the positive and negative frequency components of $\psi_d(n)$ as $\psi^+_d(n)$ and $\psi^-_d(n)$. 
Based on \eqref{eq:p'p} and the commutation relations \eqref{psi+-1}-\eqref{psi+-4}, we can further derive the following commutation relations:
\begin{equation}
	\begin{split}
		\label{eq:uu}	
		\left\{ (\psi^-_u(n))^\dagger,\psi^-_u(n+m)\right\}
		=\frac{1}{2}\Bigg[\delta_{m,0}
		-i\frac{1}{2N}\cot\left[\left(m+\frac{1}{2}\right)\frac{\pi}{N}\right]
		-i\frac{1}{2N}\cot\left[\left(m-\frac{1}{2}\right)\frac{\pi}{N}\right]
		\Bigg]
		\; ,
	\end{split}
\end{equation}
\begin{equation}
	\begin{split}	
		\label{eq:dd}	
		\left\{ (\psi^-_d(n))^\dagger,\psi^-_d(n+m)\right\}
		=\frac{1}{2}\Bigg[\delta_{m,0}
		+i\frac{1}{2N}\cot\left[\left(m+\frac{1}{2}\right)\frac{\pi}{N}\right]
		+i\frac{1}{2N}\cot\left[\left(m-\frac{1}{2}\right)\frac{\pi}{N}\right]
		\Bigg]
		\; ,
	\end{split}
\end{equation}
\begin{equation}\label{eq:du}
	\begin{split}	
		\left\{ (\psi^-_d(n))^\dagger,\psi^-_u(n+m)\right\}
		=\frac{1}{2}\Bigg[
		i\frac{1}{2N}\cot\left[(m+\frac{1}{2})\frac{\pi}{N}\right]
		-i\frac{1}{2N}\cot\left[(m-\frac{1}{2})\frac{\pi}{N}\right]
		-\frac{1}{N}
		\Bigg]
		\; ,
	\end{split}
\end{equation}
\begin{equation}\label{eq:ud}
	\begin{split}	
		\left\{ (\psi^-_u(n))^\dagger,\psi^-_d(n+m)\right\}
		=\frac{1}{2}\Bigg[
		i\frac{1}{2N}\cot\left[(m-\frac{1}{2})\frac{\pi}{N}\right]
		-i\frac{1}{2N}\cot\left[(m+\frac{1}{2})\frac{\pi}{N}\right]
		-\frac{1}{N}
		\Bigg]
		\; .
	\end{split}
\end{equation}

Given these commutation relations \eqref{eq:uu}-\eqref{eq:ud}, we can deduce the connections between the standard operator product and the normal ordered product. To be precise, employing \eqref{eq:uu}, we can establish the following relation:
\begin{equation}\label{eq:n(n+m)}
	\begin{split}	
		\psi^\dagger_u(n)\psi_u(n+m)
		=&\left[(\psi^+_u(n))^\dagger+(\psi^-_u(n))^\dagger\right]\psi^+_u(n+m)+(\psi^+_u(n))^\dagger\psi^-_u(n+m)-\psi^-_u(n+m)(\psi^-_u(n))^\dagger\\
		&\qquad+\left\{ (\psi^-_u(n))^\dagger,\psi^-_u(n+m)\right\}\\
		=&:\psi^\dagger_u(n)\psi_u(n+m):+\left\{ (\psi^-_u(n))^\dagger,\psi^-_u(n+m)\right\}\\
		=&:\psi^\dagger_u(n)\psi_u(n+m):\\
		&+\frac{1}{2}\Bigg[\delta_{m,0}
		-i\frac{1}{2N}\cot\left[\left(m+\frac{1}{2}\right)\frac{\pi}{N}\right]
		-i\frac{1}{2N}\cot\left[\left(m-\frac{1}{2}\right)\frac{\pi}{N}\right]
		\Bigg]
		\;.
	\end{split}
\end{equation}
Here, the :: symbol denotes the normal ordered product, indicating that we arrange the creation operators to the left and the annihilation operators to the right within the product of creation and annihilation operators. Similarly, utilizing the remaining commutation relations from \eqref{eq:dd}, \eqref{eq:du}, and \eqref{eq:ud}, we can arrive at
\begin{equation}\label{eq:n(n+m)dd}
	\begin{split}	
		\psi^\dagger_d(n)\psi_d(n+m)
		=&:\psi^\dagger_d(n)\psi_d(n+m):\\
		&+\frac{1}{2}\Bigg[\delta_{m,0}
		+i\frac{1}{2N}\cot\left[\left(m+\frac{1}{2}\right)\frac{\pi}{N}\right]
		+i\frac{1}{2N}\cot\left[\left(m-\frac{1}{2}\right)\frac{\pi}{N}\right]
		\Bigg]
		\; ,
	\end{split}
\end{equation}
\begin{equation}\label{eq:n(n+m)du}
	\begin{split}	
		\psi^\dagger_d(n)\psi_u(n+m)
		=&:\psi^\dagger_d(n)\psi_u(n+m):\\
		&+\frac{1}{2}\Bigg[
		i\frac{1}{2N}\cot\left[\left(m+\frac{1}{2}\right)\frac{\pi}{N}\right]
		-i\frac{1}{2N}\cot\left[\left(m-\frac{1}{2}\right)\frac{\pi}{N}\right]
		-\frac{1}{N}
		\Bigg]
		\; ,
	\end{split}
\end{equation}
\begin{equation}\label{eq:n(n+m)ud}
	\begin{split}	
		\psi^\dagger_u(n)\psi_d(n+m)
		=&:\psi^\dagger_u(n)\psi_d(n+m):\\
		&+\frac{1}{2}\Bigg[
		i\frac{1}{2N}\cot\left[\left(m-\frac{1}{2}\right)\frac{\pi}{N}\right]
		-i\frac{1}{2N}\cot\left[\left(m+\frac{1}{2}\right)\frac{\pi}{N}\right]
		-\frac{1}{N}
		\Bigg]
		\; .
	\end{split}
\end{equation}

In the regime of a small lattice distance $a$, we can utilize \eqref{eq:phipi} to deduce that $(\phi^+_{n+1}-\phi^+_n)^i(\phi^-_{n+1}-\phi^-_n)^j=O(a^{i+j})$. Moreover, through the relation between the discrete and continuous fermionic field operators given by \eqref{eq:psi(n)}, we observe that $\psi_u(n)=O(a^{\frac{1}{2}})$ and $\psi_u(n)\psi_u(n+m):=O(a)$. Therefore, referring to \eqref{pP} and \eqref{eq:n(n+m)}, we can calculate the product $\Psi^\dagger_u(n)\Psi_u(n+1)$ to the next-to-leading order of the lattice distance $a$:
\begin{equation}\label{eq:Psi(n)Psi(n+1)}
	\begin{split}			
		&\Psi^\dagger_u(n)\Psi_u(n+1)\\
		&={\mathrm e}^{ig\phi_{n+1}-ig\phi_n}\psi^\dagger_u(n)\psi_u(n+1)\\
		&={\mathrm e}^{-[g\phi^+_n,g\phi^-_n]+[g\phi^+_{n+1},g\phi^-_n]}
		{\mathrm e}^{ig\phi^-_{n+1}-ig\phi^-_n}{\mathrm e}^{ig\phi^+_{n+1}-ig\phi^+_n}\psi^\dagger_u(n)\psi_u(n+1)\\
		&={\mathrm e}^{-[g\phi^+_n,g\phi^-_n]+[g\phi^+_{n+1},g\phi^-_n]}
		\Biggl\{-i\left(
		\frac{1}{4N}\cot\left[\frac{1}{2}\frac{\pi}{N}\right]
		+\frac{1}{4N}\cot\left[\frac{3}{2}\frac{\pi}{N}\right]
		\right)
		+:\psi^\dagger_u(n)\psi_u(n+1):\\
		&\qquad\qquad+g(\phi_{n+1}-\phi_n)
		\left(
		\frac{1}{4N}\cot\left[\frac{1}{2}\frac{\pi}{N}\right]
		+\frac{1}{4N}\cot\left[\frac{3}{2}\frac{\pi}{N}\right]\right)+O(a^2)\Bigg\}\\
		&=C^{-1}
		\Biggl\{-i\left(
		\frac{1}{4N}\cot\left[\frac{1}{2}\frac{\pi}{N}\right]
		+\frac{1}{4N}\cot\left[\frac{3}{2}\frac{\pi}{N}\right]
		\right)
		+:\psi^\dagger_u(n)\psi_u(n+1):\\
		&\qquad\qquad+g(\phi_{n+1}-\phi_n)
		\left(
		\frac{1}{4N}\cot\left[\frac{1}{2}\frac{\pi}{N}\right]
		+\frac{1}{4N}\cot\left[\frac{3}{2}\frac{\pi}{N}\right]\right)+O(a^2)\Bigg\}
		\;,
	\end{split}
\end{equation}
where $C$ is defined as \begin{equation}\label{1C}
	\begin{split}
		C={\mathrm e}^{[g\phi^+_n,g\phi^-_n]-[g\phi^+_{n+1},g\phi^-_n]}={\mathrm e}^{g^2[f(0)-f(-1)]}={\mathrm e}^{g^2[f(0)-f(1)]}\;.
	\end{split}
\end{equation}
As we consider the large $N$ limit, the behavior of \eqref{eq:Psi(n)Psi(n+1)} is
\begin{equation}\label{eq:limuU}
	\begin{split}	
		\lim\limits_{N\to\infty}C\Psi^\dagger_u(n)\Psi_u(n+1)
		=\lim\limits_{N\to\infty}\left[\psi^\dagger_u(n)\psi_u(n+1)+\frac{2}{3\pi}g(\phi_{n+1}-\phi_n)+O(a^2)\right]
		\; .
	\end{split}
\end{equation}
Similarly, we have
\begin{equation}\label{eq:limUu}
	\begin{split}	
		\lim\limits_{N\to\infty}C\Psi^\dagger_u(n+1)\Psi_u(n)		=\lim\limits_{N\to\infty}\left[\psi^\dagger_u(n+1)\psi_u(n)+\frac{2}{3\pi}g(\phi_{n+1}-\phi_n)+O(a^2)\right]
		\; ,
	\end{split}
\end{equation}
\begin{equation}\label{eq:limdD}
	\begin{split}	
		\lim\limits_{N\to\infty}C\Psi^\dagger_d(n)\Psi_d(n+1)		=\lim\limits_{N\to\infty}\left[\psi^\dagger_d(n)\psi_d(n+1)+\frac{2}{3\pi}g(\phi_{n+1}-\phi_n)+O(a^2)\right]
		\; ,
	\end{split}
\end{equation}
\begin{equation}\label{eq:limDd}
	\begin{split}	
		\lim\limits_{N\to\infty}C\Psi^\dagger_d(n+1)\Psi_d(n)		=\lim\limits_{N\to\infty}\left[\psi^\dagger_d(n+1)\psi_d(n)+\frac{2}{3\pi}g(\phi_{n+1}-\phi_n)+O(a^2)\right]
		\; .
	\end{split}
\end{equation}
With these useful limits, we are ready to deduce the equation of motion for the bosonic field $\phi$ from the lattice Hamiltonian \eqref{eq:Hamiltonian}. We can derive the canonical equation for the canonical momentum $\pi_n$ using the Heisenberg equations of motion:
\begin{equation}
	\begin{split}	
		\label{eq:pi1}
		\partial_0\pi_n&=\frac{1}{i}[\pi_n,H]\\
		&=
		F\left[a(\frac{1}{a})^2(\phi_{n+1}-\phi_{n})
		-a(\frac{1}{a})^2(\phi_{n}-\phi_{n-1})\right]
		-m^2a\phi_n\\
		&+\frac{1}{2a}\Bigg[\pi_n,
		\sum\limits_n \Big\{ 
		{\mathrm e}^{ig\phi_{n+1}-ig\phi_n}\Psi^\dagger_u(n+1)\Psi_u(n)  -{\mathrm e}^{ig\phi_n-ig\phi_{n+1}}\Psi^\dagger_u(n)\Psi_u(n+1) \\
		&
		-{\mathrm e}^{ig\phi_n-ig\phi_{n+1}}\Psi^\dagger_d(n+1)\Psi_d(n)
		+{\mathrm e}^{ig\phi_{n+1}-ig\phi_n}\Psi^\dagger_d(n)\Psi_d(n+1) \Big\}\Bigg]\\
		&
		+\frac{1}{2a}\Bigg[\pi_n,\\
		&\sum\limits_n \Big\{
		-{\mathrm e}^{-ig\phi_{n+1}-ig\phi_n}\Psi^\dagger_d(n+1)\Psi_u(n) -{\mathrm e}^{-ig\phi_n-ig\phi_{n+1}}\Psi^\dagger_d(n)\Psi_u(n+1)
		+2{\mathrm e}^{-2ig\phi_n}\Psi^\dagger_d(n)\Psi_u(n) \\
		&
		+{\mathrm e}^{ig\phi_{n+1}+ig\phi_n}\Psi^\dagger_u(n+1)\Psi_d(n) +{\mathrm e}^{ig\phi_n+ig\phi_{n+1}}\Psi^\dagger_u(n)\Psi_d(n+1)
		-2{\mathrm e}^{2ig\phi_n}\Psi^\dagger_u(n)\Psi_d(n)\Big\}\Bigg]
		\; .
	\end{split}
\end{equation}
In the previous equation \eqref{eq:pi1}, we encounter two rather intricate commutation relations that need to be computed. In the subsequent analysis, our attention will be directed towards evaluating these complex commutation relations. Let's begin by examining the first commutation relation in \eqref{eq:pi1}:
\begin{equation}
	\begin{split}
		\label{eq:pi2}	
		&\Bigg[\pi_n,
		\sum\limits_n \Big\{ 
		{\mathrm e}^{ig\phi_{n+1}-ig\phi_n}\Psi^\dagger_u(n+1)\Psi_u(n)  -{\mathrm e}^{ig\phi_n-ig\phi_{n+1}}\Psi^\dagger_u(n)\Psi_u(n+1) \\
		&
		-{\mathrm e}^{ig\phi_n-ig\phi_{n+1}}\Psi^\dagger_d(n+1)\Psi_d(n)
		+{\mathrm e}^{ig\phi_{n+1}-ig\phi_n}\Psi^\dagger_d(n)\Psi_d(n+1) \Big\} \Bigg]\\
		=&g\Bigg\{-{\mathrm e}^{ig\phi_{n+1}-ig\phi_n}\Psi^\dagger_u(n+1)\Psi_u(n) 
		+{\mathrm e}^{ig\phi_n-ig\phi_{n-1}}\Psi^\dagger_u(n)\Psi_u(n-1)\\
		& -{\mathrm e}^{ig\phi_n-ig\phi_{n+1}}\Psi^\dagger_u(n)\Psi_u(n+1) 
		+{\mathrm e}^{ig\phi_{n-1}-ig\phi_n}\Psi^\dagger_u(n-1)\Psi_u(n) \\
		&
		-{\mathrm e}^{ig\phi_n-ig\phi_{n+1}}\Psi^\dagger_d(n+1)\Psi_d(n)
		+{\mathrm e}^{ig\phi_{n-1}-ig\phi_n}\Psi^\dagger_d(n)\Psi_d(n-1)\\
		&-{\mathrm e}^{ig\phi_{n+1}-ig\phi_n}\Psi^\dagger_d(n)\Psi_d(n+1)
		+{\mathrm e}^{ig\phi_n-ig\phi_{n-1}}\Psi^\dagger_d(n-1)\Psi_d(n) \Bigg\}\\	
		&=-g\Bigg\{\psi^\dagger_u(n+1)\psi_u(n)-\psi^\dagger_u(n)\psi_u(n-1)                            +\psi^\dagger_u(n)\psi_u(n+1)-\psi^\dagger_u(n-1)\psi_u(n) \\
		&
		+\psi^\dagger_d(n+1)\psi_d(n)-\psi^\dagger_d(n)\psi_d(n-1)
		+\psi^\dagger_d(n)\psi_d(n+1)-\psi^\dagger_d(n-1)\psi_d(n) \Bigg\}
		\; .
	\end{split}
\end{equation}
In the above steps, we have inserted the field redefinition \eqref{pP} to reach the concluding equality in \eqref{eq:pi2}. Utilizing equations \eqref{eq:limuU}--\eqref{eq:limDd}, we can deduce the behavior of \eqref{eq:pi2} in the limit of large $N$:
\begin{equation}\label{eq:pi3}	
	\begin{split}		
		&\lim\limits_{N\to\infty}\Bigg[\pi_n,
		\sum\limits_n \Big\{ 
		{\mathrm e}^{ig\phi_{n+1}-ig\phi_n}\Psi^\dagger_u(n+1)\Psi_u(n)  -{\mathrm e}^{ig\phi_n-ig\phi_{n+1}}\Psi^\dagger_u(n)\Psi_u(n+1) \\
		&\qquad\qquad\qquad
		-{\mathrm e}^{ig\phi_n-ig\phi_{n+1}}\Psi^\dagger_d(n+1)\Psi_d(n)
		+{\mathrm e}^{ig\phi_{n+1}-ig\phi_n}\Psi^\dagger_d(n)\Psi_d(n+1) \Big\} \Bigg]\\
		&=-gC\Bigg\{\Psi^\dagger_u(n+1)\Psi_u(n)-\Psi^\dagger_u(n)\Psi_u(n-1)                            +\Psi^\dagger_u(n)\Psi_u(n+1)-\Psi^\dagger_u(n-1)\Psi_u(n) \\
		&\qquad\qquad
		+\Psi^\dagger_d(n+1)\Psi_d(n)-\Psi^\dagger_d(n)\Psi_d(n-1)
		+\Psi^\dagger_d(n)\Psi_d(n+1)-\Psi^\dagger_d(n-1)\Psi_d(n) \Bigg\}\\
		&\qquad
		+\frac{8}{3\pi}g^2\left[(\phi_{n+1}-\phi_n)-(\phi_n-\phi_{n-1})\right]+O(a^3)
		\; .
	\end{split}
\end{equation}
With the completion of the calculation for the first commutation relation in \eqref{eq:pi1} and the inclusion of considerations regarding the large $N$ limit, we now move forward to compute the second commutation relation presented in \eqref{eq:pi1}:
\begin{equation}
	\begin{split}	
		\label{eq:pi5}
		&\Bigg[\pi_n,
		\sum\limits_n \Big\{
		-{\mathrm e}^{-ig\phi_{n+1}-ig\phi_n}\Psi^\dagger_d(n+1)\Psi_u(n) -{\mathrm e}^{-ig\phi_n-ig\phi_{n+1}}\Psi^\dagger_d(n)\Psi_u(n+1)
		+2{\mathrm e}^{-2ig\phi_n}\Psi^\dagger_d(n)\Psi_u(n) \\
		&
		+{\mathrm e}^{ig\phi_{n+1}+ig\phi_n}\Psi^\dagger_u(n+1)\Psi_d(n) +{\mathrm e}^{ig\phi_n+ig\phi_{n+1}}\Psi^\dagger_u(n)\Psi_d(n+1)
		-2{\mathrm e}^{2ig\phi_n}\Psi^\dagger_u(n)\Psi_d(n)\Big\} \Bigg]\\
		=
		&g\Bigg[
		{\mathrm e}^{-ig\phi_{n+1}-ig\phi_n}\Psi^\dagger_d(n+1)\Psi_u(n)
		+{\mathrm e}^{-ig\phi_n-ig\phi_{n-1}}\Psi^\dagger_d(n)\Psi_u(n-1)\\
		&+{\mathrm e}^{-ig\phi_n-ig\phi_{n+1}}\Psi^\dagger_d(n)\Psi_u(n+1)
		+{\mathrm e}^{-ig\phi_{n-1}-ig\phi_n}\Psi^\dagger_d(n-1)\Psi_u(n)
		-4{\mathrm e}^{-2ig\phi_n}\Psi^\dagger_d(n)\Psi_u(n) \\
		&
		+{\mathrm e}^{ig\phi_{n+1}+ig\phi_n}\Psi^\dagger_u(n+1)\Psi_d(n)
		+{\mathrm e}^{ig\phi_n+ig\phi_{n-1}}\Psi^\dagger_u(n)\Psi_d(n-1)\\ &+{\mathrm e}^{ig\phi_n+ig\phi_{n+1}}\Psi^\dagger_u(n)\Psi_d(n+1)
		+{\mathrm e}^{ig\phi_{n-1}+ig\phi_n}\Psi^\dagger_u(n-1)\Psi_d(n)
		-4{\mathrm e}^{2ig\phi_n}\Psi^\dagger_u(n)\Psi_d(n)\Bigg]\\	
		=
		&g\Bigg[
		\psi^\dagger_d(n+1)\psi_u(n)+\psi^\dagger_d(n)\psi_u(n-1)
		+\psi^\dagger_d(n)\psi_u(n+1)+\psi^\dagger_d(n-1)\psi_u(n)
		-4\psi^\dagger_d(n)\psi_u(n) \\
		&
		+\psi^\dagger_u(n+1)\psi_d(n)+\psi^\dagger_u(n)\psi_d(n-1)
		+\psi^\dagger_u(n)\psi_d(n+1)+\psi^\dagger_u(n-1)\psi_d(n)
		-4\psi^\dagger_u(n)\psi_d(n)\Bigg]\\
		=
		&g\Bigg[
		:\psi^\dagger_d(n+1)\psi_u(n):+:\psi^\dagger_d(n-1)\psi_u(n):
		-2:\psi^\dagger_d(n)\psi_u(n):\\
		&+:\psi^\dagger_d(n)\psi_u(n-1):+:\psi^\dagger_d(n)\psi_u(n+1):
		-2:\psi^\dagger_d(n)\psi_u(n): \\
		&
		+:\psi^\dagger_u(n+1)\psi_d(n):+:\psi^\dagger_u(n-1)\psi_d(n):
		-2:\psi^\dagger_u(n)\psi_d(n):\\
		&+:\psi^\dagger_u(n)\psi_d(n-1):+:\psi^\dagger_u(n)\psi_d(n+1):
		-2:\psi^\dagger_u(n)\psi_d(n):\Bigg]
		\; ,	
	\end{split}
\end{equation}
where the second equality used \eqref{pP} and the final equality relied on \eqref{eq:n(n+m)du} and \eqref{eq:n(n+m)ud}.
According to the definition of the normal product, the first line after the last equals sign in \eqref{eq:pi5} can be expressed as
\begin{equation}
	\begin{split}
		&:\psi^\dagger_d(n+1)\psi_u(n):+:\psi^\dagger_d(n-1)\psi_u(n):
		-2:\psi^\dagger_d(n)\psi_u(n):\\
		&=\left[(\psi^+_d(n+1)-\psi^+_d(n))^\dagger+(\psi^-_d(n+1)-\psi^-_d(n))^\dagger\right]\psi^+_u(n)\\
		&+(\psi^+_d(n+1)-\psi^+_d(n))^\dagger\psi^-_u(n)-\psi^-_u(n)(\psi^-_d(n+1)-\psi^-_d(n))^\dagger\\
		&-\left[(\psi^+_d(n)-\psi^+_d(n-1))^\dagger+(\psi^-_d(n)-\psi^-_d(n-1))^\dagger\right]\psi^+_u(n)\\
		&+(\psi^+_d(n)-\psi^+_d(n-1))^\dagger\psi^-_u(n)-\psi^-_u(n)(\psi^-_d(n)-\psi^-_d(n-1))^\dagger\\
		&=\Bigg\{\Big[(\psi^+_d(n+1)-\psi^+_d(n))-(\psi^+_d(n)-\psi^+_d(n-1))\Big]^\dagger\\
		&+\Big[(\psi^-_d(n+1)-\psi^-_d(n))-(\psi^-_d(n)-\psi^-_d(n-1))\Big]^\dagger\Bigg\}\psi^+_u(n)\\
		&+\Big[(\psi^+_d(n+1)-\psi^+_d(n))-(\psi^+_d(n)-\psi^+_d(n-1))\Big]^\dagger\psi^-_u(n)\\
		&-\psi^-_u(n)\Big[(\psi^-_d(n+1)-\psi^-_d(n))-(\psi^-_d(n)-\psi^-_d(n-1))\Big]^\dagger\\
		&=O(a^3)
		\; ,
	\end{split}
\end{equation}
Similarly, it can be proven that the other lines following the last equals sign in \eqref{eq:pi5} are also of $O(a^3)$ order. Thus, \eqref{eq:pi5} can be written as
\begin{equation}\label{eq:pi6}	
	\begin{split}		
		&\Bigg[\pi_n,
		\sum\limits_n \Big\{
		-{\mathrm e}^{-ig\phi_{n+1}-ig\phi_n}\Psi^\dagger_d(n+1)\Psi_u(n) -{\mathrm e}^{-ig\phi_n-ig\phi_{n+1}}\Psi^\dagger_d(n)\Psi_u(n+1)
		+2{\mathrm e}^{-2ig\phi_n}\Psi^\dagger_d(n)\Psi_u(n) \\
		&
		+{\mathrm e}^{ig\phi_{n+1}+ig\phi_n}\Psi^\dagger_u(n+1)\Psi_d(n) +{\mathrm e}^{ig\phi_n+ig\phi_{n+1}}\Psi^\dagger_u(n)\Psi_d(n+1)
		-2{\mathrm e}^{2ig\phi_n}\Psi^\dagger_u(n)\Psi_d(n)\Big\} \Bigg]
		=O(a^3)	
		\; .
	\end{split}
\end{equation}
Therefore, it is evident that the second commutation relation in \eqref{eq:pi1} is of the order $a^3$. Consequently, by substituting \eqref{eq:pi3} and \eqref{eq:pi6} into \eqref{eq:pi1}, we can deduce the canonical equation for $\pi_n$ in the context of the large $N$ limit:
\begin{equation}\label{eq:pi-final}
	\begin{split}	
		\partial_0\pi_n
		&=
		\left(F+\frac{4}{3\pi}g^2\right)a(\frac{1}{a})^2(\phi_{n+1}-\phi_{n})
		-\left(F+\frac{4}{3\pi}g^2\right)a(\frac{1}{a})^2(\phi_{n}-\phi_{n-1})
		-m^2a\phi_n\\
		&-g\frac{1}{2}C\Bigg[
		\frac{\Psi^\dagger_u(n+1)\Psi_u(n)-\Psi^\dagger_u(n)\Psi_u(n-1)}{a}                            +\frac{\Psi^\dagger_u(n)\Psi_u(n+1)-\Psi^\dagger_u(n-1)\Psi_u(n)}{a} \\
		&
		+\frac{\Psi^\dagger_d(n+1)\Psi_d(n)-\Psi^\dagger_d(n)\Psi_d(n-1)}{a}
		+\frac{\Psi^\dagger_d(n)\Psi_d(n+1)-\Psi^\dagger_d(n-1)\Psi_d(n)}{a}
		\Bigg]+O(a^2)
		\; .
	\end{split}
\end{equation}
For the sake of simplicity in the equations, we omit the explicit notation for $\lim\limits_{N\to\infty}$ sometimes.

To establish the equation of motion for $\phi_n$, we require additional relationships between $\phi_n$ and $\pi_n$. Starting from the lattice Hamiltonian \eqref{eq:Hamiltonian} and applying the Heisenberg equations, we can obtain the time derivative of the operator $\phi_n$ as follows:
\begin{equation}\label{eq:phicanonical}
	\begin{split}		
		\partial_0\phi_n&=\frac{1}{i}[\phi_n,H]\\
		&=\frac{1}{F}\frac{1}{2}
		\Big[2\frac{\pi_n}{a}-\frac{1}{a}2g
		(\Psi^\dagger_u(n)\Psi_u(n)-\Psi^\dagger_d(n)\Psi_d(n) )
		\Big]\\
		&=\frac{1}{F}\frac{1}{a}
		\Big[\pi_n-g(\Psi^\dagger_u(n)\Psi_u(n)-\Psi^\dagger_d(n)\Psi_d(n) )
		\Big]
		\;.
	\end{split}
\end{equation}
Based on the expression given in \eqref{eq:phicanonical}, we can reconfigure the canonical momentum $\pi_n$ in relation to $\phi_n$ and $\Psi(n)$:
\begin{equation}\label{eq:phi1}
	\begin{split}	
		\pi_n=aF\partial_0\phi_n
		+g\left[\Psi^\dagger_u(n)\Psi_u(n)-\Psi^\dagger_d(n)\Psi_d(n)\right]
		\; .
	\end{split}
\end{equation}
In principle, we could substitute \eqref{eq:phi1} into \eqref{eq:pi-final} to derive the equation of motion for $\phi_n$. However, the operator products involving fermionic fields as shown in \eqref{eq:phi1} require careful evaluation. It would be convenient to calculate these products before directly substituting \eqref{eq:phi1} into \eqref{eq:pi-final}. Given the relationship between $\psi$ and $\Psi$ in \eqref{pP}, along with the properties of the operator products as stated in \eqref{eq:n(n+m)} and \eqref{eq:n(n+m)dd}, we can reformulate the operator products in \eqref{eq:phi1} as follows:
\begin{equation}\label{eq:uu-dd}
	\begin{split}	
		&\Psi^\dagger_u(n)\Psi_u(n)-\Psi^\dagger_d(n)\Psi_d(n)\\
		&=	\psi^\dagger_u(n)\psi_u(n)-\psi^\dagger_d(n)\psi_d(n)\\
		&=	:\psi^\dagger_u(n)\psi_u(n):-:\psi^\dagger_d(n)\psi_d(n):\\
		&=\frac{1}{2}\left[:\psi^\dagger_u(n+1)\psi_u(n):+:\psi^\dagger_u(n)\psi_u(n+1):        -:\psi^\dagger_d(n+1)\psi_d(n):-:\psi^\dagger_d(n)\psi_d(n+1):\right]+O(a^2)
		\;.
	\end{split}
\end{equation}
Here, we have employed the relation $:\psi^\dagger_u(n)\psi_u(n):=:\psi^\dagger_u(n+1)\psi_u(n):+O(a^2)=:\psi^\dagger_u(n)\psi_u(n+1):+O(a^2)$ to arrive at the final equality. Additionally, utilizing \eqref{eq:limuU}--\eqref{eq:limDd}, we can determine the large $N$ limit of \eqref{eq:uu-dd}:
\begin{equation}\label{eq:phi2}
	\begin{split}	
		&\lim\limits_{N\to\infty}\left[\Psi^\dagger_u(n)\Psi_u(n)-\Psi^\dagger_d(n)\Psi_d(n)\right]\\
		&=\lim\limits_{N\to\infty}\frac{1}{2}C\left[\Psi^\dagger_u(n+1)\Psi_u(n)+\Psi^\dagger_u(n)\Psi_u(n+1)-\Psi^\dagger_d(n+1)\Psi_d(n)-\Psi^\dagger_d(n)\Psi_d(n+1)\right]+O(a^2)
		\; .
	\end{split}
\end{equation}
By inserting \eqref{eq:phi2} into \eqref{eq:phi1}, we arrive at the expression for $\pi_n$ in the context of the large $N$ limit:
\begin{equation}\label{eq:phi3}
	\begin{split}	
		\pi_n=&aF\partial_0\phi_n\\
		&+g\frac{1}{2}C\left[\Psi^\dagger_u(n+1)\Psi_u(n)+\Psi^\dagger_u(n)\Psi_u(n+1)-\Psi^\dagger_d(n+1)\Psi_d(n)-\Psi^\dagger_d(n)\Psi_d(n+1)\right]+O(a^2)
		\; .
	\end{split}
\end{equation}
Substituting Eq. \eqref{eq:phi3} into Eq. \eqref{eq:pi-final}, we obtain the equations of motion for the bosonic field operator $\phi_n$ in the limit $N\to\infty$:
\begin{equation}
	\begin{split}	
		\label{eq:phi-final}
		&aF\partial^0\partial_0\phi_n
		-a\left(F+\frac{4}{3\pi}g^2\right)\frac{1}{a}\left(\frac{\phi_{n+1}-\phi_{n}}{a}
		-\frac{\phi_{n}-\phi_{n-1}}{a}\right)
		+am^2\phi_n\\
		=&-gC\frac{1}{2}\partial_0
		\left(\Psi^\dagger_u(n+1)\Psi_u(n)+\Psi^\dagger_u(n)\Psi_u(n+1)
		-\Psi^\dagger_d(n+1)\Psi_d(n)-\Psi^\dagger_d(n)\Psi_d(n+1) \right)\\
		&-gC\frac{1}{2}\Bigg(
		\frac{\Psi^\dagger_u(n+1)\Psi_u(n)-\Psi^\dagger_u(n)\Psi_u(n-1)}{a}                            +\frac{\Psi^\dagger_u(n)\Psi_u(n+1)-\Psi^\dagger_u(n-1)\Psi_u(n)}{a} \\
		&
		+\frac{\Psi^\dagger_d(n+1)\Psi_d(n)-\Psi^\dagger_d(n)\Psi_d(n-1)}{a}
		+\frac{\Psi^\dagger_d(n)\Psi_d(n+1)-\Psi^\dagger_d(n-1)\Psi_d(n)}{a}
		\Bigg)
		+O(a^2).
	\end{split}
\end{equation}

From Eq. \eqref{eq:psi(n)}, we know that the variable $\Psi[n]\equiv\frac{1}{\sqrt{a}}\Psi(n)$ corresponds to the continuous field $\Psi_{c}(x)$ at $x=na$ in the continuum limit. To compare the lattice equations of motion with those of the original RS model, we rewrite Eq. \eqref{eq:phi-final} in terms of $\Psi[n]$:
\begin{equation}
	\begin{split}	
		\label{eq:phi-final2}
		&F\partial^0\partial_0\phi_n
		-\left(F+\frac{4}{3\pi}g^2\right)\frac{1}{a}\left(\frac{\phi_{n+1}-\phi_{n}}{a}
		-\frac{\phi_{n}-\phi_{n-1}}{a}\right)
		+m^2\phi_n\\
		=&-gC\frac{1}{2}\partial_0
		\left(\Psi^\dagger_u[n+1]\Psi_u[n]+\Psi^\dagger_u[n]\Psi_u[n+1]
		-\Psi^\dagger_d[n+1]\Psi_d[n]-\Psi^\dagger_d[n]\Psi_d[n+1] \right)\\
		&-gC\frac{1}{2}\Bigg(
		\frac{\Psi^\dagger_u[n+1]\Psi_u[n]-\Psi^\dagger_u[n]\Psi_u[n-1]}{a}                            +\frac{\Psi^\dagger_u[n]\Psi_u[n+1]-\Psi^\dagger_u[n-1]\Psi_u[n]}{a} \\
		&
		+\frac{\Psi^\dagger_d[n+1]\Psi_d[n]-\Psi^\dagger_d[n]\Psi_d[n-1]}{a}
		+\frac{\Psi^\dagger_d[n]\Psi_d[n+1]-\Psi^\dagger_d[n-1]\Psi_d[n]}{a}
		\Bigg)
		+O(a).
	\end{split}
\end{equation}
The bare parameters in the above equation are not yet the bare parameters of the original RS model.
From Eq. \eqref{eq:lcs}, we know that if we want the continuum limit of the lattice bare bosonic field $\lim\limits_{a\to0}\lim\limits_{N\to\infty}\phi_n=\phi$ to exactly match the bare bosonic field $\phi_0$ of the RS model, i.e., $\phi=\phi_0$, then the relationship between the lattice bare parameters and the original RS model's bare parameters is: $F=1-\frac{g_0^2}{\pi}, m=m_0, g=g_0$. Therefore, Eq. \eqref{eq:phi-final2} can be expressed in terms of the original RS model's bare parameters as
\begin{equation}\label{eq:phi-final3}
	\begin{split}			
		&\left(1-\frac{g_0^2}{\pi}\right)\partial^0\partial_0\phi_n
		-\left(1+\frac{g_0^2}{3\pi}\right)\frac{1}{a}\left(\frac{\phi_{n+1}-\phi_{n}}{a}
		-\frac{\phi_{n}-\phi_{n-1}}{a}\right)
		+m_0^2\phi_n\\
		=&-g_0C\frac{1}{2}\partial_0
		\left(\Psi^\dagger_u[n+1]\Psi_u[n]+\Psi^\dagger_u[n]\Psi_u[n+1]
		-\Psi^\dagger_d[n+1]\Psi_d[n]-\Psi^\dagger_d[n]\Psi_d[n+1] \right)\\
		&-g_0C\frac{1}{2}\Bigg(
		\frac{\Psi^\dagger_u[n+1]\Psi_u[n]-\Psi^\dagger_u[n]\Psi_u[n-1]}{a}                            +\frac{\Psi^\dagger_u[n]\Psi_u[n+1]-\Psi^\dagger_u[n-1]\Psi_u[n]}{a} \\
		&
		+\frac{\Psi^\dagger_d[n+1]\Psi_d[n]-\Psi^\dagger_d[n]\Psi_d[n-1]}{a}
		+\frac{\Psi^\dagger_d[n]\Psi_d[n+1]-\Psi^\dagger_d[n-1]\Psi_d[n]}{a}
		\Bigg)\\
		&+\mathrm O(a)\;.
	\end{split}
\end{equation}
Let's further require that the continuum limit of the lattice bare fermionic field, denoted as $\Psi_c$, is related to the bare fermionic field $\Psi_0$ of the original RS model as $\Psi_c=\left(\lim\limits_{a\to0}\lim\limits_{N\to\infty}C^{-\frac{1}{2}}\right)\Psi_0$.  In this case, the continuous limit of $\Psi_0[n]\equiv C^{-\frac{1}{2}}\Psi[n]$ becomes the original RS model bare Fermi field $\Psi_0$. Therefore, we express the above equation in terms of $\Psi_0[n]$ as
\begin{equation}
	\begin{split}	
		\label{eq:phi-final4}
		&\partial^0\partial_0\phi_{n}
		-\frac{1}{a}\left(\frac{\phi_{n+1}-\phi_{n}}{a}
		-\frac{\phi_{n}-\phi_{n-1}}{a}\right)
		+m_0^2\phi_{n}\\
		=&-g_0\frac{1}{2}\partial_0
		\left(\Psi^\dagger_{0u}[n+1]\Psi_{0u}[n]+\Psi^\dagger_{0u}[n]\Psi_{0u}[n+1]
		-\Psi^\dagger_{0d}[n+1]\Psi_{0d}[n]-\Psi^\dagger_{0d}[n]\Psi_{0d}[n+1] \right)\\
		&-g_0\frac{1}{2}\Bigg(
		\frac{\Psi^\dagger_{0u}[n+1]\Psi_{0u}[n]-\Psi^\dagger_{0u}[n]\Psi_{0u}[n-1]}{a}                            +\frac{\Psi^\dagger_{0u}[n]\Psi_{0u}[n+1]-\Psi^\dagger_{0u}[n-1]\Psi_{0u}[n]}{a} \\
		&
		+\frac{\Psi^\dagger_{0d}[n+1]\Psi_{0d}[n]-\Psi^\dagger_{0d}[n]\Psi_{0d}[n-1]}{a}
		+\frac{\Psi^\dagger_{0d}[n]\Psi_{0d}[n+1]-\Psi^\dagger_{0d}[n-1]\Psi_{0d}[n]}{a}
		\Bigg)\\
		&+\frac{g_0^2}{\pi}\partial^0\partial_0\phi_{n}
		+\frac{g_0^2}{3\pi}\frac{1}{a}\left(\frac{\phi_{n+1}-\phi_{n}}{a}
		-\frac{\phi_{n}-\phi_{n-1}}{a}\right)
		+O(a)
		\; .
	\end{split}
\end{equation}
To correspond to the lattice regularization established in the time-slicing framework, we set the regularization parameters of the original RS model to be at equal time intervals, i.e., $\epsilon^0=0$ and $\epsilon^1=\epsilon$.
In this case, the coefficient in front of the term $\partial_1\partial_1\phi_0$ in the motion equation of the original RS model, given by \eqref{eq:phi0}, is zero.
By using the specific representation of the gamma matrices provided in \eqref{gamma0}, we can observe that the only difference between the lattice bosonic field's motion equation \eqref{eq:phi-final4} and the continuum limit of the original RS model's equation \eqref{eq:phi0} lies in the last term of \eqref{eq:phi-final4}. In the lattice equation, this term is given by $\lim\limits_{a\to0}\frac{g_0^2}{3\pi}\frac{1}{a}\left(\frac{\phi_{0 n+1}-\phi_{0 n}}{a}-\frac{\phi_{0 n}-\phi_{0 n-1}}{a}\right)
=\frac{g_0^2}{3\pi}\partial_1\partial_1\phi_0 $, from which we can clearly see that the coefficient in front of the $\partial_1\partial_1\phi_0$ term is $\frac{g_0^2}{3\pi}$. In contrast, the corresponding term in the original RS model’s equation \eqref{eq:phi0} has a coefficient of 0.

Let's discuss why the  continuum limit of the lattice motion equation \eqref{eq:phi-final4} includes an additional term $\frac{g_0^2}{3\pi}\partial_1\partial_1\phi_0$ compared to the original RS model's equation \eqref{eq:phi0}. The direct reason for this difference is that in \eqref{eq:phi-final}, the coefficient in front of $\frac{1}{a}\left(\frac{\phi_{0 n+1}-\phi_{0 n}}{a}-\frac{\phi_{0 n}-\phi_{0 n-1}}{a}\right)$ is given as $F+\frac{4}{3\pi}g^2$. If we were to change this coefficient to $F+\frac{g^2}{\pi}$, the continuum limit of the lattice motion equation would exactly match the original RS model's equation.
 To understand the origin of the coefficient $\frac{4}{3\pi}g^2$ in \eqref{eq:phi-final}, let's focus on the specific term in \eqref{eq:n(n+m)} and \eqref{eq:n(n+m)dd} that contributes to it:
$$f(m) \equiv \frac{1}{2}\Bigg[\frac{1}{2N}\cot\left[\left(m+\frac{1}{2}\right)\frac{\pi}{N}\right]
+\frac{1}{2N}\cot\left[\left(m-\frac{1}{2}\right)\frac{\pi}{N}\right]\Bigg]
\; ,$$
Recalling the procedure of taking the continuum limit in our calculations, we initially set $m=1$ and subsequently perform the continuum limit, yielding $\lim\limits_{a\to0}\lim\limits_{N\to\infty} f(m=1)=\frac{2}{3\pi}$, which leads to the coefficient $\frac{4}{3\pi}g^2$ in \eqref{eq:phi-final}. However, if we instead fix $\epsilon \equiv ma$ initially and subsequently take the continuum limit, we get
\begin{equation}\label{eq:fc}
	\begin{split}	
		f_c(\epsilon) &\equiv \lim\limits_{a\to0}\lim\limits_{N\to\infty} f(m)\\
		&= \lim\limits_{a\to0}\lim\limits_{N\to\infty}\frac{1}{2}\Bigg[\frac{1}{2N}\cot\left[\left(\frac{\epsilon}{a}+\frac{1}{2}\right)\frac{\pi}{N}\right]
		+\frac{1}{2N}\cot\left[\left(\frac{\epsilon}{a}-\frac{1}{2}\right)\frac{\pi}{N}\right]\Bigg]\\
		&= \lim\limits_{a\to0}\frac{1}{4\pi}
		\left(\frac{1}{\frac{\epsilon}{a}+\frac{1}{2}}
		+\frac{1}{\frac{\epsilon}{a}-\frac{1}{2}}\right)\\
		&= \lim\limits_{a\to0}\frac{1}{2\pi}\frac{a}{\epsilon}
		\; .
	\end{split}
\end{equation}
In this case, if we further require $m=1$, meaning $\epsilon=a$, then formally, the above expression becomes $\left[\lim\limits_{a\to0}\lim\limits_{N\to\infty} f(m)\right]\Big|_{m=1}=f_c(\epsilon=a)=\frac{1}{2\pi}$, rather than $\frac{2}{3\pi}$. This would cause the coefficient in front of $\frac{1}{a}\left(\frac{\phi_{0 n+1}-\phi_{0 n}}{a}-\frac{\phi_{0 n}-\phi_{0 n-1}}{a}\right)$ in equation \eqref{eq:phi-final} to change from $F+\frac{4}{3\pi}g^2$ to $F+\frac{g^2}{\pi}$, ultimately yielding the same motion equation as the original RS model.
However, when taking the continuum limit, we keep $\epsilon$ fixed, so $\epsilon=a$ is not a realistic scenario. A more rigorous derivation proceeds as follows.

Let's define the variable $x \equiv na$. Referring to the derivation of \eqref{eq:Psi(n)Psi(n+1)} and utilizing \eqref{eq:fc}, we can obtain the following result:
\begin{equation}\label{eq:fc1}
	\begin{split}	
		C'\Psi^\dagger_{cu}(x)\Psi_{cu}(x+\epsilon)
		&=\lim\limits_{a\to0}\frac{1}{a}\lim\limits_{N\to\infty}
		C'\Psi^\dagger_u(n)\Psi_u(n+m)\\       
		&=\lim\limits_{a\to0}\lim\limits_{N\to\infty}
		\left[\psi^\dagger_u(n)\psi_u(n+m)+\frac{1}{2\pi}g
		\frac{\phi_{n+m}-\phi_n}{\epsilon}+O(a^2)\right]\\
		&=\psi^\dagger_{cu}(x)\psi_{cu}(x+\epsilon)
		+\frac{1}{2\pi}g
		\frac{\phi(x+\epsilon)-\phi(x)}{\epsilon}
		\; ,
	\end{split}
\end{equation}
where $C'$ is defined as 
\begin{equation}\label{eq:1C'}
	\begin{split}
		C'=\lim\limits_{a\to0}\lim\limits_{N\to\infty}{\mathrm e}^{[g\phi^+_n,g\phi^-_n]-[g\phi^+_{n+m},g\phi^-_n]}
		=\lim\limits_{a\to0}\lim\limits_{N\to\infty} {\mathrm e}^{g^2[f(0)-f(m)]}
		=\lim\limits_{a\to0}\lim\limits_{N\to\infty}{\mathrm e}^{g^2[f(0)-f(\frac{\epsilon}{a})]}
		\; .
	\end{split}
\end{equation}
The subsequent step involves considering the limit as $\epsilon\to 0$. Utilizing \eqref{eq:fc1}, we obtain the regularization for the field operator product of $\Psi_c$ as
\begin{equation}\label{eq:fc2}
	\begin{split}	
		\lim\limits_{\epsilon\to0}C'\Psi^\dagger_{cu}(x)\Psi_{cu}(x+\epsilon)
		&=\lim\limits_{\epsilon\to0}\psi^\dagger_{cu}(x)\psi_{cu}(x+\epsilon)
		+\frac{1}{2\pi}g
		\lim\limits_{\epsilon\to0}\frac{\phi(x+\epsilon)-\phi(x)}{\epsilon}
		\; .
	\end{split}
\end{equation}
Likewise, for the $d$ components of the fermionic field, we also obtain
\begin{equation}\label{eq:fc3}
	\begin{split}	
		\lim\limits_{\epsilon\to0}C'\Psi^\dagger_{cd}(x)\Psi_{cd}(x+\epsilon)
		&=\lim\limits_{\epsilon\to0}\psi^\dagger_{cd}(x)\psi_{cd}(x+\epsilon)
		+\frac{1}{2\pi}g
		\lim\limits_{\epsilon\to0}\frac{\phi(x+\epsilon)-\phi(x)}{\epsilon}.
	\end{split}
\end{equation}
The sum of \eqref{eq:fc2} and \eqref{eq:fc3} is precisely the axial-vector current formula in the original RS model (see Equation (3.35) in \cite{Rothe1975}):
\begin{equation}\label{eq:fc0}
	\begin{split}	
		N \left[\bar\Psi_0(x)\gamma^5\gamma^\mu\Psi_0(x)\right]
		=j^\mu_{5f}(x)+\frac{g_0}{\pi}\partial^\mu\phi_0(x)
		\equiv j^\mu_5(x)
		\; ,
	\end{split}
\end{equation}
where $\mu=1$. 
The original RS model indeed used \eqref{eq:fc0} to derive the equations of motion \eqref{eq:phi0}, demonstrating that our lattice theory is consistent with the original RS model.


To further clarify the reason for the coefficient difference between the lattice motion equation and the original RS model motion equation, let's analyze the continuum limit of \eqref{eq:limuU}:
\begin{equation}\label{eq:fcuu}
	\begin{split}	
		\lim\limits_{a\to0}\frac{1}{a}\lim\limits_{N\to\infty}
		C\Psi^\dagger_u(n)\Psi_u(n+1)
		=\lim\limits_{a\to0}\frac{1}{a}\lim\limits_{N\to\infty}
		\psi^\dagger_u(n)\psi_u(n+1)
		+\frac{2}{3\pi}g
		\lim\limits_{a\to0}\lim\limits_{N\to\infty}\frac{\phi_{n+1}-\phi_n}{a}\; .
	\end{split}
\end{equation}
In Equation \eqref{eq:fcuu}, the coefficient in front of the coupling constant $g$ is $\frac{2}{3\pi}$, while in Equation \eqref{eq:fc2}, the coefficient in front of the coupling constant $g$ is $\frac{1}{2\pi}$. The reason for this difference in coefficients between the two equations lies in the order of taking limits. To be more specific, in deriving Equation \eqref{eq:fc2}, we took three limits in total, denoted as $\lim\limits_{\epsilon\to0}\lim\limits_{a\to0}\lim\limits_{N\to\infty}$. It's important to note that in this sequence, we first let $a$ tend to zero and then let $\epsilon$ tend to zero. However, if we require $\epsilon$ and $a$ to simultaneously approach zero, i.e., $\lim\limits_{\epsilon=a\to0}\lim\limits_{N\to\infty}$, we will obtain Equation \eqref{eq:fcuu}, and the coefficient in front of the coupling constant $g$ will change from $\frac{1}{2\pi}$ to $\frac{2}{3\pi}$.
This leads to an additional term in the continuum limit of the lattice motion equation \eqref{eq:phi-final4} compared to the original RS model's motion equation \eqref{eq:phi0}. This additional term is precisely the one mentioned repeatedly before, i.e., $\lim\limits_{a\to0}\frac{g_0^2}{3\pi}\frac{1}{a}\left(\frac{\phi_{0 n+1}-\phi_{0 n}}{a}-\frac{\phi_{0 n}-\phi_{0 n-1}}{a}\right)
=\frac{g_0^2}{3\pi}\partial_1\partial_1\phi_0 $.

Physically, this difference fundamentally originates from the lattice's ultraviolet behavior.
After taking the continuum limit $\lim\limits_{a\to0}\lim\limits_{N\to\infty}$ of the lattice theory, we obtain a continuous theory, which captures the infrared behavior of the lattice. Roughly speaking this is because, for any non-zero field separation $\epsilon$, the distance between $\Psi^\dagger_{cu}(x)$ and $\Psi_{cu}(x+\epsilon)$ already encompasses infinitely many lattice sites in the continuum limit, and the lattice's ultraviolet behavior has been eliminated in this limit, leaving only the infrared behavior. In this context, even as the field separation $\epsilon$ tends to zero, it does not touch upon the lattice's ultraviolet behavior.
However, if we set the field spacing $\epsilon=a$ while the continuum limit $\lim\limits_{a\to0}$ is not yet complete, it means that even later, no matter how small the lattice spacing becomes, the two fields will always be on adjacent lattice points. This implies that the lattice ultraviolet behavior continues to affect the system, even after taking the ``continuum limit". 
The lattice theory's motion equation incorporates lattice ultraviolet behavior that the original RS model's motion equation lacks, which is why there are slight differences in their equations of motion.
However, this does not imply that the continuum limit of the lattice theory is not the original RS model. 
By comparing \eqref{eq:fc2}, \eqref{eq:fc3}, and \eqref{eq:fc0}, we observe that the lattice theory aligns with the continuous field theory of the RS model if we first take the continuum limit and subsequently let the interval $\epsilon$ vanish.

Lastly, it is important to highlight that we've established relationships between the bare field operators on the lattice in the continuum limit and the bare field operators within the continuous field theory. Additionally, we have derived connections between the bare parameters of these two theories:
\begin{equation}\label{eq:bare}
	\begin{split}	
		\phi=\phi_0 &\quad , \qquad
		\Psi_c=\left[\lim\limits_{a\to0}\lim\limits_{N\to\infty}C^{-\frac{1}{2}}\right]\Psi_0\; ,\\
		m=m_0 &\quad,\qquad
		g=g_0\quad,\qquad
		F=1-\frac{g_0^2}{\pi}
		\; ,
	\end{split}
\end{equation}
where the continuum limit for the factor $C$ can be obtained using equation \eqref{eq:f-f}: $\lim\limits_{a\to0}\lim\limits_{N\to\infty}C=\lim\limits_{a\to0}\lim\limits_{N\to\infty}{\mathrm e}^{g^2[f(0)-f(1)]}= {\mathrm e}^{g^2\frac{1}{F\pi}}$.

\section{The Equation of Motion For the fermionic field}
\label{eqomF}
Before deriving the equation of motion for $\Psi(n)$, let's first calculate a useful commutation relation.
By using \eqref{eq:phi}, we can express the time derivative of the field $\phi_n$ in terms of creation and annihilation operators as follows:
\begin{equation}
	\begin{split}	
		\partial_0\phi_n(t)
		=F^{-\frac{1}{2}}\sum_q i \sqrt{\frac{\omega_q}{2L}}
		(a^\dagger_q {\mathrm e}^{i\omega_q t-inaq}-a_q {\mathrm e}^{-i\omega_q t+inaq})    
		\;.
	\end{split}
\end{equation}
Further utilizing the commutation relations of creation and annihilation operators, we can derive the following commutation relation:
\begin{equation}
	\begin{split}	
		\label{eq:f[partial phi]}
		\left[\phi^+_n,\partial_0\phi^-_m\right]
		&=F^{-1}i\sum_{k=-\frac{N-1}{2}}^{\frac{N-1}{2}}  \frac{1}{2L} {\mathrm e}^{i(n-m)\frac{2\pi k}{N}}
		=i\frac{1}{2aF}\delta_{nm}
		\; .
	\end{split}
\end{equation}
This commutation relation will be used repeatedly in the subsequent derivations.

In a manner analogous to the approach taken for the bosonic field's equation of motion, we will follow a similar procedure to derive the equation of motion for the fermionic field. We begin with the lattice Hamiltonian \eqref{eq:Hamiltonian} and derive the canonical equation for the operator $\Psi(n)$ using the Heisenberg equation:
\begin{equation}
	\begin{split}	
		\label{eq:psi1}
		i\partial_0\Psi_u(n)=&[\Psi_u(n),H]\\
		=&-i\frac{1}{2}\Bigg\{ 
		\Big[\Psi_u(n+1)-\Psi_u(n) \Big] \frac{1}{a}
		+\Big[\Psi_u(n)-\Psi_u(n-1) \Big] \frac{1}{a}\Bigg\}\\
		& +\frac{1}{2F}
		\Bigg\{-\frac{1}{a}2g \Psi_u(n)\pi_n\\
		&+\frac{1}{a}g^2\Big[
		\Psi_u(n)(\Psi_u^\dagger(n)\Psi_u(n)-\Psi_d^\dagger(n)\Psi_d(n) )
		+
		(\Psi_u^\dagger(n)\Psi_u(n)-\Psi_d^\dagger(n)\Psi_d(n) )\Psi_u(n)\Big]	
		\Bigg\}\\
		&+\frac{1}{a}\frac{1}{2}i \Bigg\{ 
		\Big[{\mathrm e}^{ig\phi_n-ig\phi_{n-1}}-1\Big]\Psi_u(n-1)  -\Big[{\mathrm e}^{ig\phi_n-ig\phi_{n+1}}-1\Big]\Psi_u(n+1)\Bigg\} \\
		&
		+\frac{1}{a}\frac{1}{2}i\Bigg\{
		{\mathrm e}^{ig\phi_{n}+ig\phi_{n-1}}\Psi_d(n-1) +{\mathrm e}^{ig\phi_n+ig\phi_{n+1}}\Psi_d(n+1)
		-2{\mathrm e}^{2ig\phi_n}\Psi_d(n)\Bigg\}
		\; .
	\end{split}
\end{equation}
Let's analyze the contents of the last three sets of curly braces in the above expression one by one.

First, consider the terms within the curly brace after the coefficient $\frac{1}{F}$ in \eqref{eq:psi1}. Utilizing the relationship between $\pi_n$ and $\partial_0\phi$ as given in \eqref{eq:phi1}, we can express them as follows:
\begin{equation}
	\begin{split}	
		\label{eq:1/F}
		-\frac{1}{a}&2g \Psi_u(n)\pi_n
		+\frac{1}{a}g^2\Big(
		\Psi_u(n)(\Psi_u^\dagger(n)\Psi_u(n)-\Psi_d^\dagger(n)\Psi_d(n) )
		+
		(\Psi_u^\dagger(n)\Psi_u(n)-\Psi_d^\dagger(n)\Psi_d(n) )\Psi_u(n)\Big)	\\	
		=&-2gF \Psi_u(n)\partial_0\phi_n-2gF \partial_0\phi_n\Psi_u(n)\\
		=&-2gF \Psi_u(n)\partial_0\phi^+_n-2gF \partial_0\phi^-_n\Psi_u(n)
		\;,
	\end{split}
\end{equation}
where the last equality is based on \eqref{pP} and the commutation relation \eqref{eq:f[partial phi]}.

Next, let's examine the terms inside the penultimate curly brace in \eqref{eq:psi1}. According to \eqref{pP}, we can calculate that the first term inside the penultimate curly brace in \eqref{eq:psi1} is
\begin{equation}\label{eq:e1}
	\begin{split}			
		&\left({\mathrm e}^{ig\phi_n-ig\phi_{n-1}}-1\right)\Psi_u(n-1) \\ 
		&=K
		{\mathrm e}^{ig\phi^-_n-ig\phi^-_{n-1}}
		{\mathrm e}^{ig\phi^-_{n-1}}	{\mathrm e}^{ig\phi^+_{n-1}}\psi_u(n-1)
		{\mathrm e}^{ig\phi^+_n-ig\phi^+_{n-1}}
		-K{\mathrm e}^{ig\phi^-_{n-1}}	{\mathrm e}^{ig\phi^+_{n-1}}\psi_u(n-1)\\
		&=K\left[(ig\phi^-_n-ig\phi^-_{n-1}){\mathrm e}^{ig\phi^-_{n-1}}	{\mathrm e}^{ig\phi^+_{n-1}}\psi_u(n-1)
		+{\mathrm e}^{ig\phi^-_{n-1}}	{\mathrm e}^{ig\phi^+_{n-1}}\psi_u(n-1)(ig\phi^+_n-ig\phi^+_{n-1})
		+\mathrm O(a^\frac{5}{2})\right]\\
		&=K\left[(ig\phi^-_n-ig\phi^-_{n-1}){\mathrm e}^{ig\phi^-_{n}}	{\mathrm e}^{ig\phi^+_{n}}\psi_u(n)
		+{\mathrm e}^{ig\phi^-_{n}}	{\mathrm e}^{ig\phi^+_{n}}\psi_u(n)(ig\phi^+_n-ig\phi^+_{n-1})
		+\mathrm O(a^\frac{5}{2})\right]\\
		&=
		ig(\phi^-_n-\phi^-_{n-1})\Psi_u(n)
		+ig\Psi_u(n)(\phi^+_n-\phi^+_{n-1})
		+K\mathrm O(a^\frac{5}{2})
		\; ,
	\end{split}
\end{equation}
where $K\equiv {\mathrm e}^{-\frac{1}{2}[\phi_n^+,\phi_n^-]}={\mathrm e}^{-\frac{1}{2}f(0)}$. 
Similarly, we can compute the other term inside the penultimate curly brace in \eqref{eq:psi1}:
\begin{equation}\label{eq:e2}
	\begin{split}			
		&\left({\mathrm e}^{ig\phi_n-ig\phi_{n+1}}-1\right)\Psi_u(n+1) \\ 
		&=K
		{\mathrm e}^{ig\phi^-_n-ig\phi^-_{n+1}}
		{\mathrm e}^{ig\phi^-_{n+1}}	{\mathrm e}^{ig\phi^+_{n+1}}\psi_u(n+1)
		{\mathrm e}^{ig\phi^+_n-ig\phi^+_{n+1}}
		-K{\mathrm e}^{ig\phi^-_{n+1}}	{\mathrm e}^{ig\phi^+_{n+1}}\psi_u(n+1)\\
		&=K\left[(ig\phi^-_n-ig\phi^-_{n+1}){\mathrm e}^{ig\phi^-_{n+1}}	{\mathrm e}^{ig\phi^+_{n+1}}\psi_u(n+1)
		+{\mathrm e}^{ig\phi^-_{n+1}}	{\mathrm e}^{ig\phi^+_{n+1}}\psi_u(n+1)(ig\phi^+_n-ig\phi^+_{n+1})
		+\mathrm O(a^\frac{5}{2})\right]\\
		&=K\left[(ig\phi^-_n-ig\phi^-_{n+1}){\mathrm e}^{ig\phi^-_{n}}	{\mathrm e}^{ig\phi^+_{n}}\psi_u(n)
		+{\mathrm e}^{ig\phi^-_{n}}	{\mathrm e}^{ig\phi^+_{n}}\psi_u(n)(ig\phi^+_n-ig\phi^+_{n+1})
		+\mathrm O(a^\frac{5}{2})\right]\\
		&=K\Bigg\{\left[(ig\phi^-_n-ig\phi^-_{n+1})-(ig\phi^-_{n-1}-ig\phi^-_{n})\right]{\mathrm e}^{ig\phi^-_{n}}	{\mathrm e}^{ig\phi^+_{n}}\psi_u(n)\\
		&\qquad\quad
		+{\mathrm e}^{ig\phi^-_{n}}	{\mathrm e}^{ig\phi^+_{n}}\psi_u(n)\left[(ig\phi^+_n-ig\phi^+_{n+1})-(ig\phi^+_{n-1}-ig\phi^+_{n})\right]
		\\&
		\qquad\quad
		+(ig\phi^-_{n-1}-ig\phi^-_{n}){\mathrm e}^{ig\phi^-_{n}}	{\mathrm e}^{ig\phi^+_{n}}\psi_u(n)
		+{\mathrm e}^{ig\phi^-_{n}}	{\mathrm e}^{ig\phi^+_{n}}\psi_u(n)(ig\phi^+_{n-1}-ig\phi^+_{n})
		+\mathrm O(a^\frac{5}{2})\Bigg\}\\
		&=K\left[(ig\phi^-_{n-1}-ig\phi^-_{n}){\mathrm e}^{ig\phi^-_{n}}	{\mathrm e}^{ig\phi^+_{n}}\psi_u(n)
		+{\mathrm e}^{ig\phi^-_{n}}	{\mathrm e}^{ig\phi^+_{n}}\psi_u(n)(ig\phi^+_{n-1}-ig\phi^+_{n})
		+\mathrm O(a^\frac{5}{2})\right]\\
		&=
		ig(\phi^-_{n-1}-\phi^-_{n})\Psi_u(n)
		+ig\Psi_u(n)(\phi^+_{n-1}-\phi^+_{n})
		+K\mathrm O(a^\frac{5}{2})
		\; .
	\end{split}
\end{equation}

Finally, analyzing the term inside the last curly brace in \eqref{eq:psi1}, after some straightforward calculations, we obtain
\begin{equation}
	\begin{split}	
		\label{eq:last row}
		&{\mathrm e}^{ig\phi_{n}+ig\phi_{n-1}}\Psi_d(n-1)+{\mathrm e}^{ig\phi_n+ig\phi_{n+1}}\Psi_d(n+1)
		-2{\mathrm e}^{2ig\phi_n}\Psi_d(n)\\
		&={\mathrm e}^{ig\phi_{n}}\psi_d(n-1)+{\mathrm e}^{ig\phi_n}\psi_d(n+1)-2{\mathrm e}^{ig\phi_n}\psi_d(n)\\
		&={\mathrm e}^{ig\phi_{n}}
		\Big[\Big(\psi_d(n+1)-\psi_d(n)\Big)-\Big(\psi_d(n)-\psi_d(n-1)\Big)\Big]\\
		&=K\mathrm O(a^\frac{5}{2})
		\; .
	\end{split}
\end{equation}

Substituting the results from \eqref{eq:1/F}, \eqref{eq:e1}, \eqref{eq:e2}, and \eqref{eq:last row} into equation \eqref{eq:psi1} yields the new form of the equation of motion for $\Psi$:
\begin{equation}
	\begin{split}	
		\label{eq:psi2}
		i\partial_0\Psi_u(n)
		&=-i\frac{1}{2}\Bigg\{ 
		\Big[\Psi_u(n+1)-\Psi_u(n) \Big] \frac{1}{a}
		+\Big[\Psi_u(n)-\Psi_u(n-1) \Big] \frac{1}{a}\Bigg\}\\
		& -g\Psi_u(n)\partial_0\phi^+_n-g\partial_0\phi^-_n\Psi_u(n)\\
		&- g(\phi^-_n-\phi^-_{n-1})\frac{1}{a} \Psi_u(n)
		-g\Psi_u(n)(\phi^+_n-\phi^+_{n-1})\frac{1}{a}
		+K\mathrm O(a^\frac{3}{2})
		\; .
	\end{split}
\end{equation}
The equation \eqref{eq:psi2} corresponds precisely to the equation of motion shown in the original RS model reference \cite{Rothe1975}, represented as (3.24a):
\begin{equation}\label{eq:opsi1}
	\begin{split}	
		i\gamma\cdot\partial\Psi_0(x)
		=g_0\gamma^5\gamma^\mu  N\left[\partial_\mu\phi_0(x)	\Psi_0(x)\right]
		\; ,
	\end{split}
\end{equation}
where the regularization $N[\cdots]$ corresponds to equation (3.24b) in the original RS model reference \cite{Rothe1975}:
\begin{equation}\label{eq:opsi2}
	\begin{split}	
		N\left[\partial_\mu\phi_0(x)	\Psi_0(x)\right]
		\equiv\partial_\mu\phi_0^-(x)	\Psi_0(x)+	\Psi_0(x)\partial_\mu\phi_0^+(x)
		\; .
	\end{split}
\end{equation}

However, the above equation is expressed in terms of the positive and negative frequency components of $\phi$. To represent the equation of motion \eqref{eq:psi2} using the field $\phi$ itself, further analysis is required. Utilizing \eqref{[phi]}, it can be easily calculated that the commutation relation $\left[\phi_n^-,{\mathrm e}^{ig\phi_m}\right]=-ig{\mathrm e}^{ig\phi_m}f(n-m)$ holds. Using this commutation relation, we can rewrite the first term in the last line of equation \eqref{eq:psi2} as
\begin{equation}
	\begin{split}	
		\label{eq:e3}
		(\phi^-_n-\phi^-_{n-1})\Psi_u(n)
		&=\Psi_u(n)(\phi^-_n-\phi^-_{n-1})	
		-ig[f(0)-f(1)]\Psi_u(n)
		\; .
	\end{split}
\end{equation}
On the other hand, since the commutation relation \eqref{eq:f[partial phi]}, we also have $\left[\partial_0\phi_n^-,{\mathrm e}^{ig\phi_m^+}\right]=\frac{g}{2aF}{\mathrm e}^{ig\phi_m^+}\delta_{nm}$. With this result, we can express the penultimate line of \eqref{eq:psi2} as follows:
%
\begin{equation}
	\begin{split}	
		\label{eq:1/F2}
		&\Psi_u(n)\partial_0\phi^+_n+\partial_0\phi^-_n\Psi_u(n)\\
		=&\Psi_u(n)\partial_0\phi^+_{n-1}
		+\Psi_u(n)\Big[\partial_0\phi^+_n-\partial_0\phi^+_{n-1}\Big]
		+\partial_0\phi^-_{n-1}\Psi_u(n)
		+\Big[\partial_0\phi^-_n-\partial_0\phi^-_{n-1}\Big]\Psi_u(n)\\
		=&\Psi_u(n)\partial_0\phi^+_{n-1}
		+\partial_0\phi^-_{n-1}\Psi_u(n)+KO(a^\frac{3}{2})\\
		=&\Psi_u(n)\partial_0\phi_{n-1}+KO(a^\frac{3}{2})
		\; .
	\end{split}
\end{equation}
By substituting \eqref{eq:e3} and \eqref{eq:1/F2} into \eqref{eq:psi2}, we can derive the equation of motion for $\Psi$ in terms of the original fundamental degrees of freedom as follows:
\begin{equation}
	\begin{split}	
		\label{eq:psiu-final}
		i\partial_0\Psi_u(n)
		&=-i\frac{1}{2}\Bigg\{ 
		\Big[\Psi_u(n+1)-\Psi_u(n) \Big] \frac{1}{a}
		+\Big[\Psi_u(n)-\Psi_u(n-1) \Big] \frac{1}{a}\Bigg\}\\
		& -g\Psi_u(n)\partial_0\phi_{n-1}
		-g\Psi_u(n)(\phi_n-\phi_{n-1})\frac{1}{a}
		+ig^2[f(0)-f(1)]\frac{1}{a}\Psi_u(n)
		+KO(a^\frac{3}{2})
		\; .
	\end{split}
\end{equation}
In a similar manner, we can derive the equation of motion for the other component $\Psi_d$ of the fermionic field:
\begin{equation}
	\begin{split}	
		\label{eq:psid-final}
		i\partial_0\Psi_d(n)
		&=i\frac{1}{2}\Bigg\{ 
		\Big[\Psi_d(n+1)-\Psi_d(n) \Big] \frac{1}{a}
		+\Big[\Psi_d(n)-\Psi_d(n-1) \Big] \frac{1}{a}\Bigg\}\\
		& +g\Psi_d(n)\partial_0\phi_{n-1}
		-g\Psi_d(n)(\phi_n-\phi_{n-1})\frac{1}{a}
		+ig^2[f(0)-f(1)]\frac{1}{a}\Psi_d(n)
		+KO(a^\frac{3}{2})
		\; .
	\end{split}
\end{equation}

Using \eqref{eq:f-f} and the specific expression of the $\gamma$ matrices in \eqref{gamma0}, we can express Eq. \eqref{eq:psiu-final} and \eqref{eq:psid-final} in a unified form as follows:
\begin{equation}
	\begin{split}	
		\label{fpsi-final}
		&i\gamma^0\partial_0\Psi(n)
		+i\gamma^1
		\frac{1}{2}\Bigg\{ 
		\Big[\Psi(n+1)-\Psi(n) \Big] \frac{1}{a}
		+\Big[\Psi(n)-\Psi(n-1) \Big] \frac{1}{a}\Bigg\}\\
		&=
		g\gamma^5\gamma^0\Psi(n)\partial_0\phi_{n-1}
		+g\gamma^5\gamma^1\Psi(n)(\phi_n-\phi_{n-1})\frac{1}{a}
		-ig^2\frac{1}{F\pi}\gamma^5\frac{a\gamma_1}{-a^2}\Psi(n)
		+KO(a^\frac{3}{2})
		\; .
	\end{split}
\end{equation}

Letting the lattice spacing $\epsilon=a$ in Eq. \eqref{czh0}, and based on Eq \eqref{eq:bare}, \eqref{[phi]}, \eqref{eq:phi+-}, and the definition of $K$ as $K \equiv {\mathrm e}^{-\frac{1}{2}g^2f(0)}$, we find that $\Psi[n]_R \equiv \frac{1}{\sqrt a}{\mathrm e}^{\frac{1}{2}g^2f(0)}\Psi(n) = K^{-1}\frac{1}{\sqrt a}\Psi(n)$ has the continuum limit of the renormalized continuous fermionic field $\Psi_r(x)$.
On the other hand, the renormalized lattice fermionic field given in Section \ref{5} is $\Psi(n)_R ={\mathrm e}^{\frac{1}{2}g^2f(0)}\Psi(n)=K^{-1}\Psi(n)$, which exactly matches $\Psi[n]_R = \frac{1}{\sqrt a}\Psi(n)_R$. This implies that the fact that the continuum limit of $\Psi[n]_R$ is $\Psi_r(x)$ is consistent with the correspondence between lattice and continuous fermionic fields discussed in Section \ref{5}.
Based on the relation between renormalized parameters and bare parameters \eqref{rb}, we can rewrite Eq. \eqref{fpsi-final} as
%
%
%
\begin{equation}
	\begin{split}	
		\label{fpsi-final2}
		&\text{i}\gamma^0\partial_0\Psi[n]_R
		+\text{i}\gamma^1
		\frac{1}{2}\Bigg\{ 
		\Big[\Psi[n+1]_R-\Psi[n]_R \Big] \frac{1}{a}
		+\Big[\Psi[n]_R-\Psi[n-1]_R \Big] \frac{1}{a}\Bigg\}\\
		&=
		g_r\gamma^5\left[\gamma^0\Psi[n]_R\partial_0\phi_{n-1}
		+\gamma^1\Psi[n]_R(\phi_n-\phi_{n-1})\frac{1}{a}
		-\text{i}\frac{g_r}{\pi}\frac{a\gamma_1}{-a^2}\Psi[n]_R
		\right]
		+\mathrm O(a)
		\; .
	\end{split}
\end{equation}
Comparing the continuum limit of \eqref{fpsi-final2} with the fermionic field equation of the original RS model \eqref{eq:psir} (with regularization parameters set as equal time intervals $\epsilon^0=0,\epsilon^1=\epsilon$), we notice that the two equations are nearly identical. The only difference lies in the coefficient of the last term: it is $\frac{g_r}{\pi}$ in Eq. \eqref{fpsi-final2}, whereas in Eq. \eqref{eq:psir}, it's $\frac{g_r}{2\pi}$.

We've previously analyzed the reasons for the differences between the lattice bosonic field equation and the original RS model equation. Here, the same reasons lead to slight differences in the coefficients of the lattice fermionic field equation compared to the original RS model equation. Specifically, the mathematical origin of these differences lies in the distinct orders of taking limits. The correct limit to obtain the original RS equation \eqref{eq:psir} can be effectively thought of as $\lim\limits_{\epsilon\to0}\lim\limits_{a\to0}\lim\limits_{N\to\infty}$, whereas the continuum limit of the lattice equation \eqref{fpsi-final2} is $\lim\limits_{\epsilon=a\to0}\lim\limits_{N\to\infty}$.
From a physical perspective, the limit $\lim\limits_{\epsilon=a\to0}\lim\limits_{N\to\infty}$ makes \eqref{fpsi-final2} exhibit the lattice's ultraviolet behavior. Therefore, \eqref{fpsi-final2} differs slightly from \eqref{eq:psir}, which only contains lattice infrared behavior. However, this does not imply that the continuum limit of the lattice theory is not the original RS model. It's worth noting that the alternative form of the lattice fermionic field equation \eqref{eq:psi2} doesn't directly manifest the lattice's ultraviolet behavior, and its continuum limit is consistent with the original RS model's fermionic field equations \eqref{eq:opsi1} and \eqref{eq:opsi2} (see Eq. (3.24) in \cite{Rothe1975}).

\newpage

\bibliographystyle{JHEP}
\bibliography{RS9}

\end{document}